\documentclass[manuscript]{emulateapj}
\usepackage{natbib}
\usepackage{graphicx}
\newcommand{\um}{~$\mu$m}

\newcommand{\hubble}{$H_0=72$ km s$^{-1}$ Mpc$^{-1}$}
\newcommand{\etal}{{\em et al.}}
\bibpunct{(}{)}{,}{a}{}{}

\shorttitle{SIGS Paper I} 
\shortauthors{Lanz et al.}

\begin{document}

\title{Global Star Formation Rates and Dust Emission Over the Galaxy Interaction Sequence }
\author{Lauranne Lanz\altaffilmark{1},  
Andreas Zezas\altaffilmark{1,2,3}, Nicola Brassington\altaffilmark{4}, Howard A. Smith\altaffilmark{1}, 
Matthew L. N. Ashby\altaffilmark{1},
  Elisabete da Cunha\altaffilmark{5}, Giovanni G. Fazio\altaffilmark{1},
 Christopher C. Hayward\altaffilmark{6}, Lars Hernquist\altaffilmark{1},
Patrik Jonsson\altaffilmark{7}}

\altaffiltext{1}{Harvard-Smithsonian Center for Astrophysics, 60 Garden St., 
Cambridge, MA 02138, USA; llanz@head.cfa.harvard.edu}
\altaffiltext{2}{University of Crete, Physics Department \& Institute of Theoretical \& Computational Physics, 71003 Heraklion, Crete, Greece}
\altaffiltext{3}{Foundation for Research and Technology-Hellas, 71110 Heraklion, Crete, Greece}
\altaffiltext{4}{School of Physics, Astronomy and Mathematics, University of Hertfordshire, College Lane, Hatfield, AL10 9AB, UK}
\altaffiltext{5}{Max Planck Institute for Astronomy (MPIA), K\"{o}nigstuhl 17, 69117, Heidelberg, Germany}
\altaffiltext{6}{Heidelberger Institut f\"{u}r Theoretische Studien, Schloss-Wolfsbrunnenweg 35, 69118, Heidelberg, Germany}
\altaffiltext{7}{Space Exploration Technologies, 1 Rocket Road, Hawthorne, CA, 90250}

\begin{abstract}
We measured and modeled the spectral energy distributions (SEDs) in 28 bands from the ultraviolet 
to the far-infrared (FIR) for 31 interacting galaxies in 14 systems. The sample is drawn from the \emph{Spitzer}
Interacting Galaxy Survey, which probes a range of galaxy interaction parameters at multiple wavelengths 
with an emphasis on the infrared bands. The subset presented in this paper consists of all galaxies for 
which FIR \emph{Herschel} SPIRE observations are publicly available. Our SEDs combine the \emph{Herschel} 
photometry with multi-wavelength data from \emph{Spitzer},  \emph{GALEX},  \emph{Swift} UVOT, and 2MASS. 
While the shapes of the SEDs are broadly similar across our sample, strongly interacting galaxies typically 
have more mid-infrared emission relative to their near-infrared and FIR emission than weakly or moderately 
interacting galaxies. We modeled the full SEDs to derive host galaxy star formation rates (SFR), specific star 
formation rates (sSFR), stellar masses, dust temperatures, dust luminosities, and dust masses. We find 
increases in the dust luminosity and mass, SFR, and cold (15-25 K) dust temperature as the interaction progresses 
from \emph{moderately} to \emph{strongly} interacting and between \emph{non-interacting} and \emph{strongly}
 interacting galaxies. We also find  increases in the SFR between \emph{weakly} and \emph{strongly} interacting galaxies. In contrast, the 
sSFR remains unchanged across all the interaction stages. The ultraviolet photometry is crucial for 
constraining the age of the stellar population and the SFR, while dust mass is primarily 
determined by SPIRE photometry. The SFR derived from the SED modeling agrees well with rates 
estimated by proportionality relations that depend on infrared emission. 

\end{abstract}


\section{INTRODUCTION}
Galaxy evolution is believed to be heavily influenced by interactions between galaxies, 
both for local systems and for distant objects at earlier cosmological times. In the canonical view, 
interactions between galaxies have three primary observable effects. In the most dramatic cases, 
interactions stimulate star formation in a burst of activity that is presumed to power the high infrared 
(IR) luminosities typically seen in such systems.  Many local ultra-luminous IR galaxies  (L $\ge 
10^{12}$~L$_{\odot}$; ULIRGs) and luminous IR galaxies ($10^{11} $~L$_{\odot}\le$ L $\le 
10^{12}$~L$_{\odot}$; LIRGs) show evidence of galaxy interactions \citep[e.g.,][]{vei02}. Similarly, 
their high-redshift counterparts, sub-millimeter galaxies, first detected by SCUBA and now studied 
extensively by the Spectral and Photometric Imaging Receiver (SPIRE) on the \emph{Herschel 
Space Observatory}\footnote{\emph{Herschel} is an ESA space observatory with science instruments 
provided by European-led Principal Investigator consortia and with important participation from NASA.}, 
are thought to be predominantly mergers \citep[e.g.,][]{bla99}, although the relative contribution 
of mergers of different stages to their numbers is still an open question (e.g., Hayward \etal~2012a, 2012b). 

The second effect is that interactions significantly affect the subsequent evolution of 
 galaxies, which may lead to significant changes in their morphology. Disturbed galaxies have 
long been associated with mergers \citep[e.g.,][]{too72}. Numerical simulations of interactions 
(e.g., Hopkins \etal~2006; Hopkins 2012; Mihos \& Hernquist 1994, 1996; Barnes 1992; 
Barnes \& Hernquist 1996; Sanders 1999) show a variety of morphological distortions as well 
as variable amounts of star formation. The simulations also demonstrate the complexity of the 
problem: the degree of induced activity and distortion varies greatly with the parameters of the 
encounter, the phase of the interactions, the molecular gas content (``wetness"), and the mass 
of the progenitor galaxies among many other properties.

Third, the canonical picture, as seen in many simulations \citep[e.g.,][]{dim05, spr05b}, involves 
merger-driven gas inflow to the central regions, resulting in heightened activity of the central 
supermassive black hole as well as starburst activity due to the increased central gas density and 
possibly turbulence.  The process in principle converts a low-luminosity nucleus into an active 
galactic nucleus (AGN) but one whose luminosity might range widely depending on the stage of 
the interaction. Indeed, observations of merging galaxies over the years have tended to provide 
evidence supporting the conclusion that, at least on a statistical level, interactions trigger an 
enhancement in the formation of stars as well as nuclear activity. However, the recent literature 
includes works that argue both for and against a strong connection between nuclear activity 
and mergers \citep[e.g.,][]{li08, koc12, ell11, scu12, silv11}.

Therefore, despite many previous studies \citep[e.g.,][]{dah85,san88, kew01, lam03}, both observational 
and through simulations, our understanding of the evolution of the physical activity during the 
course of a galaxy-galaxy interaction remains incomplete.  In the past decade, two new 
developments have dramatically changed our understanding of star formation and accretion 
activity around galactic nuclei, which are the two dominant processes at work in controlling 
the observed emission. The first is the success of space missions, in particular, the \emph{Spitzer 
Space Telescope}\footnote{\emph{Spitzer} is operated by the Jet Propulsion Laboratory, 
California Institute of Technology under a contract with NASA.} \citep{wer04} and the \emph{Herschel 
Space Observatory} \citep{pil10} in the IR, as well as the \emph{Galaxy Evolution 
Explorer}\footnote{\emph{GALEX} is operated for NASA by the California Institute of Technology 
under NASA contract NAS5-98034.} (\emph{GALEX}; Martin \etal~2005) and \emph{Swift} 
\citep{geh04} in the ultraviolet (UV), providing photometry across the complete spectral range 
from UV to millimeter wavelengths. Most importantly, detailed imaging and high sensitivity 
photometry now available at the critical far-infrared (FIR) emission peak resulting from warm, 
luminous dust heated by starbursts provides crucial information regarding dust heating and 
embedded star formation. The combination of UV and IR observations is essential to obtain 
a complete census of recent and ongoing star formation by capturing both the unobscured 
and obscured emission from young stars. 

The second development has been the success of computational codes. We have new tools for 
the derivation of galaxy properties including masses, star formation rates, and interstellar 
medium (ISM) parameters from global fits, which allow self-consistent measurements of critical 
parameters combining stellar evolution models \citep[e.g.,][]{bru03} with radiative transfer 
through a dusty ISM \citep[e.g.,][]{cha00}.  A second set of tools uses sophisticated 
hydrodynamic computational codes to simulate interactions \citep[e.g., GADGET - ][]{spr05}, 
while simultaneously new radiative transfer models can compute the predicted emission 
from these evolving interacting systems \citep[e.g., SUNRISE - ][]{jon06}. 

It is important to recognize that observational biases can be significant. Due to the long 
timescales of an interaction (typically 10$^{8-9}$ years), observers rely on studies of a 
range of interacting systems to reconstruct a likely sequence of events.  Moreover, determining 
the exact phase of any particular observed interaction from its morphology is uncertain at best, 
because the appearance of a system at a given interaction phase also depends on the 
specific geometry of the encounter, the masses of the galaxies, metallicity, molecular gas content, 
and (not least) previous interaction histories \citep[e.g.,][]{dim07}. Systems are ordered into 
an evolutionary sequence  using intuition provided by simulations and physical models, which 
are themselves based on observations of particular systems. Selection criteria, however, can 
introduce a bias for more luminous, morphologically disturbed systems and, hence, towards 
the most active phases of interactions. Therefore, a selection criterion not associated with 
either morphological disturbance or degree of activity is critical for obtaining a sample 
containing systems throughout the full interaction sequence.

We have undertaken a program to take advantage of all these developments: full, 
multi-wavelength datasets of an interacting galaxy sample selected with few biases; 
hydrodynamic simulations; and radiative transfer modeling, in a systematic effort to better 
understand  systems across a range of interaction stages and to iterate refinements to the 
various modeling and radiative transfer codes. We have chosen a  representative sample of 
objects spanning the interaction sequence, obtained their full spectral energy distributions 
(SEDs), and are comparing the results against a variety of models - based on both 
templates/stellar evolution/radiative transfer and on diagnostic features. 

This first paper of the project presents results and conclusions for a sample of thirty-one 
interacting galaxies in fourteen systems for which there are currently complete multi-wavelength 
data that can be used to study the variations in their star formation and dust heating. This paper is organized as 
follows. We describe  the full \emph{Spitzer} Interacting Galaxy Survey (SIGS)  sample in Section 
2 and the classification of each of the sources in the interaction sequence. Section 3 describes the 
wide range of observational photometry used to construct the SEDs. It also describes the issues 
associated with obtaining reliable photometry from the diverse datasets. In Section 4, we model 
the SEDs of these objects. Section 5 discusses the variations seen across the interaction 
sequence and constraints imposed by photometry from different instruments and compares star 
formation rates derived using the entire SED to those from relations depending on one or two 
wavelengths. In Section 6, we summarize our results.

\begin{deluxetable*}{llrrccrrr}
\tabletypesize{\scriptsize}
\tablecaption{Sample Description\label{sample}}
\tablewidth{0.95\linewidth}
\tablehead{
\colhead{} & \colhead{} & \colhead{R.A.} & \colhead{Decl.} 
& \colhead{Distance} & \colhead{Interaction} & \multicolumn{3}{c}{Aperture}\\
\colhead{Group} & \colhead{Galaxy} & \colhead{(J2000)} & \colhead{(J2000)} 
& \colhead{(Mpc)} & \colhead{Stage}  & \colhead{Size} & \colhead{Angle}& \colhead{From}\\
\colhead{(1)} & \colhead{(2)} & \colhead{(3)} & \colhead{(4)} & \colhead{(5)} & \colhead{(6)} & \colhead{(7)} & \colhead{(8)} & \colhead{(9)}
}
\startdata
 1 	& NGC2976		     	& 	09 47 16.3	& +67 54 52.0	&	  3.75	& 2.0$\pm$0.0 & $3\farcm57\times1\farcm77$	 & $51\fdg8$	 & 3.6\um	 \\  
  	& NGC3031		     	& 	09 55 33.2	& +69 03 57.9	&	  3.77	& 2.0$\pm$0.4 & $10\farcm11\times5\farcm82$& $64\fdg0$	 & 3.6\um	 \\
	& NGC3034		     	& 	09 55 52.2	& +69 40 47.8   &	  3.89	& 2.0$\pm$0.4 & $2\farcm87\times1\farcm07$	 & $336\fdg4$	 & 3.6\um	 \\
	& NGC3077		     	&	10 03 19.8	& +68 44 01.5	&	  3.93	& 2.0$\pm$0.5 & $2\farcm12\times1\farcm62$	 & $318\fdg5$	 & 3.6\um	 \\
2	& NGC3185		     	&	10 17 38.7	& +21 41 16.2	&	22.6~~~	& 2.0$\pm$0.5 & $1\farcm84\times0\farcm99$	 & $41\fdg9$	 &NUV	 \\     
	& NGC3187		     	&	10 17 48.4	& +21 52 30.9	&	26.1~~~	& 3.0$\pm$0.5 & $2\farcm25\times1\farcm04$	 & $338\fdg7$	 &NUV	 \\
	& NGC3190		     	& 	10 18 05.7	& +21 49 57.0	&	22.5~~~	& 3.0$\pm$0.5 & $2\farcm14\times0\farcm97$	 & $28\fdg4$	 &3.6\um	 \\   
3	& NGC3226		     	&	10 23 27.0	& +19 53 53.2	&	23.3~~~	& 4.0$\pm$0.5 & $1\farcm29\times1\farcm00$	 & $302\fdg5$	 &3.6\um	 \\
	& NGC3227		     	&	10 23 30.5	& +19 51 55.1	&	20.6~~~   	& 4.0$\pm$0.5 & $1\farcm89\times1\farcm03$	 & $60\fdg4$	 &3.6\um	 \\  
4	& NGC3395		 	&	10 49 50.0	& +32 58 55.2	&	27.7~~~	& 4.0$\pm$0.5 & $1\farcm46\times0\farcm89$	 & $278\fdg9$	 &3.6\um	 \\ 
	& NGC3396  			&	10 49 55.2	& +32 59 25.7	&	27.7~~~	& 4.0$\pm$0.5 & $1\farcm38\times0\farcm60$	 & $9\fdg6$	 &3.6\um	 \\ 
5	& NGC3424		     	&	10 51 46.9	& +32 54 04.1	&	26.1~~~	& 2.0$\pm$0.4 & $1\farcm81\times0\farcm59$	 & $17\fdg4$	 &NUV	 \\   
	& NGC3430		   	&	10 52 11.5	& +32 57 05.0	&	26.7~~~	& 2.0$\pm$0.4 & $2\farcm69\times1\farcm46$	 & $301\fdg3$	 &NUV	 \\ 
6	& NGC3448		     	&	10 54 38.7	& +54 18 21.0	&	24.4~~~	& 3.0$\pm$0.0 & $1\farcm57\times0\farcm59$	 & $338\fdg6$	 &3.6\um	 \\ 
	& UGC6016		     	&	10 54 13.4	& +54 17 15.5	& $27.2^{*}$~~ & 3.0$\pm$0.0 & $1\farcm28\times0\farcm67$	 & $329\fdg3$	 &3.6\um	 \\ 
7	& NGC3690/IC694		&	11 28 31.2	& +58 33 46.7	& $48.1^{*}$~~ & 4.0$\pm$0.4 & $1\farcm20\times0\farcm93$	 & $40\fdg6$	 &3.6\um	 \\ 
8 	& NGC3786		     	&	11 39 42.5	& +31 54 34.2	&	41.7~~~	& 3.0$\pm$0.5 & $1\farcm04\times0\farcm57$	 & $340\fdg7$	 &3.6\um	 \\
	& NGC3788		     	&	11 39 44.6	& +31 55 54.3  	&	36.5~~~	& 3.0$\pm$0.5 & $1\farcm32\times0\farcm41$	 & $84\fdg8$	 &3.6\um	 \\
9	& NGC4038/4039		&	12 01 53.9	& -18 52 34.8	&     	25.4~~~	& 4.0$\pm$0.0 & $3\farcm00\times2\farcm33$	 & $304\fdg3$	 &3.6\um	 \\ 
10	& NGC4618		     	&	12 41 32.8	& +41 08 44.4	&	  7.28	& 3.0$\pm$0.5 & $2\farcm69\times2\farcm08$	 & $284\fdg1$	 &3.6\um	 \\ 
 	& NGC4625		     	&	12 41 52.6	& +41 16 20.6	&	  8.20	& 3.0$\pm$0.5 & $1\farcm88\times1\farcm49$	 & $296\fdg5$	 &NUV	 \\   
11	& NGC4647		     	&	12 43 32.6	& +11 34 53.9	& 	16.8~~~	& 3.0$\pm$0.5 & $1\farcm53\times1\farcm24$	 & $18\fdg6$	 &NUV	 \\ 
	& NGC4649		     	&	12 43 40.0	& +11 33 09.8	&	17.3~~~   	& 3.0$\pm$0.5 & $1\farcm81\times1\farcm33$	 & $34\fdg2$	 &3.6\um	 \\ 
12	& M51A			     	&	13 29 54.1	& +47 11 41.2	&	  7.69	& 3.0$\pm$0.5 & $6\farcm86\times4\farcm42$	 & $293\fdg5$	 &NUV	 \\
	& M51B			     	&	13 29 59.7	& +47 15 58.5	&	  7.66	& 3.0$\pm$0.5 & $2\farcm68\times1\farcm95$	 & $18\fdg3$	 &3.6\um	 \\
13	& NGC5394			&	13 58 33.7	& +37 27 14.4	& $56.4^{*}$~~ & 4.0$\pm$0.5 & $0\farcm89\times0\farcm50$	 & $84\fdg2$	 &NUV	 \\
	& NGC5395			& 	13 58 37.6	& +37 25 41.2	& $56.4^{*}$~~ & 4.0$\pm$0.5 & $2\farcm88\times1\farcm08$	 & $87\fdg9$	 &NUV	 \\
14	& M101			     	&	14 03 09.8	& +54 20 37.3 	&	  6.70	& 3.0$\pm$0.5 & $10\farcm00\times8\farcm53$& $156\fdg4$	 &3.6\um	 \\
	& NGC5474		     	&	14 05 01.2	& +53 39 11.6	&	  5.94	& 3.0$\pm$0.5 & $2\farcm53\times2\farcm24$	 & $290\fdg2$	 &3.6\um		     
\enddata	
\tablecomments{Distance moduli were obtained from \citet{tully2008}, \citet{tully1988}, and the Extra-galactic Distance Database. The distances in column (5)
marked with $^{*}$ did not have distance moduli and were calculated based on heliocentric velocities, corrected per \citet{mould2000} 
and assuming \hubble. The determination of interaction stage is described in Section 2.2. In column (6) we give the median and standard deviation
of the classifications by the co-authors. The parameters of the elliptical apertures are given in columns (7) and (8) and we note whether it was determined
on the \emph{GALEX} NUV or IRAC 3.6\um~image. The angle is given degrees north of west. }
\end{deluxetable*}

\section{THE \emph{SPITZER} INTERACTING GALAXY SURVEY (SIGS) SAMPLE}

\subsection{Sample Description}

The SIGS sample was designed to span the full range of galaxy interaction parameters by using 
a sample selected strictly on the basis of interaction probability rather than morphology, activity, 
luminosity, or other derivative indicators. The catalog includes interactions of all types, not just 
those that give rise to obvious morphological peculiarities and/or nuclear/starburst activity, thus 
minimizing morphological biases so we can address the relationships between interactions and 
activity. A selection criterion not dependent on visible signs of tidal interactions is important 
because of the dependence of the response of interacting galaxies on the relative inclinations of 
disks (e.g., Toomre \& Toomre 1972; D'Onghia \etal~2010) and the uncertain distribution of dark matter around the galaxies 
(e.g., Dubinski \etal~1996, 1999). The SIGS sample was based on the Keel-Kennicutt visibly selected 
catalog of interacting spiral galaxies \citep[][hereafter K85]{keel85}, which selected galaxies based 
on the local density of nearby neighbors and consists of bright spiral galaxies having neighbors 
with typical projected separations of 4-5 effective radii. A criterion based on the relative recessional velocities 
$|\Delta{\rm v}|< 600~{\rm km~s^{-1}}$ was imposed to exclude non-associated, projected pairs. 

In order to resolve structures on scales of a few hundred pc, we limited the original sample to sources 
closer than cz$~<4000~{\rm km~s^{-1}}$. To investigate the effects of tidal interaction, we added a  
complementary set to the prime sample: the K85 ``Arp Sample" with the same maximum distance as the 
Keel-Kennicutt complete sample. This set is based on the Arp catalog of peculiar galaxies from which K85 
selected all objects showing evidence of tidal interaction not strong enough to disrupt the galactic disks (i.e., 
it does not bias against late stage mergers). Although K85 excluded some fainter members of the interacting 
groups (their selection criteria required a B band magnitude of $B_{T}\leq13.0$), we include 
them in order to obtain a complete picture of the activity in the different interacting systems. 

The total SIGS sample consists of 103 individual interacting galaxies in 48 systems. The combined 
galaxies  span the range of interaction types, luminosities, and galaxy types. SIGS is comprised 
primarily of spiral-spiral interactions, with some spiral-elliptical and spiral-irregular interactions. Its set of systems
contains both major and minor mergers, ranging from systems likely to be in first approach 
(e.g. NGC 3424/NGC 3430) through close passages (e.g. M51) to final collision (NGC 3690/IC 694), and
span a luminosity range from 1.3$\times10^{10}-5.1\times10^{14}$~L$_{\odot}$. From this complete sample, 
which has a sufficiently large number of objects to allow us to study statistically the activity in interacting 
galaxies across a wide range of encounter parameters, we will be able to study the increase of star 
formation and AGN activity in interacting disk galaxies. As discussed in section 1, while there have 
been a significant number of studies probing star formation rate (SFR) enhancement and nuclear activity, the importance of 
the different interaction parameters in triggering these events is not well understood. The SIGS sample 
provides us with the opportunity to observe a large range of galaxies, including very early interaction 
stages. The level and distribution of star-formation in such early stage interactions has not been 
systematically studied before, therefore our sample will allow us to identify the initial increase in 
SFR caused by the interaction, as well as identify where this enhancement is located in the 
galaxies (i.e. in the central region of the galaxy, along the disk, or within tidal features). 
Additionally, the size of our sample also 
provides us with the ability to probe these enhancements for all systems as a function of different 
interaction parameters, such as galaxy mass, mass ratio and gas content.  A detailed description of the 
SIGS sample along with the analysis of the \emph{Spitzer} data and a presentation of the images and the 
photometric results is given in Brassington \etal~(2013, in preparation).

There are currently fourteen interacting systems from the SIGS set which have publicly available observations by all the 
facilities: \emph{Herschel} (SPIRE and partial coverage with the Photodetector Array Camera and
Spectrometer (PACS)), \emph{Spitzer}, 2 Micron All Sky Survey (2MASS), and either 
\emph{GALEX} or \emph{Swift}, enabling us to model their emission from far-UV (FUV) to FIR 
in 28 filters when ancillary archival measurements are added. Not all galaxies have photometric
data in all filters; we used as many photometric data as available, generally 15-25. 
These galaxies comprise the sample we examine in this paper and were selected from the SIGS 
sample on the basis of available SPIRE observations. They are listed in Table \ref{sample} along 
with key parameters.

{\begin{turnpage}
 \begin{deluxetable*}{lrlrlrlrlrrc}
\tabletypesize{\scriptsize}
\tablecaption{Description of \emph{Spitzer} IRAC and MIPS Observations\label{spitzerobs}}
\tablewidth{0pt}
\tablehead{
\colhead{Galaxy} & \multicolumn{4}{c}{IRAC} & \multicolumn{5}{c}{MIPS~24\um} \\
\colhead{} & \colhead{PID} &\colhead{Date}& \colhead{Exposure/Band}&\colhead{Pipeline}& \colhead{PID} &\colhead{Date}&\colhead{Mode}& \colhead{Exposure Time} &\colhead{Pipeline}
}
\startdata
NGC2976 				     & 159     	& 2004 Oct 29-30    		&30$\times$30 s     	& 13.2.0	& 159  	& 2004 Oct 16 		&  Scan	& 169.8 s	&	14.4.0	 \\   
NGC3031 				     & 159     	& 2004 May 1			&240$\times$30 s  	& 13.0.2	& 159	& 2003 Nov 24		&  Scan	& 175.8 s	&   	14.4.0	\\   
NGC3034 				     & 159     	& 2005 May 6-9, Oct 25   	&120$\times$30 s   	& 14.0.0	& 159 	& 2004 Nov 11		&  Scan	& 152.5 s	&  	14.4.0	\\ 
NGC3077				  	     & 59		 & 2004 Mar 8			&8$\times$12 s  	& 18.18.0	&   59	& 2004 Mar 16		&  Phot	& 159.3 s	&	18.13.0	 \\ 
						     & 40204 	& 2007 Nov 15			&30$\times$30 s  	& 18.18.0	&   	& 	&  	  	& 		&				\\
NGC3185				  	     & 40936  	& 2007 Dec 23	 		&8$\times$12 s  	& 18.18.0	&  & &			&		&				 \\ 
NGC3185/3187/3190 		     & 159     	& 2004 Apr 28           		& 48$\times$30 s  	& 13.0.2	& 159	& 2004 Dec 28 		& Scan 	& 173.3/156.2/176.6 s	&	14.4.0	 \\ 
NGC3226/3227		    	     & 3269   	& 2004 Dec 21			& 2$\times$12 s   	& 13.2.0	& 1054	& 2003 Nov 24	 	& Phot	& 593.4 s	&	14.4.0	 \\
						     &  1054   	& 2003 Nov 26			&48$\times$12 s   	& 13.2.0	& &&  			&			&	&		 \\
NGC3395/3396			     &20671	& 2006 Dec 29			& 24$\times$12 s 	& 18.7.0	& 20140 	& 2005 Dec 3 		& Phot  	&  226.4 s	&  	14.4.0	   \\
NGC3424/3430			     &20140	& 2006 Jun 1		   	& 30$\times$12 s 	& 14.0.0	& 50696 	& 2008 Jun 21-23 	& Phot     	&  220.1/542.3 s    	& 	18.13.0	  \\ 
NGC3448/UGC6016		     &3247 		& 2004 Dec 16			& 72$\times$12 s 	& 14.0.0	& 3247 	& 2007 Jun 19 		& Phot 	& 557.8 s	&	14.4.0	 \\
NGC3690/IC694			     &32   		& 2003 Dec 18 			& 120$\times$12 s 	& 13.2.0	& 32 		& 2005 Jan 2 		& Phot 	& 79.6 s	&	14.4.0	\\ 
NGC3786/3788			     & 3247		& 2004 Dec 17			&  46$\times$12 s 	& 14.0.0	& 3247 	& 2005 May 12 	& Phot 	& 557.8 s	&	14.4.0	  \\
NGC4038/4039			     & 32		& 2003 Dec 24			&100$\times$12 s	& 13.2.0	& 32 		& 2005 Jan 25 		& Scan 	& 87.1 s	&	14.4.0	  \\
NGC4618/4625			     & 69    		& 2004 May 21		  	& 10$\times$30 s  	& 13.2.0	& 69 		& 2004 Jun 3 		& Phot 	& 754.6/278.9 s	&	14.4.0	  \\
				 	     	     &159   		& 2004 May 18, May 21	& 16$\times$30 s  	& 13.2.0	& 159	& 2004 Dec 26-Jan 2 & Scan 	& 176.6/165.7 s	&	14.4.0	\\ 
NGC4647/4649			     & 69    		& 2004 Jun 10	 		& 10$\times$12 s 	& 13.2.0	& 69 		& 2005 Jun 26 		& Phot 	& 139.4/278.9 s	&	18.12.0	 \\ 
M51						     &159    		& 2004 May 18, May 22	& 108$\times$30 s 	& 13.2.0	& 159	& 2004 Jun 22		& Scan 	& 175.8/174.5 s	&	14.4.0	  \\ 
NGC5394/5395			     & 3672 	& 2005 Jan 21			& 10$\times$30 s 	& 18.7.0	& 3247 	& 2005 Jan 25 		& Phot 	& 557.8 s	&	14.4.0	  \\ 
M101					     &60      		& 2004 Mar 8			& 338$\times$12 s 	&  13.2.0	& 60		& 2007 Jun 19		& Scan 	& 176.5 s	&	 14.4.0  \\
NGC5474					     &159   		& 2004 May 18, May 22	& 62$\times$30 s 	&  13.2.0	& 159 	& 2004 Dec 26 		& Scan      & 162.1 s    &	18.12.0
\enddata	
\tablecomments{MIPS exposures are determined differently based on the observing mode. For galaxies observed in the Phot mode, we give the 
total exposure time of the frames covering the galaxy. For galaxies observed in the Scan mode, we give the average observing time on the galaxy. }
\end{deluxetable*}
\end{turnpage}}

\subsection{Estimating the Interaction Phase}

\citet{too72} were the first to systematically model and describe the morphological characteristics of interacting 
galaxies. Using simple simulations, they showed that tails and bridges could result from tidal forces and 
reconstructed the orbits that could produce the tidal features seen in some of the best
known interacting systems including M51, the Mice (NGC 4676), and the Antennae (NGC 4038/4039). 
Their work also highlighted the close connection between observations and modeling: our 
classification of the interaction stages in our sample is based on theoretical descriptions of how such interactions 
are expected to proceed.

As \citet{rich12} have shown, projected distance alone is an unreliable indicator of interaction stage.
We therefore used the \citet{dopita02} five-stage scheme to classify the interaction stage of our galaxies. 
By construction, our sample does not include any Stage 1 galaxies (isolated, non-interacting galaxies). 
Stage 2 galaxies are described as weakly interacting systems, which are close on the sky, but show
minimal morphological distortion. These systems could be either before or after the first passage. Stage 3
galaxies, which we call moderately interacting, show stronger signs of morphological distortion and often
tidal tails. Depending on the geometry of the encounter, these systems could be in the midst of the first
or a subsequent passage. Stage 4 (strongly interacting) galaxies show strong signs of disturbance and are
therefore in more evolved stage of interaction. Our sample falls into these three categories. While the 
SIGS sample has a Stage 5 (coalescence/post-merger systems), the sample presented in this paper 
does not. The SIGS sample is roughly equally divided between Stages 2-4, while the sample
presented in this paper has 7, 14, and 7. 

This classification method is clearly a statistical scheme in the sense that, for each individual
galaxy, the classification stage does not translate directly to an interaction phase. However, since the 
scheme is based on morphological appearance of galaxies, it provides a direct picture of the
effect of the interaction on the distributions of the stellar component of the galaxies and their
star formation activity. The classification was carried out independently for each galaxy in the 
SIGS sample by six collaborators on the basis of appearance alone in Digitized Sky Survey 
(DSS) images. Stage 2 galaxies show little morphological distortion, while Stage 4 galaxies are 
strongly distorted. Stage 3 galaxies show some distortion in the form of tidal features, although 
their disks remain undisturbed. Visible 
DSS images are best suited for this purpose, since they trace on-going star 
formation as well as older stellar populations in a single image. In Appendix A, we show representative examples 
of the galaxies in Stages 2-4. Galaxy groups in which classifications differed by more than one 
stage were re-examined; the median of the  classifications is used for each galaxy. Table 
\ref{sample} lists the interaction stage for all of the galaxies in our sample.

\begin{deluxetable*}{lllllcl}
\tabletypesize{\scriptsize}
\tablecaption{Description of \emph{Herschel}  Observations\label{hershobs}}
\tablewidth{0pt}
\tablehead{
\colhead{Galaxy} &\colhead{Instrument} &  \colhead{ObsID} &\colhead{Date} & \colhead{Obs. Mode} 
& \colhead{Exposure (s) } & \colhead{PACS Bands} }
\startdata
NGC2976 				    & SPIRE	& 1342192106        	& 2010 Mar 11       & Large Map & 1076 & \\
						    & PACS		& 1342207170-73  	& 2010 Oct 26     & Scan Map & 3624 & 75\um, 110\um, 170\um \\
NGC3031 				    & SPIRE    	& 1342185538 		& 2009 Oct  6      & Large Map & 5042 & \\
						    & PACS		& 1342186085-86   	& 2009 Oct 17    & Scan Map & 22208 & 75\um, 170\um \\
NGC3034 			             & SPIRE	& 1342185537 		& 2009 Oct  6      & Large Map & 2418 & \\
						    & PACS		& 1342209350-51  	& 2010 Nov 10 & Scan Map  & 6542 & 75\um, 170\um \\
NGC3077 				    & SPIRE	& 1342193015 		& 2010 Mar 28        & Large Map & 2095 & \\
						    & PACS		& 1342216507-10   	& 2011 Mar 21       & Scan Map & 7356 & 75\um, 110\um, 170\um \\
NGC3185/3187/3190	             & SPIRE	& 1342196668 		& 2010 May 18           & Large Map & 1035 & \\
NGC3187/3190			    & PACS		& 1342207145-48 	& 2010 Oct 25     & Scan Map  & 2708 & 75\um, 110\um, 170\um \\
NGC3226/3227                   	    & SPIRE	& 1342197318 		& 2010 May 30          & Large Map & 2624 & \\
						    & PACS		& 1342221146-47    & 2010 May 16           & Scan Map & 1178 & 75\um, 170\um \\
NGC3395/3396			    & SPIRE	& 1342209286 		& 2010 Nov 9 & Large Map & 999 & \\
						    & PACS		& 1342221104-07 	& 2011 May 16          & Scan Map  & 2564 & 75\um, 110\um, 170\um \\
NGC3424/3430			    & SPIRE	& 1342195946 		& 2010 May 8             & Large Map & 1618  &\\
NGC3448/UGC6016		    & SPIRE	& 1342185539 		& 2009 Oct 6       & Large Map & 1833 & \\
NGC3690/IC694			    & SPIRE	& 1342199344 		& 2010 Jun 29          & Large Map & 459 & \\
						    & PACS 	& 1342210600-05	& 2010 Nov 30 & Scan Map & 2462 &75\um, 110\um, 170\um \\
						    & PACS		& 134211104-05 	& 2010 Dec 13 & Scan Map & 524 & 110\um, 170\um \\
NGC3786/3788		 	    & SPIRE	& 1342223233		& 2011 Jun 28	  & Small Map & 169 & \\
						    & PACS		&  1342223319-20	& 2011 Jun 29	  & Scan Map  & 104 & 75\um, 170\um \\
NGC4038/4039			    & SPIRE	& 1342188686 		& 2009 Dec 29 & Large Map & 710 & \\
						    & PACS 	& 1342187836-39	& 2009 Dec 8	  & Scan Map & 2662 & 75\um, 110\um, 170\um \\
NGC4618/4625	 		    & SPIRE	& 1342188755 		& 2009 Dec 31 & Large Map & 1052 &\\
NGC4625					    & PACS 	& 1342210468-71 	& 2010 Nov 19 & Scan Map & 2708 & 75\um, 110\um, 170\um \\
NGC4647/4649 			    & SPIRE	& 1342188778 		& 2009 Dec 31 & Large Map & 4295 & \\
M51						    & SPIRE	& 1342188589 		& 2009 Dec 26 & Large Map & 1577 & \\
						    & PACS		& 1342188328-29 	& 2009 Dec 20 & Scan Map & 4422 & 75\um,  170\um \\
NGC5394/5395			    & SPIRE	& 1342236140		& 2012 Jan 1	  & Large Map & 1253 & \\
						    & PACS		& 1342211285-88	& 2010 Dec 17 & Scan Map  & 2200 & 75\um, 110\um, 170\um \\
M101					    & SPIRE   	& 1342188750  	& 2009 Dec 30  & Large Map & 9443 & \\
						    & PACS		& 1342198471-74 	& 2010 Jun 16-17	   & Scan Map  & 38077 & 75\um, 110\um, 170\um \\
NGC5474 			             & SPIRE	& 1342188751		& 2009 Dec 26  & Large Scan& 1052 & \\
						    & PACS 	& 13422077178-81  & 2010 Oct 26 	   &  Scan Map   & 2708 & 75\um, 110\um, 170\um
\enddata	
\end{deluxetable*}

\begin{deluxetable*}{lllrlrc}
\tabletypesize{\scriptsize}
\tablecaption{Description of \emph{GALEX} Observations\label{galexobs}}
\tablewidth{0pt}
\tablehead{
\colhead{Galaxy} &\colhead{Tilename} & \multicolumn{2}{c}{NUV} & \multicolumn{2}{c}{FUV}\\
\colhead{} & \colhead{} &\colhead{Date}& \colhead{Exposure (s) }&\colhead{Date}& \colhead{Exposure (s)}
}
\startdata
NGC2976 				&GI2\_024002\_NGC2976\_stream 	& 2006 Jan 04 	       & 18113.55 & 2006 Jan 04 & 17212.50 \\
NGC3031 				&GI1\_071001\_M81			& 2005 Jan 12 	       & 29421.55 & 2006 Jan 05 & 14706.70\\
NGC3034 			         &NGA\_M82					& 2009 Jan 31 	       & 17311.95 & 2009 Jan 31 & 11527.35\\
NGC3185/3187/3190		&NGA\_NGC3190				&2004 Jan 30  	       & 3545.80	 & 2005 Feb 19	& 1299.15 \\
NGC3395/3396/3424/3430	&GI1\_078004\_NGC3395		&2006 Mar 23         & 2666.15	 & 2006 Mar 23	& 1500.10 \\
NGC3448/UGC6016		&AIS\_92						&2004 Feb 4  	       & 423.00 	 & 2004 Apr 21	& 143.00 \\
NGC3690/IC694			&AIS\_99						&2007 Feb 13	       & 211.00 	 & 2007 Feb 13	& 211.00 \\
NGC3786/3788			& AIS\_111					&2007 Feb 20	       & 103.05	 & 2007 Feb 20 & 103.05\\
NGC4038/4039		    	&NGA\_Antennae				&2004 Feb 22	       &1541.30	 & 3004 Feb 22& 1541.30 \\
NGC4618/4625	 	    	&NGA\_NGC4625				&2004 Apr 5	       &3259.00	 & 2004 Apr 5		&3259.00\\
NGC4647/4649 		    	&GI1\_109003\_NGC4660		&2005 Apr 30	       &3113.25	 & 2008 Apr 23	& 1624.10 \\
M51						&GI3\_050006\_NGC5194		&2007 May 29	       &10216.20	 & 2007 May 29	& 10216.20 \\
NGC5394/5395			&GI1\_026018\_Arp84			&2006 Apr 12	       & 4268.65	& 2007 May 30		& 2811.40 \\
M101					&GI3\_050008\_NGC5457		&2008 Apr 4	       & 13294.40	 & 2008 Apr 4		& 13293.4 \\
NGC5474 			         	&NGA\_NGC5474				&2003 Jun 19	       & 1610.00   & 2003 Jun 19	& 1610.10 
\enddata	
\end{deluxetable*}

 \begin{deluxetable*}{lccccc}[h]
\tabletypesize{\scriptsize}
\tablecaption{Description of \emph{Swift} UVOT Observations\label{swiftobs}}
\tablewidth{0pt}
\tablehead{
\colhead{} &\colhead{} &\colhead{} & \multicolumn{3}{c}{Exposure Times (s)}\\
\colhead{Galaxy} & \colhead{ObsID} & \colhead{Date} & \colhead{uvw2}& \colhead{uvm2}& \colhead{uvw1}
}
\startdata
NGC3226/3227 & 00031280001 & 2008 Nov 4   &  342 &  249 & 352  \\  
NGC3226/3227 & 00031280002 & 2008 Nov 5   &  704 &  511 &  346 \\
NGC3226/3227 & 00031280003 & 2008 Nov 12 &  692 &  424 &  372 \\
NGC3226/3227 & 00031280004 & 2008 Nov 13 &  744 &  538 &  372 \\
NGC3226/3227 & 00031280005 & 2008 Nov 21 &  744 &  522 &  381 \\
NGC3226/3227 & 00031280006 & 2008 Nov 22 &  763 &  137 &  381 \\
NGC3226/3227 & 00031280007 & 2008 Nov 25 &  763 &  531 &  381 \\
NGC3226/3227 & 00031280008 & 2008 Nov 27 &  763 &  196 &  246 \\
NGC3226/3227 & 00031280009 & 2008 Dec 2   &  0     &      0   & 293 \\
NGC3226/3227 & 00031280010 & 2008 Dec 3   &  274 &  349 & 126 \\
\hline
NGC3424/3430 & 00091132001 & 2011 Apr 16 & 0      & 0        & 1976  \\
NGC3424/3430 & 00091132003 & 2011 Jun 28 & 0      & 0        & 0         \\
NGC3424/3430 & 00091132004 & 2011 Jul 4     & 0      & 80     &1315 \\
NGC3424/3430 & 00091132005 & 2011 Jul 7     & 302 & 0        &  0       \\
NGC3424/3430 & 00091132006 & 2011 Jul 8     & 0     & 1877 & 0         \\
NGC3424/3430 & 00091132007 & 2011 Oct 7    & 750 & 988   &  0        \\
NGC3424/3430 & 00091132008 & 2011 Oct 10  & 0      &  0      &  0        
\enddata	
\tablecomments{The {\em Swift} observation ID number (Col. 2) and the start date of each observation 
(Col. 3) are given for each observation of each object for which observations with minimal coincidence losses
exist. Exposure times in the each filter are given in Col. 4-6.}
\end{deluxetable*}

\begin{figure*}
\centerline{\includegraphics[width=0.75\linewidth]{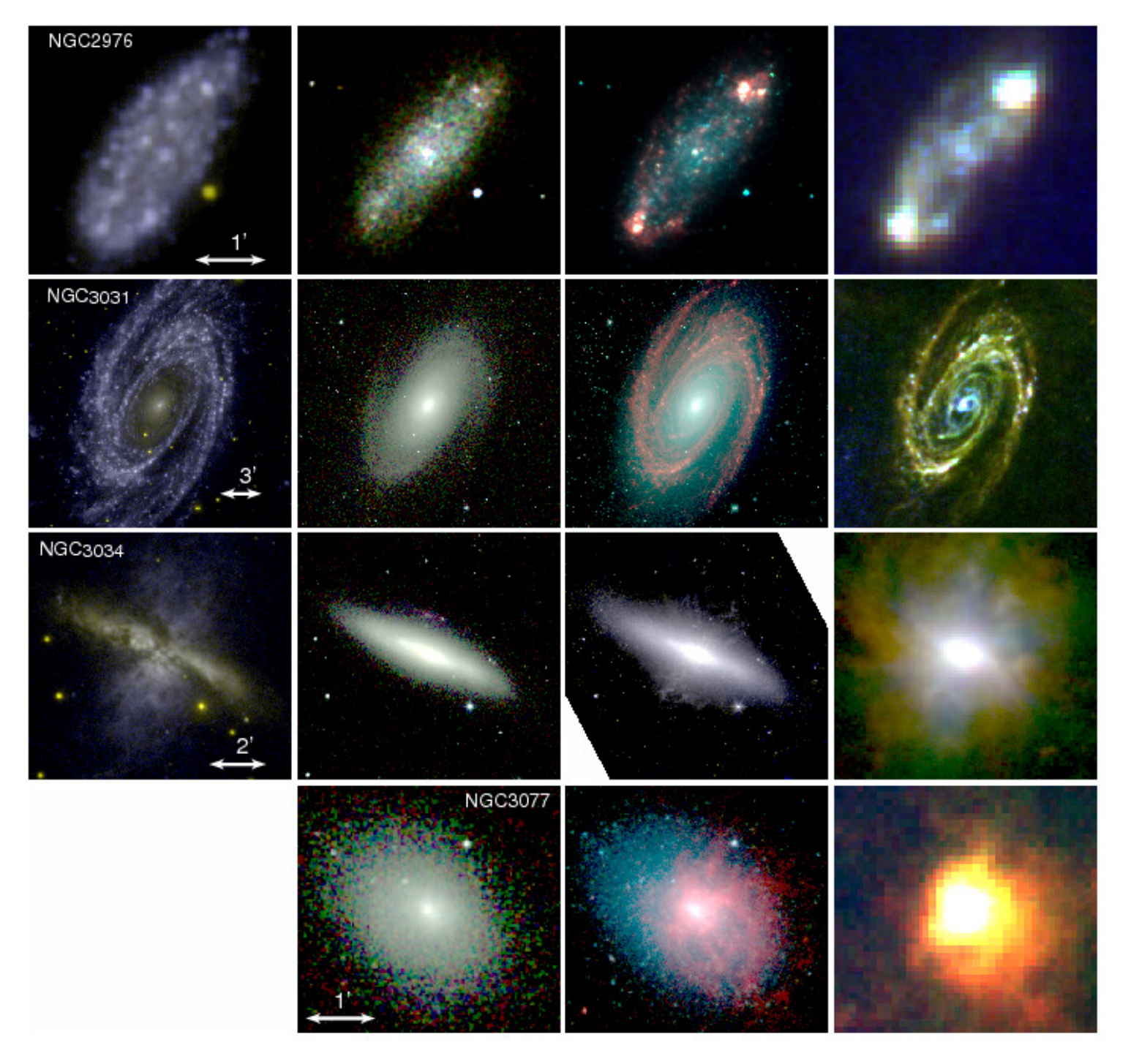}}
\caption{NGC 2976, NGC 3031, NGC 3034, and NGC 3077 (from top to bottom) as observed, from 
left to right, by \emph{GALEX} (NUV in yellow; FUV in blue), 2MASS (\emph{J} in blue, \emph{H} in 
green, and \emph{Ks} in red), IRAC (3.6\um~in blue, 4.5\um~in green, and 8.0\um~in red), and 
\emph{Herschel} (PACS 75\um~in blue, PACS 170\um~in green, and SPIRE 250\um~in red).  The 
longer wavelength IRAC observations of NGC 3034 were saturated, so 4.5\um~is shown in 
yellow instead. NGC 3077 was not observed by either \emph{GALEX} or\emph{ Swift}. At the 
distance of these galaxies, 1'$\approx$1.1 kpc.}
\label{n2976img}
\end{figure*}

\begin{figure*}[h]
\centerline{\includegraphics[width=0.85\linewidth]{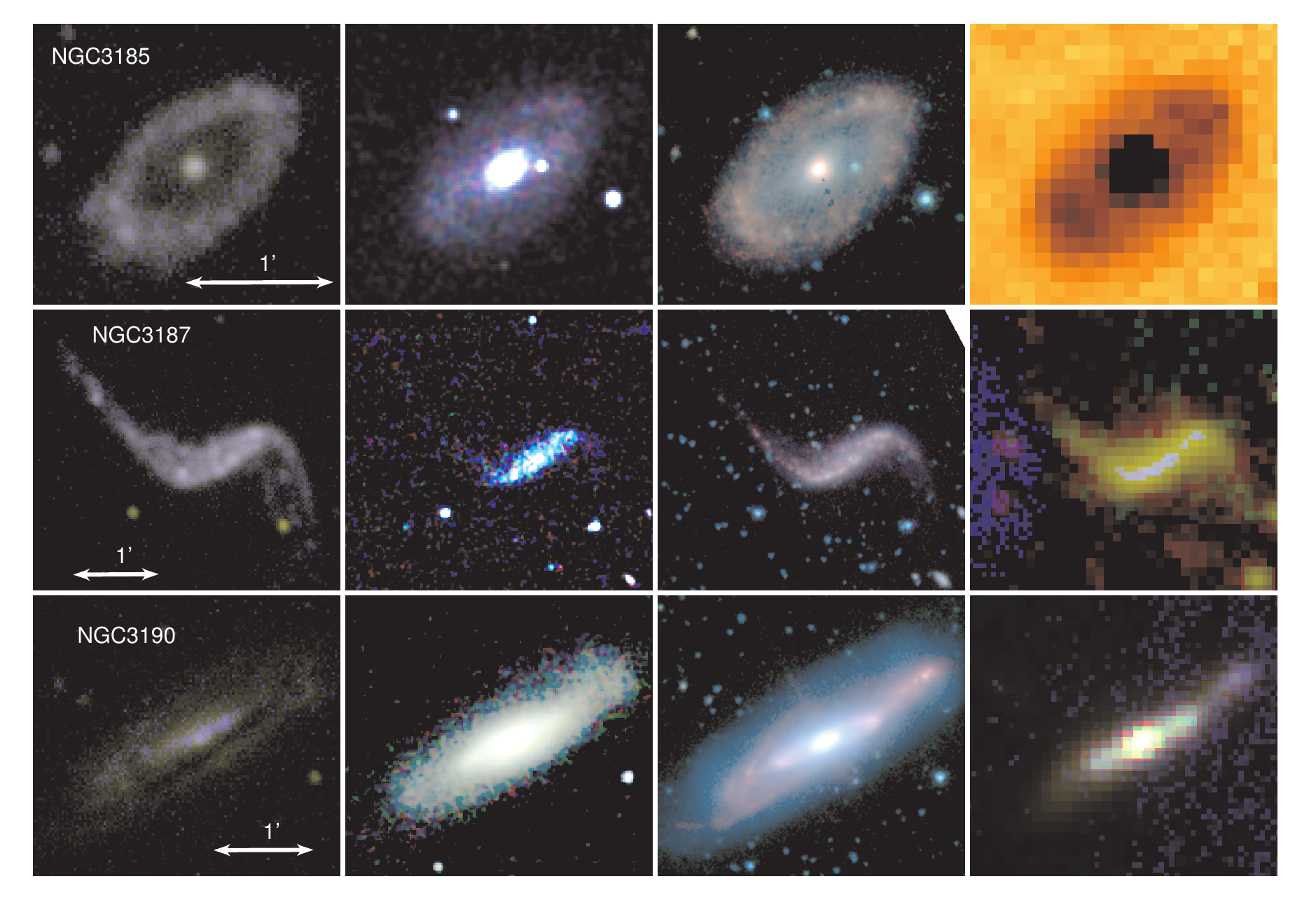}}
\caption{As Figure \ref{n2976img}, but for NGC 3185, NGC 3187, and NGC 3190. NGC 3185 
was not observed by PACS, the right image only shows the SPIRE 250\um~image in which 
darker pixels have higher flux.. At the distance of these galaxies, 1' is approximately 6-7 kpc.}
\label{grp2img}
\end{figure*}

\begin{figure*}
\centerline{\includegraphics[width=0.85\linewidth]{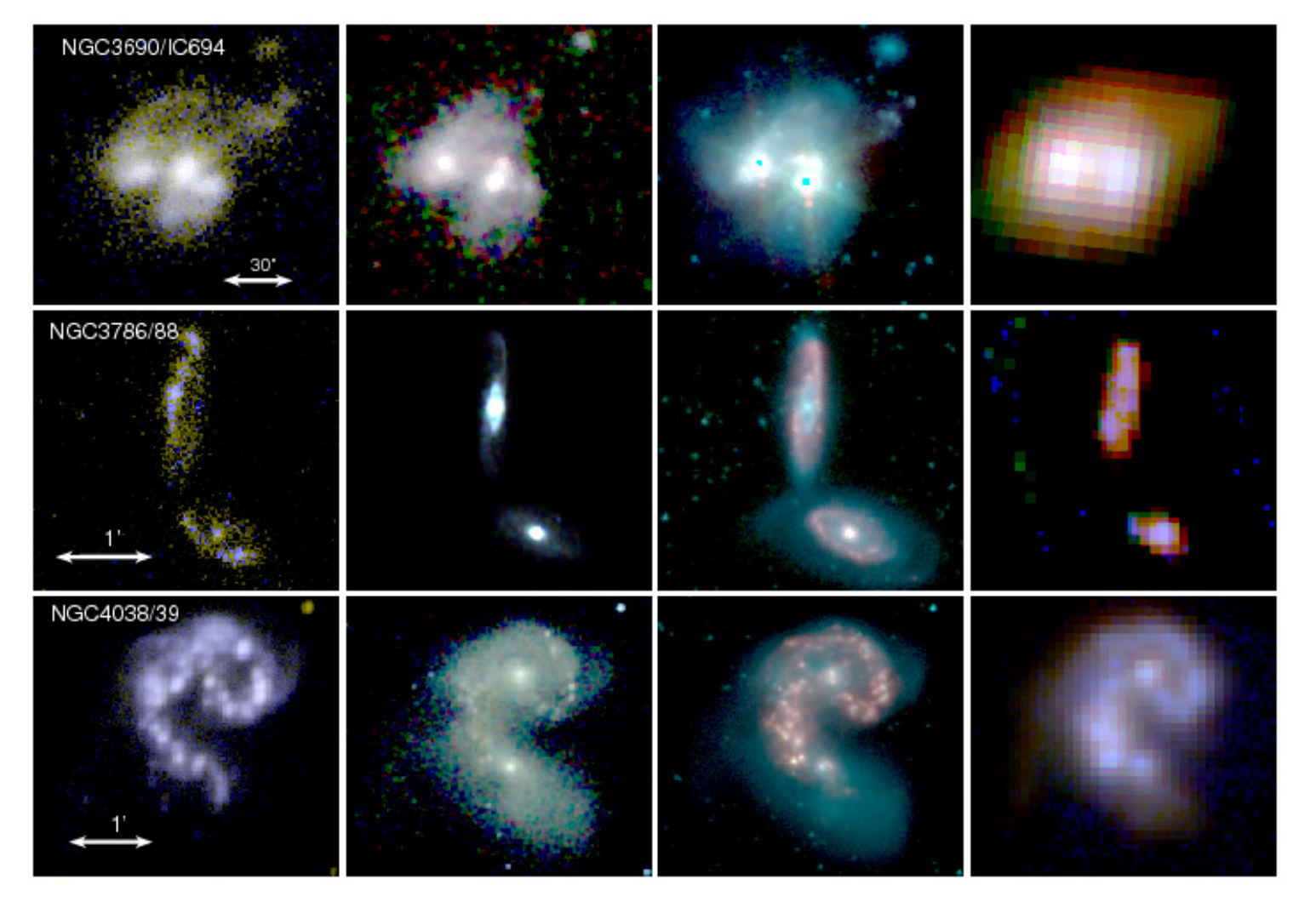}}
\caption{As Figure \ref{n2976img}, but for NGC 3690/IC694, NGC 3786 (bottom)/NGC 3788 
(top), and NGC 4038/4039. The 8\um~IRAC image of NGC 3690/IC 694 is saturated in the 
nuclei of the two galaxies, resulting in the blue-green artifacts. At the distance of these galaxies, 
1' is approximately 14 kpc (NGC 3690), 11-12 kpc (NGC 3786/3788), and 7.4 kpc (NGC 4038/4039).}
\label{grp789img}
\end{figure*}

\begin{figure*}
\centerline{\includegraphics[width=\linewidth]{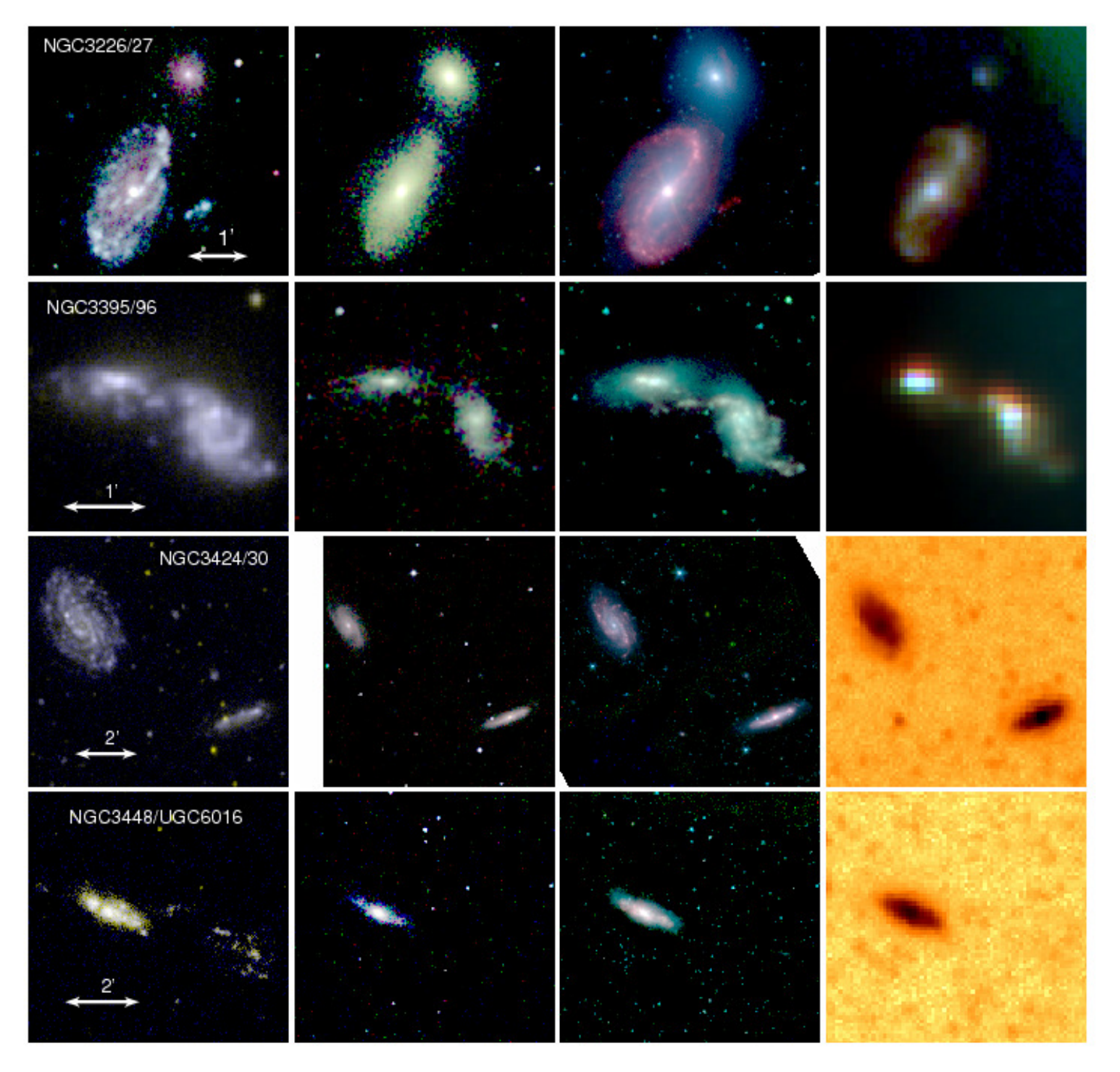}}
\caption{As Figure \ref{n2976img}, but for NGC 3226 (upper)/NGC 3227 (lower), NGC 3395 (right)/NGC 
3396 (left), NGC 3424 (right)/NGC 3430 (left), and NGC 3448 (left)/UGC 6016 (right). NGC 3226/3227 
was not observed by \emph{GALEX} but by \emph{Swift}. Their left image show the \emph{Swift} 
observations through the UVW1 filter in blue, the UVM2 filter in green, and the UVW2 filter in red. 
NGC 3424/30 and NGC 3448/UGC 6016 were not observed with PACS, so the right image only shows 
the SPIRE 250\um~image as in Figure \ref{grp2img}. UGC 6016, while having significant extended 
diffuse emission in the UV, is not well detected in the IR bands. At the distance of these galaxies 
1' is approximately 6-8 kpc.}
\label{grp3456img}
\end{figure*}

\subsection{Comparison Non-Interacting Sample} 

As a comparison sample of non-interacting galaxies, we used a subset of the ``normal" galaxy sample
of Smith \etal~(2007). Smith \etal~(2007) identified 42 galaxies from the \emph{Spitzer} Infrared Nearby Galaxies
Survey (SINGS; Kennicutt \etal~2003; Dale \etal~2005) of which 26 were spirals, which had not been 
subject to strong distortions. We were more conservative in our definition of non-interacting, by removing galaxies 
associated with clusters or radial-velocity groups, and we removed the three that were not observed with SPIRE as part of 
the Key Insights on Nearby Galaxy: a Far Infrared Survey with Herschel (KINGFISH; Kennicutt \etal~2011). 
Our comparison sample is comprised of 15 galaxies: NGC 925, NGC 1291, 
NGC 2841, NGC 3049, NGC3184, NGC 3521, NGC 3621, NGC 3938, NGC 4236, NGC 4559, NGC 4594, NGC 4736, 
NGC 4826, NGC 5055, and NGC 6946. We used the distances provided in Smith \etal~(2007) and the UV-MIR
photometry given in Dale \etal~2007) and the FIR photometry given in Dale \etal~(2012).

\section{OBSERVATIONS AND DATA REDUCTION}
\begin{figure*}
\centerline{\includegraphics[width=\linewidth]{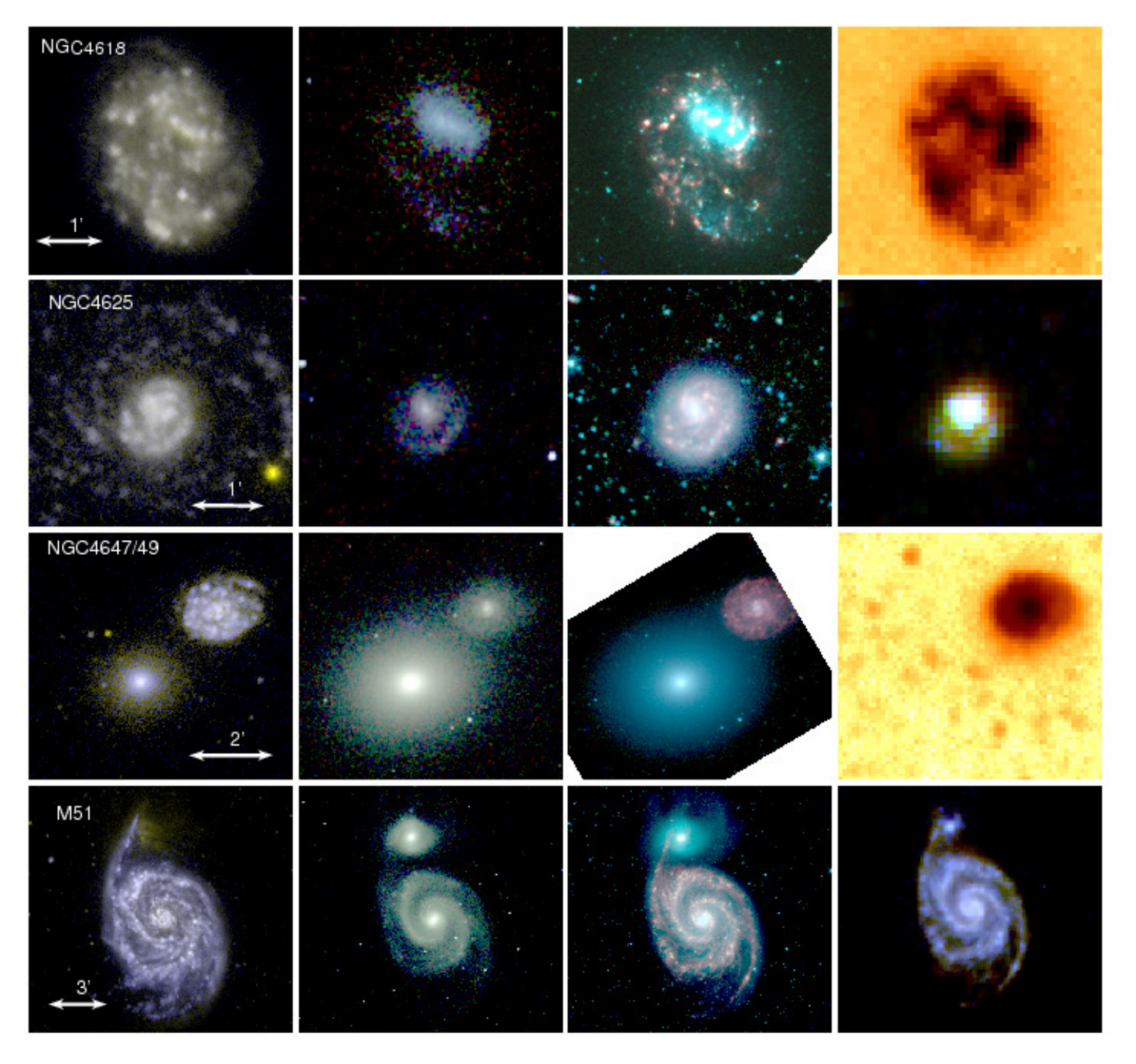}}
\caption{As Figure \ref{n2976img}, but for NGC 4618, NGC 4625, NGC 4647 (right)/NGC 
4649 (left), and M51. The right images of NGC 4618 and NGC 4647/49 only show the 
SPIRE 250\um~image. At the distance of these galaxies, 1' is approximately 2.1-2.4 kpc 
(NGC 4618/4625), 5 kpc (NGC 4647/4649), and 2.2 kpc (M51).}
\label{grp10-12img}
\end{figure*}

\begin{figure*}
\centerline{\includegraphics[width=\linewidth]{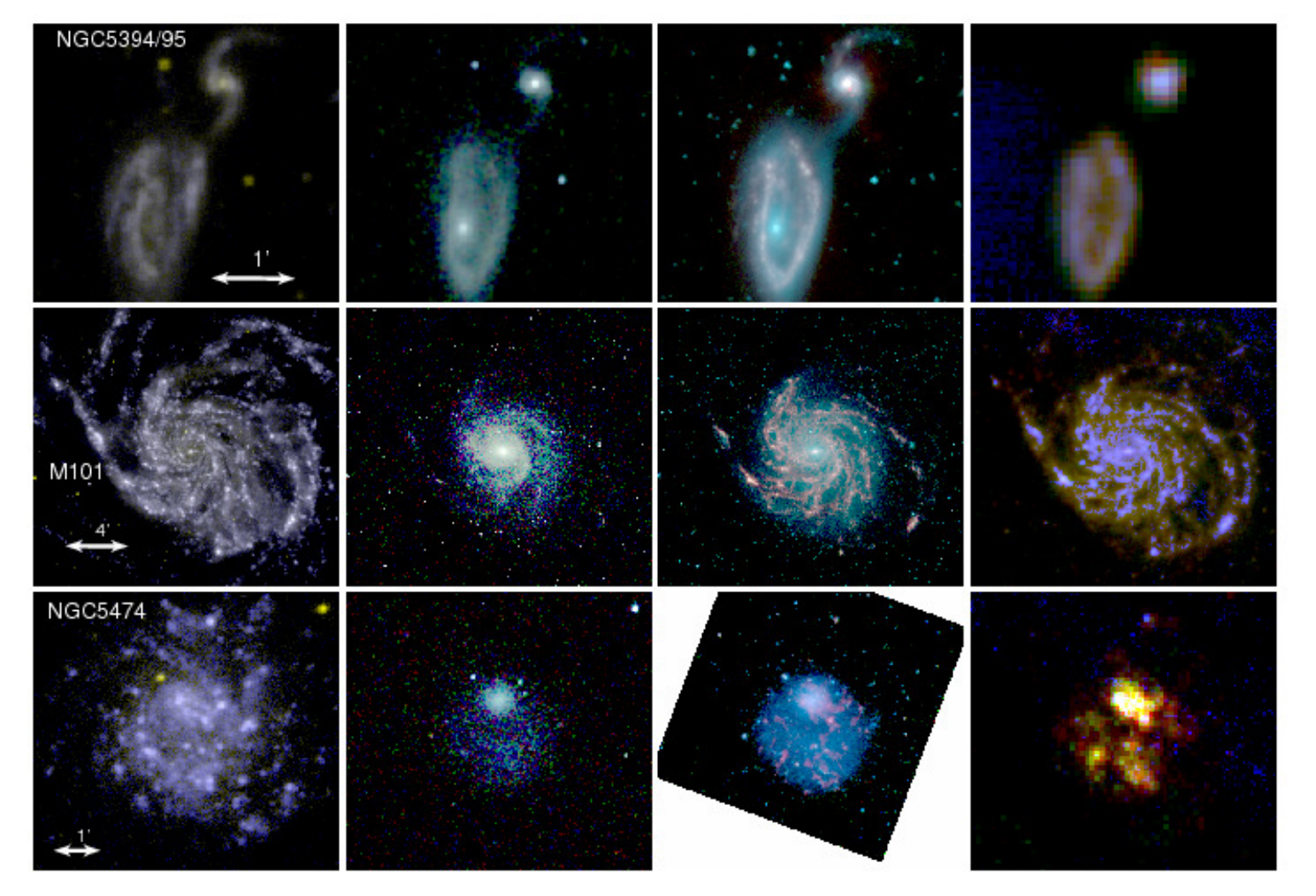}}
\caption{As Figure \ref{n2976img}, but for NGC 5394 (top)/NGC 5395 (bottom), 
M101, and NGC 5474.  At the distances of these galaxies, 1' is approximately 
16 kpc (NGC 5394/5395) and 1.7-2.0 kpc (M101/NGC 5474).  }
\label{n5474img}
\end{figure*}

The sample presented here has a complete set of near-infrared (NIR) to FIR photometry observed by 2MASS, \emph{Spitzer}, and
 \emph{Herschel} respectively,  as well as near-UV (NUV) and FUV photometry observed primarily by \emph{GALEX} and 
 completed by the Ultraviolet/Optical Telescope (UVOT) on \emph{Swift}. In the next sections, we describe the 
 observations and their reduction. The observations were supplemented with mid-infrared (MIR) to FIR 
fluxes measured by the \emph{Infrared Astronomical Satellite} (\emph{IRAS}), the 70\um~and 160\um~Multiband 
Imaging Photometer (MIPS) detectors on  \emph{Spitzer}, and UBV fluxes from the Third Reference Catalog 
\citep[RC3;][]{dev91} where available in the literature through the NASA Extragalactic Database (NED). 
 The MIPS 24\um~fluxes from these sources typically agree within the uncertainties with those we measure.  
 Figures \ref{n2976img}$-$\ref{n5474img} show from left to right the \emph{GALEX}, 2MASS, \emph{Spitzer} 
 Infrared Array Camera (IRAC), and \emph{Herschel} observations of each galaxy. Some galaxies (e.g. NGC 
3031 and M51A) have similar morphology from UV to FIR. In contrast, others have distinct morphological 
differences between the UV and IR, such as the FIR bright spots of NGC 2976 or the extended UV disk of NGC 3430.
Appendix C contains notes on the individual galaxies.

 \subsection{Galaxy Distances}
 
 All of the galaxies in our sample are nearby (within 60 Mpc) and can therefore have peculiar velocities that contribute
 significantly to their recessional velocities. \citet{tully2008} recently compiled redshift-independent distances for nearby
 galaxies with velocities less than 3000 km s$^{-1}$ using alternate methods including Cepheids \citep{free01}, the 
 luminosity of stars at the tip of the red giant branch \citep{kar06}, surface brightness fluctuations \citep{ton01}, and the 
 Tully-Fisher relation \citep{tully1977}. Distances to additional galaxies based on their group or cluster association 
 are given in the Extra-galactic Distance Database\footnote{http://edd.ifa.hawaii.edu} (EDD; Tully, 2010, private 
communication). Twenty-six of our galaxies have distance moduli given by either \citet{tully2008}, \citet{tully1988}, or 
EDD. For the five galaxies lacking distance moduli, we obtained heliocentric velocities from the PSCz catalog 
(Saunders et al. 2000; NGC 3690/IC 694, NGC 5394, and NGC 5395) and RC3 (UGC 6016), which we corrected to 
account for the velocity field of Virgo, the Great Attractor, and the Shapley supercluster, following \citet{mould2000}. 
Distances were then calculated assuming $H_{0} = $72 km s$^{-1}$ Mpc$^{-1}$. The distances are given in 
Table \ref{sample}.

\subsection{Infrared Photometry}

\subsubsection{\emph{Spitzer} Observations}

The IRAC \citep{fazio04} and MIPS \citep{rieke04} 24\um~observations were taken as 
part of a variety of programs, including the main SIGS program (PID 20140; P.I. A. Zezas), 
which also observed galaxy groups that had not previously been observed. The observation parameters are 
given in Table \ref{spitzerobs}. The IRAC Basic Calibrated Data (BCD) were retrieved from the Spitzer 
archive and cleaned before being coadded into mosaics with 0.6$\arcsec$ pixels using 
IRACproc \citep{sch06}. The MIPS 24\um~BCDs were merged to form mosaics with 2.45$\arcsec$ 
pixels using the Mosaicker and Point Source Extractor package \citep[MOPEX;][]{makovoz05}. The
reduction of these data will be described in detail in Brassington \etal~(2013, in preparation). While the pipeline
versions range from S13-S18, the difference between the pipelines are minor and do not impact significantly the 
photometry\footnote{http://irsa.ipac.caltech.edu/data/SPITZER/docs/irac/ iracinstrumenthandbook/79/}. 
The pipeline version for each galaxy is given in Table \ref{spitzerobs}.

\subsubsection{\emph{Herschel} Observations}

The parameters for the \emph{Herschel} SPIRE \citep{griffin10} and PACS \citep{pog10} observations are given 
in Table \ref{hershobs}.  The \emph{Herschel} data were taken as part of two Science Demonstration Phase 
programs (P.I.s C. Wilson and S. Eales), four Key Project programs (P.I.s R. Kennicutt, S. Eales, C. Wilson, and E. 
Sturm), and one Guaranteed Time program (P.I. L. Spinoglio). All of the galaxies were observed by SPIRE at 
250\um, 350\um, and 500\um; this was part of the selection criteria of this sample. Approximately 50\% of the 
sample were observed in all three PACS bands and an additional $\sim$25\% were observed at 75\um~and 170\um. 

The data were retrieved from the Herschel Science Archive and processed using the calibration 
trees of version 8.0.1 of the Herschel Interactive Processing Environment \citep[HIPE;][]{ott2010}. This processing
was accomplished using the default pipeline scripts available through HIPE to make Large Map mode mosaics for
the SPIRE data and extended source mosaics with MadMap for PACS data. We discuss additional details regarding
the processing of PACS data in Appendix B1.

\subsubsection{2MASS Observations}
NIR mosaics  of the sample galaxies observed as part of the 2MASS \citep{skru06} were retrieved 
from the NASA/IPAC Infrared Science Archive\footnote{NASA/IPAC Infrared Science Archive is
operated by the Jet Propulsion Laboratory, California Institute of Technology, under contract with NASA.}, and 
from the Large Galaxy Atlas \citep{jar03} when possible. The counts measured in the images were converted to 
Janskys using the zero points of \citet{coh03}. We compared our  fluxes measured in the 
apertures described in \S 3.4 to the total fluxes given in NED from \citet{jar03} and the 2MASS 
Extended Object Catalog and found good agreement.

\subsubsection{Ancillary IRAS Photometry}
\emph{IRAS} photometry was obtained from the HIRES Atlas \citep{sur04}, 
the \emph{IRAS} Revised Bright Galaxy Sample \citep{san03}, 
the \emph{IRAS} Bright Galaxy Sample \citep{soi89}, and the Faint Source Catalogue \citep{mosh90}.
The latter three catalogs present photometry derived from the native IRAS beam size of $2'-5'$; this 
can be problematic for systems in close interaction
phases. We therefore preferentially used the HIRES Atlas, which was reprocessed with 30$''-1\farcm5$. 
In the one system where only low-resolution photometry is available and the galaxies are close enough for 
contamination to occur, we do not include the IRAS photometry in our analysis.

\subsection{Ultraviolet Photometry}
\subsubsection{\emph{GALEX} Observations}
Twenty-eight  of our sample galaxies were observed by \emph{GALEX}; three sources within the sample
(NGC 3226, NGC 3227, and NGC 3077), however, were not observed due to the presence of  
nearby bright stars. For the galaxies with  \emph{GALEX} photometry, mosaics of the longest 
observations were retrieved from the Mikulski Archive for Space Telescopes\footnote{STScI is operated
by the Association of Universities for Research in Astronomy, Inc., under NASA contract
NAS5-26555. Support for MAST for non-HST data is provided by the NASA Office of Space
Science via grant NNX09AF08G and by other grants and contracts.} using GalexView version 1.4.6. 
The details of those observations are given in Table \ref{galexobs}. The NUV observation of 
NGC 3690/IC 694 was reprocessed by D. Neill at our request to correct a masking problem. We use
the conversions from count rate to fluxes provided by \citet{god04}\footnote{http://galexgi.gsfc.nasa.gov/docs/galex/FAQ/counts\_background.html}.

\subsubsection{\emph{Swift} UVOT Observations}
Most of the gaps in the \emph{GALEX} coverage can be filled in with data from the \emph{Swift} UVOT telescope, 
which has three UV filters that bracket the \emph{GALEX} NUV filter in mean wavelength. Two of the
 three galaxies lacking \emph{GALEX} data, NGC 3226 and NGC 3227, were observed by UVOT. Unfortunately, NGC 
 3077's nearby bright star exceeded the tolerances of this telescope as well. We originally planned to use existing
 UVOT photometry for all our sample. We obtained the raw data and exposure maps from the \emph{Swift} archive 
 for the seventeen galaxies with UVOT data and coadded the observations into one mosaic and exposure map per UV filter
 per interacting system. However, as described by \citet{hover11},  the photon-counting nature of the  \emph{Swift}
 detectors makes them vulnerable to coincidence losses, which become significant when the count rate is greater than 0.007
 counts per second per pixel. We calculated count rate maps to determine where coincidence losses need to be taken
 into account. Due the difficulties associated with coincidence losses in extended sources, described in greater length in 
 Appendix B2, we opted only to use the UVOT data for the missing  \emph{GALEX} objects NGC 3226 and NGC 3227. We 
 added one test case, NGC 3424, to confirm that the UVOT data yielded fluxes consistent with \emph{GALEX} and found good 
 agreement.  The details of the observations of these three galaxies are given in Table \ref{swiftobs}. To convert the 
 count rate to fluxes, we used the conversion assuming a stellar spectrum described in \citet{bree10}. 

\subsection{Aperture and Uncertainty Determination}
For consistency, we used matched apertures across all wavebands in our photometric analysis. 
Generally, the IR emission of galaxies is more extended that their UV emission. However, some of 
the galaxies are more extended in the UV than in the IR (e.g. NGC 3430). We used the SExtractor 
algorithm \citep{ber96} to determine Kron apertures in both the NUV and the 3.6\um~IRAC images. 
In all cases, the larger of the two apertures was then used to measure the integrated galaxy flux at 
all wavelengths in order to obtain flux from a consistent area of each galaxy across our wavelength 
range. The size and position angle of each aperture as well as on which image it was determined is
given in Table \ref{sample}. Background regions were selected to mimic the content of background
and foreground objects in the aperture on the outskirts of the galaxies. Once the aperture was 
selected, flux densities in the aperture and background regions were measured using the analysis 
tools of the SAOImage DS9 \citep{joy03}.

Due to the proximity of some members of the same interacting system, their apertures can overlap. 
We dealt with these situations in one of three ways. For significantly overlapping systems (NGC 4038/4039, 
NGC 3690/IC 694, and NGC 3395/3396), separate apertures could not be robustly determined. In these 
cases, we treated the combined system as a single object. Second, there were two systems (M51 A/B and 
NGC 5394/5495) where the aperture for the smaller galaxy was mostly contained within the aperture of 
the larger galaxy, but it was clear that the emission in the overlap area came from the smaller galaxy. In 
these cases, we subtracted the emission and area of the  overlap region from that of the larger 
aperture. Third, there were three systems (NGC 3226/3227, NGC 3786/3788, and NGC 4647/4649) 
where the aperture overlapped but without significant contamination. In these cases, we extrapolated the 
expected flux in the overlap area from the surface brightness in the rest of the elliptical aperture at the same radii. 
 
The \emph{Spitzer} fluxes  required aperture corrections. We determined the effective radius of the 
elliptical aperture\footnote{r$_{{\rm eff}}=\sqrt{a b}$ for semi-major axis \emph{a} and semi-minor 
axis \emph{b}} and used the extended source flux corrections given in the IRAC Instrument 
Handbook\footnote{http://irsa.ipac.caltech.edu/data/SPITZER/docs/irac/\\ iracinstrumenthandbook/30/}. 
For the MIPS 24\um~aperture  corrections, we interpolated between the aperture corrections given in the 
MIPS Instrument Handbook\footnote{http://irsa.ipac.caltech.edu/data/SPITZER/docs/mips/\\ mipsinstrumenthandbook/50/}. 
The \emph{GALEX} data were corrected for obscuration due to  Milky Way dust using the extinction laws given by \citet{wyder05}.

Uncertainties in the absolute fluxes are the sum in quadrature of a statistical uncertainty and a 
calibration uncertainty. The \emph{Spitzer} bandpass uncertainties are typically dominated by the 
calibration uncertainty of 3\% for IRAC \citep{coh03} and 4\% for MIPS 24\um~\citep{eng07}. We used a 
calibration uncertainty of 10\% for the \emph{GALEX} data \citep{god04} and a 5-15\% uncertainty 
for the \emph{Swift} bands \citep{poole08}, and the statistical uncertainty is calculated using Poisson 
statistics. We used a 7\% calibration uncertainty for the SPIRE bandpasses \citep{swi10} and 
10\% for the PACS bandpasses \citep{pal12} and followed \citet{dale11} in calculating the statistical 
uncertainty. The  photometry results for \emph{GALEX},  \emph{Swift}, and 2MASS;  \emph{Spitzer}; and 
\emph{Herschel} are provided in Table \ref{fluxes_uv}-\ref{fluxes_fir}, respectively. When flux is not
determined significantly, we provide 3$\sigma$ upper limits, but we do not provide lower limits in 
cases of saturated images. The additional photometry from the literature is given in Table \ref{litflux}.

{\begin{turnpage}
\begin{deluxetable*}{lcccccccccc}[h]
\tabletypesize{\scriptsize}
\tablecaption{\emph{GALEX, Swift}, and 2MASS Photometry\label{fluxes_uv}}
\tablewidth{0pt}
\tablehead{
\colhead{} &\multicolumn{2}{c}{GALEX} &\multicolumn{3}{c}{Swift} &\multicolumn{3}{c}{2MASS} \\
\colhead{Galaxy} & \colhead{FUV} & \colhead{NUV} & \colhead{UVW2} & \colhead{UVM2} & \colhead{UVW1}& \colhead{J} & \colhead{H} & \colhead{Ks}  \\
\colhead{} & \colhead{(mJy)} & \colhead{(mJy)} & \colhead{(mJy)} & \colhead{(mJy)} & \colhead{(mJy)} & \colhead{(mJy)} & \colhead{(mJy)} &
 \colhead{(mJy)}  }
\startdata
NGC2976			& 24.85$\pm$2.49 	& 38.70$\pm$3.87	&  ...	&...	&...	& 766.4$\pm$36.6	& 854.5$\pm$55.6	& 678.3$\pm$60.2	\\
NGC3031			& 175.5$\pm$17.6	& 274.2$\pm$27.4	&...	&...	&...	& 19730$\pm$730	& 23760$\pm$960	& 19770$\pm$810	\\
NGC3034			& 8.57$\pm$0.86	& 35.29$\pm$3.53	&...	&...	&...	& 6614$\pm$230.	& 8972$\pm$322	& 8550.$\pm$304	\\
NGC3077			&...	&...	&...	&...	&	...						& 832.4$\pm$32.0	& 941.5$\pm$39.7	& 745.9$\pm$35.2	\\
NGC3185			& 2.05$\pm$0.21	& 2.27$\pm$0.23	&...	&...	&...	& 172.5$\pm$9.6	& 200.0$\pm$15.5	& 170.9$\pm$16.9	\\
NGC3187			& 3.28$\pm$0.33	& 4.49$\pm$0.45	&...	&...	&...	& 51.05$\pm$8.68	& 81.91$\pm$14.54	& 53.14$\pm$16.01	\\
NGC3190			& 0.40$\pm$0.04	& 1.81$\pm$0.18      &...	&...	&...	& 626.8$\pm$22.9     & 792.1$\pm$30.3	& 684.8$\pm$27.6 	\\
NGC3226			& ...	&	...						& 0.86$\pm$0.09	& 0.80$\pm$0.12	& 2.65$\pm$0.39	
				& 254.2$\pm$10.5	& 269.9$\pm$13.3	& 232.2$\pm$12.8	\\
NGC3227			&...	&	...						& 4.37$\pm$0.47	& 4.10$\pm$0.60	& 9.47$\pm$0.46	
				& 523.9$\pm$19.9	& 607.5$\pm$25.4	& 526.7$\pm$23.4	\\
NGC3395/3396	& 19.15$\pm$1.92	& 27.67$\pm$2.77	&...	&...	&...	& 155.1$\pm$10.5	& 184.1$\pm$18.8	& 159.4$\pm$15.5	\\
NGC3424			& 0.91$\pm$0.09	& 2.69$\pm$0.27	& 1.85$\pm$0.20	& 1.70$\pm$0.25	& 3.61$\pm$0.18	
					& 163.6$\pm$7.2	& 206.3$\pm$9.9	& 174.9$\pm$9.7	\\
NGC3430			& 8.38$\pm$0.84 	& 12.42$\pm$1.24	&...	&...	&...	& 224.4$\pm$13.2	& 231.7$\pm$18.2	& 165.7$\pm$18.3	\\
NGC3448			& 5.67$\pm$0.58	& 8.96$\pm$0.90	&...	&...	&...	& 106.0$\pm$6.59	& 116.4$\pm$11.1	& 103.4$\pm$9.8	\\
UGC6016			& 1.20$\pm$0.13	& 1.34$\pm$0.14	&...	&...	&...	& $<$ 17.2	& $<$ 23.1	& $<$ 20.5	\\
NGC3690/IC694	& 10.15$\pm$1.03	& 15.70$\pm$1.57	&...	&...	&...	& 222.4$\pm$9.5	& 300.2$\pm$14.5	& 285.3$\pm$13.0	\\
NGC3786			& 0.97$\pm$0.11	& 1.85$\pm$0.19	&...	&...	&...	& 110.9$\pm$5.2	& 135.4$\pm$8.8	& 118.3$\pm$7.4	\\
NGC3788			& 1.61$\pm$0.18	& 2.98$\pm$0.30	&...	&...	&...	& 117.6$\pm$5.0	& 139.8$\pm$8.3	& 120.5$\pm$7.0	\\
NGC4038/4039	& 34.15$\pm$3.42	& 55.10$\pm$5.51	&...	&...	&...	& 927.6$\pm$44.7	& 1126$\pm$62	& 915.7$\pm$76.2	\\
NGC4618			& 27.69$\pm$2.77	& 39.19$\pm$3.92	&...	&...	&...	& 326.8$\pm$30.7	& 318.0$\pm$50.7	& 245.9$\pm$45.3	\\
NGC4625			& 4.83$\pm$0.48	& 7.30$\pm$0.73	&...	&...	&...	& 90.41$\pm$14.42	& 111.5$\pm$22.3	& 76.6$\pm$24.3	\\
NGC4647			& 4.66$\pm$047	& 10.00$\pm$1.00	&...	&...	&...	& 473.4$\pm$21.2	& 528.1$\pm$25.7	& 441.9$\pm$28.0	\\
NGC4649			& 3.80$\pm$0.38	& 7.29$\pm$0.73	&...	&...	&...	& 2694$\pm$94	& 3290.$\pm$119	& 2713$\pm$99	\\
M51A			& 116.1$\pm$11.6	& 202.5$\pm$20.3	&...	&...	&...	& 4285$\pm$165	& 5343$\pm$243	& 4213$\pm$255	\\
M51B			& 6.75$\pm$0.68	& 12.65$\pm$1.27	&...	&...	&...	& 1895$\pm$69	& 2374$\pm$91	& 1915$\pm$78	\\
NGC5394			& 0.61$\pm$0.06	& 1.48$\pm$0.15	&...	&...	&...	& 62.17$\pm$2.98	& 77.30$\pm$4.45	& 66.69$\pm$5.08	\\
NGC5395			& 2.91$\pm$0.29	& 5.70$\pm$0.57	&...	&...	&...	& 250.9$\pm$11.3	& 301.5$\pm$16.7	& 247.6$\pm$18.9	\\
M101			& 338.1$\pm$33.8	& 452.9$\pm$45.3	&...	&...	&...	& 3711$\pm$338	& 4279$\pm$551	& 3893$\pm$569	\\
NGC5474			& 19.47$\pm$1.95	& 25.91$\pm$2.59	&...	&...	&...	& 228.7$\pm$32.0	& 231.9$\pm$47.7	& 172.4$\pm$57.9				
\enddata
\tablecomments{The upper limits given are 3$\sigma$ upper limits.}
\end{deluxetable*}

\begin{deluxetable*}{lccccccccccc}
\tabletypesize{\scriptsize}
\tablecaption{\emph{Spitzer} IRAC and MIPS Photometry\label{fluxes_mir}}
\tablewidth{0pt}
\tablehead{
\colhead{Galaxy} & \colhead{3.6\um} & \colhead{4.5\um} & \colhead{5.8\um} & \colhead{8.0\um} & \colhead{24\um}  \\
\colhead{} &  \colhead{(mJy)} & \colhead{(mJy)} & \colhead{(mJy)} & \colhead{(mJy)} & \colhead{(mJy)} 
}
\startdata
NGC2976			& 393.2$\pm$11.8	& 269.1$\pm$8.1	& 476.6$\pm$14.3	& 957.7$\pm$28.7	& 1454$\pm$58	\\
NGC3031			& 9936$\pm$298	& 6146$\pm$492	& 5217$\pm$417	& 6329$\pm$506 	& 6011$\pm$240	\\
NGC3034			& 6564$\pm$197	& 5223$\pm$157	& ...	&...	&...	\\
NGC3077			&373.3$\pm$11.2	& 267.3$\pm$8.0	& 298.1$\pm$9.0	& 571.1$\pm$17.1	& 1752$\pm$53	\\
NGC3185			& 76.44$\pm$2.29	& 50.53$\pm$1.52	& 52.95$\pm$1.60	& 115.0$\pm$3.5	& 192.3$\pm$7.7	\\
NGC3187			& 22.51$\pm$0.68	& 16.09$\pm$0.48	& 26.48$\pm$0.80	& 62.56$\pm$1.88	& 91.48$\pm$3.66	\\
NGC3190			& 337.0$\pm$10.1	& 213.9$\pm$6.4	& 176.1$\pm$5.3	& 288.8$\pm$8.7	& 271.9$\pm$10.9	\\
NGC3226			& 122.2$\pm$3.7	& 76.34$\pm$2.29	& 48.92$\pm$1.47	& 44.26$\pm$1.33	& 37.10$\pm$1.48	\\
NGC3227			& 287.7$\pm$8.6	& 218.7$\pm$6.6	& 256.4$\pm$7.7	& 597.0$\pm$17.9	& 1769$\pm$71	\\
NGC3395/3396	& 87.62$\pm$2.63	& 61.30$\pm$1.84	& 145.1$\pm$4.4	& 423.2$\pm$12.7	& 1190.$\pm$48	\\
NGC3424			& 103.4$\pm$3.1	& 72.75$\pm$2.18	& 148.3$\pm$4.5	& 460.7$\pm$13.8	& 776.4$\pm$31.0	\\
NGC3430			& 116.0$\pm$3.5	& 78.97$\pm$2.37	& 110.2$\pm$3.3	& 372.0$\pm$11.2	& 434.7$\pm$17.4	\\
NGC3448			& 62.03$\pm$1.86	& 44.02$\pm$1.32	& 79.53$\pm$2.39	& 193.53$\pm$5.81	& 580.7$\pm$23.2	\\
UGC6016			& 1.52$\pm$0.05	& 0.90$\pm$0.03	& $<$ 8.0			& 2.13$\pm$0.15	& 4.00$\pm$0.13	\\
NGC3690/IC694	& 293.2$\pm$8.8	& 347.6$\pm$10.4	& 841.0$\pm$25.2	&	...			& 18660$\pm$750	\\
NGC3786			& 32.61$\pm$0.98	& 21.07$\pm$0.63	& 27.69$\pm$0.83	& 66.39$\pm$1.99	& 266.5$\pm$10.7	\\
NGC3788			& 30.97$\pm$0.93	& 23.80$\pm$0.71	& 26.32$\pm$0.79	& 60.14$\pm$1.80	& 166.1$\pm$6.6	\\
NGC4038/4039	& 523.3$\pm$15.7 	& 359.1$\pm$10.8	& 706.1$\pm$21.2	& 1757$\pm$53	& 6131$\pm$245	\\
NGC4618			& 152.2$\pm$4.6	& 97.15$\pm$2.91	& 157.6$\pm$4.7	& 327.5$\pm$9.8	& 394.3$\pm$15.7	\\
NGC4625			& 43.04$\pm$1.29	& 27.67$\pm$0.83	& 45.97$\pm$1.38	& 126.3$\pm$3.8	& 124.4$\pm$5.0	\\
NGC4647			& 195.3$\pm$5.9	& 124.3$\pm$3.7	& 222.2$\pm$6.9	& 553.0$\pm$16.6	& 612.9$\pm$24.5	\\
NGC4649			& 1202$\pm$36 	& 711.6$\pm$21.4	& 449.6$\pm$13.5	& 280.0$\pm$8.4	& 126.9$\pm$5.1	\\
M51A			& 2474$\pm$78	& 1662$\pm$54	& 3637$\pm$110.	& 10790$\pm$320	& 12520$\pm$510	\\
M51B			& 965.9$\pm$37.9	& 632.9$\pm$28.0	& 667.5$\pm$25.9	& 1430.1$\pm$50.9	& 2149$\pm$94	\\
NGC5394			& 40.79$\pm$1.22	& 28.54$\pm$0.86	& 67.03$\pm$2.01	& 208.5$\pm$6.3	& 854.7$\pm$34.2	\\
NGC5395			& 141.0$\pm$4.2	& 95.16$\pm$2.85	& 143.0$\pm$4.3	& 404.4$\pm$12.1	& 444.1$\pm$17.8	\\
M101			& 2373$\pm$71	& 1593$\pm$48	& 3056$\pm$92	& 7423$\pm$223	& 10610$\pm$425	\\
NGC5474			& 98.27$\pm$2.95	& 66.25$\pm$1.99	& 75.66$\pm$2.27	& 105.9$\pm$3.2	& 151.1$\pm$7.0	
\enddata
\tablecomments{IRAC 5.8\um, IRAC 8.0\um, and MIPS 24\um~are saturated for NGC 3034, as is 8\um~for NGC 3690/IC 694. The
upper limits are 3$\sigma$ upper limits.}
\end{deluxetable*}

\begin{deluxetable*}{lccccccccccc}
\tabletypesize{\scriptsize}
\tablecaption{\emph{Herschel} PACS and SPIRE Photometry\label{fluxes_fir}}
\tablewidth{0pt}
\tablehead{
\colhead{} &   \multicolumn{3}{c}{PACS} & \multicolumn{3}{c}{SPIRE}\\
\colhead{Galaxy}&  \colhead{75\um} & \colhead{110\um} & \colhead{170\um} & \colhead{250\um} & \colhead{350\um} & \colhead{500\um} \\
\colhead{}  & \colhead{(Jy)} & \colhead{(Jy)} & \colhead{(Jy)} & \colhead{(Jy)} & \colhead{(Jy)} &  \colhead{(Jy)}  
}
\startdata
NGC2976			& 35.48$\pm$3.55	& 48.90$\pm$4.90	& 48.88$\pm$4.89	
				& 24.87$\pm$1.74	& 11.84$\pm$0.83	& 4.86$\pm$0.34	\\
NGC3031			& 67.56$\pm$6.86	& ...	& 351.5$\pm$35.2	
				& 161.5$\pm$11.3	& 78.75$\pm$5.51	& 32.98$\pm$2.31	\\
NGC3034			& 1985$\pm$198	& ...	& 1291$\pm$129	
				& 363.1$\pm$25.4	& 121.5$\pm$8.5	& 35.45$\pm$2.48	\\
NGC3077			& 22.52$\pm$3.38	& 32.12$\pm$4.82	& 23.77$\pm$3.57	
				& 8.54$\pm$0.60	& 3.36$\pm$0.24	& 1.18$\pm$0.08	\\
NGC3185			&...	&...	&	...
				& 2.50$\pm$0.21	& 1.23$\pm$0.14	& 0.38$\pm$0.09	\\
NGC3187			& 2.20$\pm$0.39	& 5.52$\pm$0.89	& 3.87$\pm$0.62	
				& 2.37$\pm$0.17	& 1.39$\pm$0.10	& 0.69$\pm$0.05	\\
NGC3190			& 6.98$\pm$1.06	& 12.29$\pm$1.87	& 16.86$\pm$2.54	
				& 8.06$\pm$0.56	& 3.45$\pm$0.24	& 1.20$\pm$0.08	\\
NGC3226			& 0.22$\pm$0.05	&...	& 2.59$\pm$0.27	
				& 0.81$\pm$0.06	& 0.30$\pm$0.02	& 0.10$\pm$0.01	\\
NGC3227			& 11.87$\pm$1.19	&...	& 22.33$\pm$2.24	
				& 10.96$\pm$0.77	& 4.43$\pm$0.31	& 1.50$\pm$0.11	\\
NGC3395/3396	& 12.94$\pm$1.45	& 16.49$\pm$1.78	& 17.19$\pm$1.75	
				& 6.95$\pm$0.49	& 2.93$\pm$0.21	& 1.06$\pm$0.08	\\
NGC3424			& ...	&...	&	...
				& 8.15$\pm$0.57	& 3.37$\pm$0.24	& 1.13$\pm$0.08	\\
NGC3430			&...	&	...&	...
				& 8.07$\pm$0.57	& 3.61$\pm$0.25	& 1.38$\pm$0.10	\\
NGC3448			&...	&	...&	...
				& 4.68$\pm$0.33	& 2.11$\pm$0.15	& 0.84$\pm$0.06	\\
UGC6016			&...	&...	&	...
				& 0.10$\pm$0.02	& 0.060$\pm$0.014	& 0.014$\pm$0.002	\\
NGC3690/IC694	& 139.3$\pm$13.9	& 126.7$\pm$12.7	& 74.19$\pm$7.42	
				& 21.34$\pm$1.49	& 7.37$\pm$0.52	& 2.22$\pm$0.16	\\
NGC3786			& 2.30$\pm$0.25	& ...	& 3.93$\pm$0.42	
				& 1.97$\pm$0.14	& 0.83$\pm$0.06	& 0.27$\pm$0.02	\\
NGC3788			& 2.02$\pm$0.22	&...	& 6.83$\pm$0.70	
				& 3.29$\pm$0.23	& 1.41$\pm$0.10	& 0.49$\pm$0.04	\\
NGC4038/4039	& 80.95$\pm$8.11	& 116.0$\pm$11.6	& 99.79$\pm$9.98	
				& 37.57$\pm$2.63	& 14.82$\pm$1.04	& 5.01$\pm$0.35	\\
NGC4618			&	...&...	&	...
				& 8.61$\pm$0.60	& 4.19$\pm$0.29	& 1.71$\pm$0.12	\\
NGC4625			& 2.94$\pm$0.31	& 2.87$\pm$0.33	& 4.86$\pm$0.50	
				& 2.40$\pm$0.17	& 1.16$\pm$0.08	& 0.47$\pm$0.40	\\
NGC4647			&...	&	...&	...
				& 11.12$\pm$0.78	& 4.60$\pm$0.32	& 1.56$\pm$0.11	\\
NGC4649			& ...	&...	&	...
				& $<$ 0.09	& $<$ 0.09	& $<$ 0.06	\\
M51A			& 181.1$\pm$18.1	& ...	& 441.4$\pm$44.1	
				& 184.2$\pm$12.9	& 74.22$\pm$5.20	& 25.38$\pm$1.78	\\
M51B			& 24.63$\pm$2.47	&...	& 53.72$\pm$5.37	
				& 20.71$\pm$1.45	& 8.22$\pm$0.58	& 2.72$\pm$0.19	\\
NGC5394			& 6.04$\pm$0.61	& 8.31$\pm$0.83	& 8.27$\pm$0.83	
				& 2.95$\pm$0.21	& 1.06$\pm$0.07	& 0.33$\pm$0.02	\\
NGC5395			& 7.30$\pm$0.76	& 11.03$\pm$1.12	& 16.28$\pm$1.64	
				& 8.73$\pm$0.61	& 3.80$\pm$0.27	& 1.39$\pm$0.10	\\
M101			& 97.05$\pm$14.56	& 265.0$\pm$39.8	& 373.4$\pm$56.0	
				& 172.6$\pm$12.1	& 79.91$\pm$5.60	& 31.66$\pm$2.22	\\
NGC5474			& 2.84$\pm$0.28	& 5.81$\pm$0.58	& 9.08$\pm$0.908	
				& 3.55$\pm$0.26	& 1.97$\pm$0.15	& 0.86$\pm$0.08	
\enddata
\end{deluxetable*}

\begin{deluxetable*}{lccccccclcc}
\tabletypesize{\scriptsize}
\tablecaption{Literature Photometry\label{litflux}}
\tablewidth{0pt}
\tablehead{
\colhead{} & \multicolumn{3}{c}{Third Reference Catalogue} & 
\multicolumn{5}{c}{IRAS}  & \multicolumn{2}{c}{MIPS} \\
 \colhead{Galaxy} & \colhead{U} & \colhead{B} & \colhead{V} & 
 \colhead{12\um}& \colhead{25\um}& \colhead{60\um}& \colhead{100\um} & \colhead{Ref.} & \colhead{70\um} & \colhead{160\um} \\
\colhead{}  & \colhead{(mJy)} & \colhead{(mJy)} & \colhead{(mJy)} & \colhead{(Jy)} & \colhead{(Jy)} & \colhead{(Jy)} &  \colhead{(Jy)}   & \colhead{} &  \colhead{(Jy)}  &  \colhead{(Jy)}  
}
\startdata
NGC2976		     	&85.10$\pm$10.90	& 200.0$\pm$25.5	 &314.0$\pm$40.1	& 
				0.92$\pm$0.02		& 1.71$\pm$0.02	 & 13.09$\pm$0.29	& 33.43$\pm$0.34 & (2)
				&20.43$\pm$4.34\tablenotemark{b}	& 53.56$\pm$12.28		\\	 
NGC3031		     	&...				& 2970.$\pm$83~~	 &6100.$\pm$180.	&    		 
				5.86$\pm$0.88		& 5.42$\pm$0.81 	 &44.73$\pm$6.71	& 174.02$\pm$26.10 & (2)
				&85.18$\pm$18.05	& 360.0$\pm$84.1	\\	 
NGC3034		     	&259.0$\pm$23.70	& 812.0$\pm$70.2&1570.$\pm$137.	& 	     
				79.43$\pm$7.94~~	& 332.6$\pm$33.3~~	&1480.$\pm$148~	& 1374.$\pm$137. & (2)
				&...	& ...	\\	 
NGC3077		     	&90.70$\pm$11.70	& 243.0$\pm$30.9	&418.0$\pm$53.3	&  
				0.76$\pm$0.02		& 1.88$\pm$0.03	&15.90$\pm$0.04	& 26.53$\pm$0.13 & (2)
				&...	& ...	\\	 
NGC3185		     	&...				& 27.10$\pm$1.81	&49.30$\pm$4.07	& 
				0.15$\pm$0.04\tablenotemark{a}	& 0.14$\pm$0.04\tablenotemark{a}&1.43$\pm$0.09	& 3.67$\pm$0.22 & (4)
				&...	& ...	\\	 
NGC3187		     	&5.94$\pm$0.73	& 11.60$\pm$1.24	&15.30$\pm$1.74	& 
				...	& ...	&...	& ... &
				&...	&...	\\	 
NGC3190		     	&16.50$\pm$1.63~	& 60.50$\pm$5.83	&126.0$\pm$12.2	& 
				0.32$\pm$0.03\tablenotemark{c}		&0.35$\pm$0.08\tablenotemark{c}	&3.19$\pm$0.35\tablenotemark{c}	& 10.11$\pm$0.51\tablenotemark{c} & (4)
				&5.66$\pm$1.20	& 15.01$\pm$3.51	\\	 
NGC3226		     	&...	&51.20$\pm$10.14	&100.0$\pm$21.0	& 
				$<$0.27	& $<$0.18 	& $<$0.60	& ... & (1) 
				&...	& ...	\\	 
NGC3227		     	&51.30$\pm$10.60	& 155.0$\pm$31.3~	&281.0$\pm$57.6	& 
				1.11$\pm$0.12	& 2.04$\pm$0.21	&9.01$\pm$1.00	& 19.11$\pm$2.00 & (1)
				&...	& ...	\\	 
NGC3395/3396  		&...	& ...	&...	& 
				0.47$\pm$0.03	& 1.47$\pm$0.04	&10.92$\pm$0.03~~ 	& 17.83$\pm$0.10 & (2)
				&...	& ...	\\	 
NGC3424		     	&...	& ...	&...	& 
				0.59$\pm$0.04	& 0.94$\pm$0.04	&9.03$\pm$0.09	& 17.03$\pm$0.21 & (1)
				&...	& ...\\	 
NGC3430		   	&21.60$\pm$4.51	& 56.20$\pm$11.40	&87.30$\pm$17.80	& 
				0.38$\pm$0.08	& 0.78$\pm$0.05	&4.36$\pm$0.07	& 10.88$\pm$1.00 & (1)
				&...	& ...	\\	 
NGC3448		     	&22.00$\pm$2.83	& 43.40$\pm$5.52~	&55.10$\pm$7.03	& 
				0.34$\pm$0.11\tablenotemark{a}	& 0.76$\pm$0.21\tablenotemark{a}	&6.74$\pm$0.35	& 12.17$\pm$0.47 & (1)
				&...	& ...	\\	 
UGC6016		     	&...	& ...	&...	& 
				$<$0.10	& $<$0.15	&$<$0.50	& $<$0.50 & (1)
				&...	& ...	\\	 
NGC3690/IC694		&...	& ...	&... 	& 
				3.90$\pm$0.40	& 24.14$\pm$2.40	& 121.64$\pm$	12.50& 122.45$\pm$12.50 & (1)
				&... & ....	\\	 
NGC3786		     	&...	& ...	&...	& 
				...	& ...	&...	& ... &
				&...	& ...	\\	 
NGC3788		     	&...	& ...	&...	& 
				...	& ...	&...	& ... &
				&...	& ...	\\	 
NGC4038/4039		&...	& ...	&... 	& 
				1.94$\pm$0.04	& 6.54$\pm$0.03	& 45.16$\pm$0.06	& 87.09$\pm$0.20~~ & (2)
				&...	& ...	\\	 
NGC4618		     	&70.10$\pm$2.95	& 139.0$\pm$5.2	&178.0$\pm$6.9	& 
				0.40$\pm0.03$	& 0.45$\pm$0.04	&4.92$\pm$0.04	& 13.05$\pm$0.17 & (3) 
				&...	& ...	\\	 
NGC4625		     	&14.50$\pm$0.88	& 28.90$\pm$1.09	&41.80$\pm$1.76	& 	
				0.12$\pm$0.03\tablenotemark{a}	&0.19$\pm$0.04	&1.20$\pm$0.13	& 3.58$\pm$0.25 & (4)
				&2.06$\pm$0.44	& 5.42$\pm$1.28	\\	 
NGC4647		     	&23.20$\pm$2.91	& 71.40$\pm$5.46	&111.0$\pm$8.8 	& 
				0.76$\pm$0.08	& 1.06$\pm$0.16	&6.04$\pm$0.24	& 17.56$\pm$0.81 & (1)
				&...	& ...	\\	 
NGC4649		     	&...	& 508.0$\pm$23.9	&1060.$\pm$509	& 
				0.33$\pm$0.05	& $<$0.32	&$<$0.24	& $<$1.05 & (1)
				&...	& ...	\\	 
M51A		     	&...	& 1110.$\pm$63~~	&1650.$\pm$95~~	&  
				7.21$\pm$0.08	&9.56$\pm$0.08\tablenotemark{b}	&97.42$\pm$0.19	& 221.2$\pm$0.3 & (2)
				&147.1$\pm$31.3	& 494.7$\pm$113.1	\\	 
M51B		     	&89.90$\pm$6.23	& 282.0$\pm$18.7	&551.0$\pm$37.1	& 
				 0.35$\pm$0.02\tablenotemark{b}	& 1.02$\pm$0.05\tablenotemark{b}	&15.22$\pm$0.80	& 31.33$\pm$0.37 & (2)
				&16.31$\pm$4.61	& 14.86$\pm$2.97\tablenotemark{b}	\\	 
NGC5394			&5.37$\pm$1.16	& 14.10$\pm$2.85	&22.80$\pm$4.66  	& 
				0.52$\pm$0.05	& 1.19$\pm$0.11	&5.62$\pm$1.41\tablenotemark{a}	& 10.43$\pm$3.10\tablenotemark{a} & (1)
				&...	& ...	\\	 
NGC5395			&23.90$\pm$5.06	& 61.60$\pm$12.50	&103.0$\pm$21.1  	&   
				0.40$\pm$0.04	& 0.48$\pm$0.06	&6.86$\pm$1.50	& 14.21$\pm$3.10 & (1)
				&...	& ...	\\	 
M101		     	&...	& 2020.$\pm$175	&2610.$\pm$248	& 
				6.20$\pm$0.93	& 11.78$\pm$1.77	& 88.04$\pm$13.21	& 253.0$\pm$38.0 & (2)
				&...	& ...	\\	 
NGC5474		     	&...	& 131.0$\pm$19.4	&176.0$\pm$27.6	&  		
				$<$0.88	& $<$0.76	&1.33$\pm$0.07	& 4.80$\pm$0.24 &(4)
				& 3.73$\pm$0.27\tablenotemark{b}	&10.56 $\pm$2.47		 
\enddata	
\tablecomments{The IRAS fluxes are from (1)the HIRES Atlas \citep{sur04}, (2) the IRAS Revised Bright Galaxy Sample 
\citep{san03}, (3) the IRAS Bright Galaxy Sample \citep{soi89}, and (4) the Faint Source Catalogue \citep{mosh90}
 obtained through NED. The MIPS 70\um~and 160\um~data come from \citet{dale07} with the exception of NGC 3031 \citep{dale09} 
and NGC 3077 \citep{eng08}. Upper limits were not used in the fits. }
 \tablenotetext{a}{Not included in fits because detection is less than 4$\sigma$.}
 \tablenotetext{b}{Not included in fits because differs significantly from other measurements.}
 \tablenotetext{c}{Not included in fits because IRAS fluxes likely contains more than one galaxy.}
\end{deluxetable*}
\end{turnpage}}

\begin{figure*}
\centerline{\includegraphics[width=0.925\linewidth]{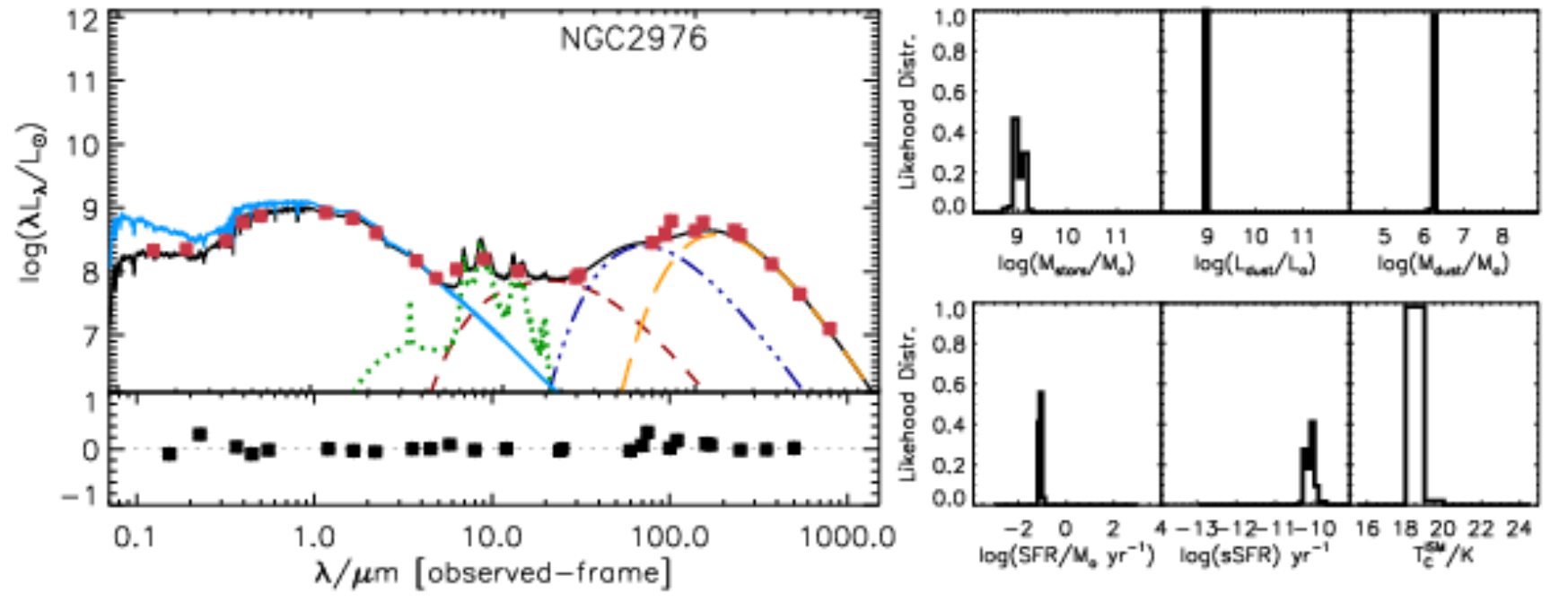}}
\centerline{\includegraphics[width=0.925\linewidth]{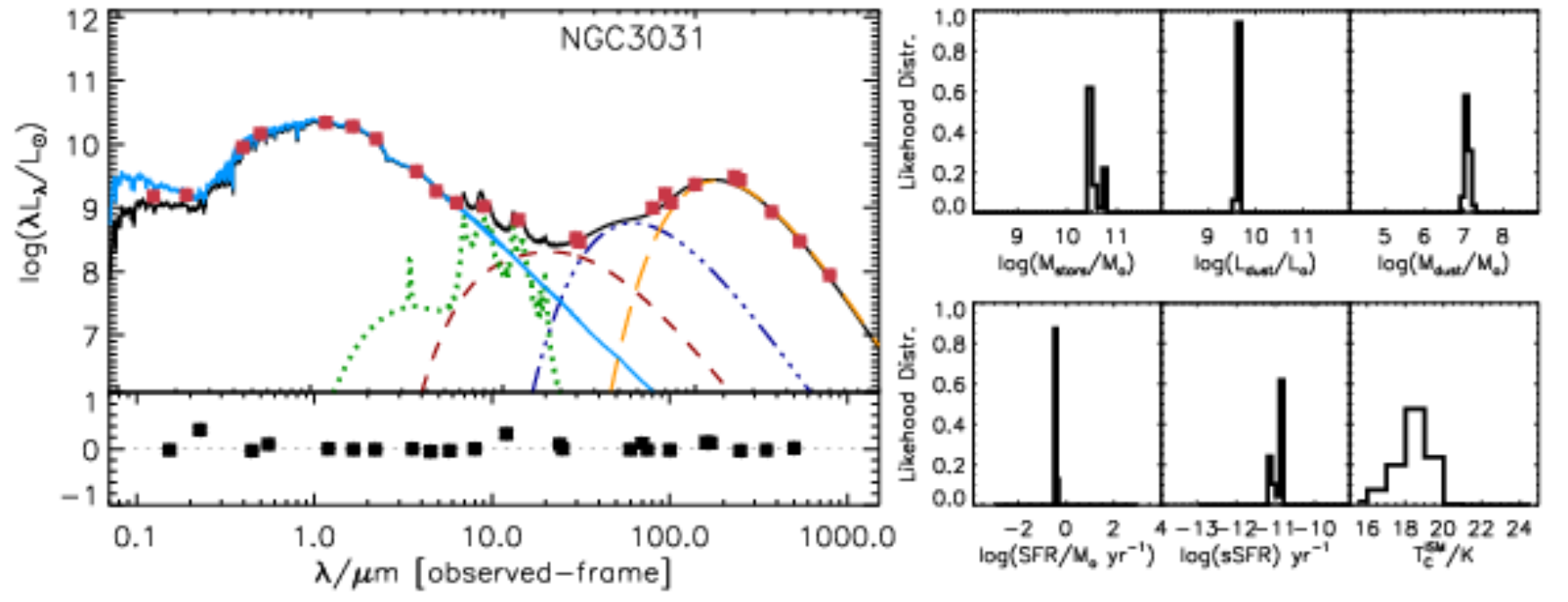}}
\caption{SEDs for NGC 2976 (top) and NGC 3031 (bottom)
 with data shown as red points, the best fit model plotted in black, and
the stellar emission in the absence of dust shown in blue. The components of 
the infrared emission are over plotted: PAH emission (dotted, green line), MIR emission
at  130 K and 250 K (red, dashed line), warm 30-60 K dust emission (dot-dashed, purple line), and
cold 15-25 K dust emission (long dashed, orange line). Below the fitted
SED is plotted the fractional difference between the model and data. To the
right of the SED, we plot a subset of the probability distribution functions (PDFs) 
of the fitted parameters for (from left to right): stellar mass,  dust luminosity, 
and dust mass (top) and SFR, sSFR, and cold dust temperature (bottom).}
\label{seds1}
\end{figure*}

\begin{figure*}
\centerline{\includegraphics[width=0.925\linewidth]{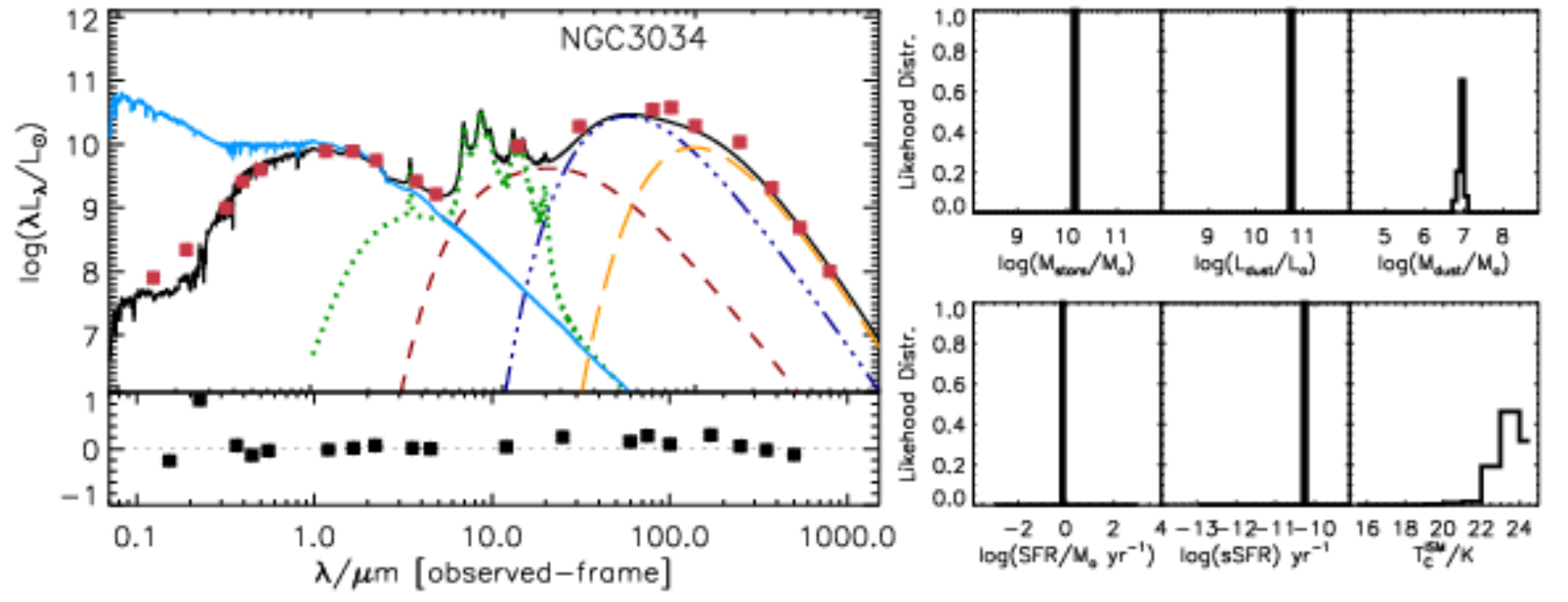}}
\centerline{\includegraphics[width=0.925\linewidth]{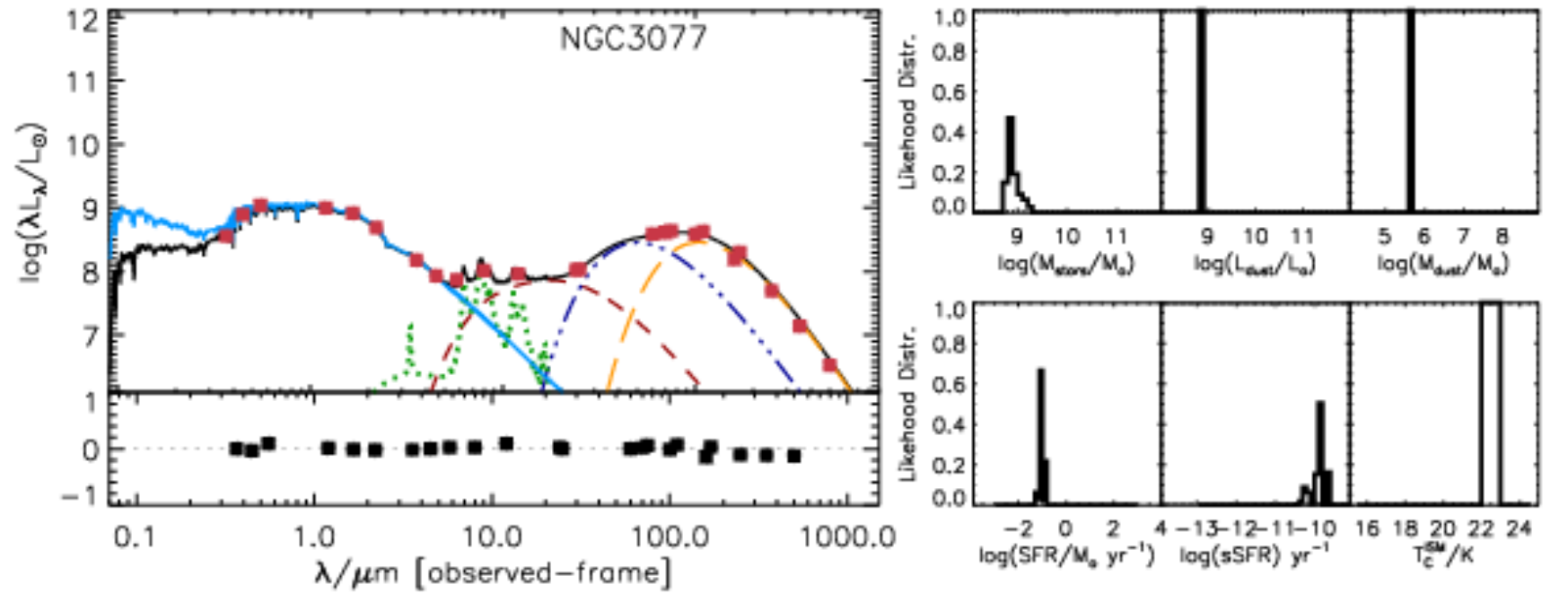}}
\centerline{\includegraphics[width=0.925\linewidth]{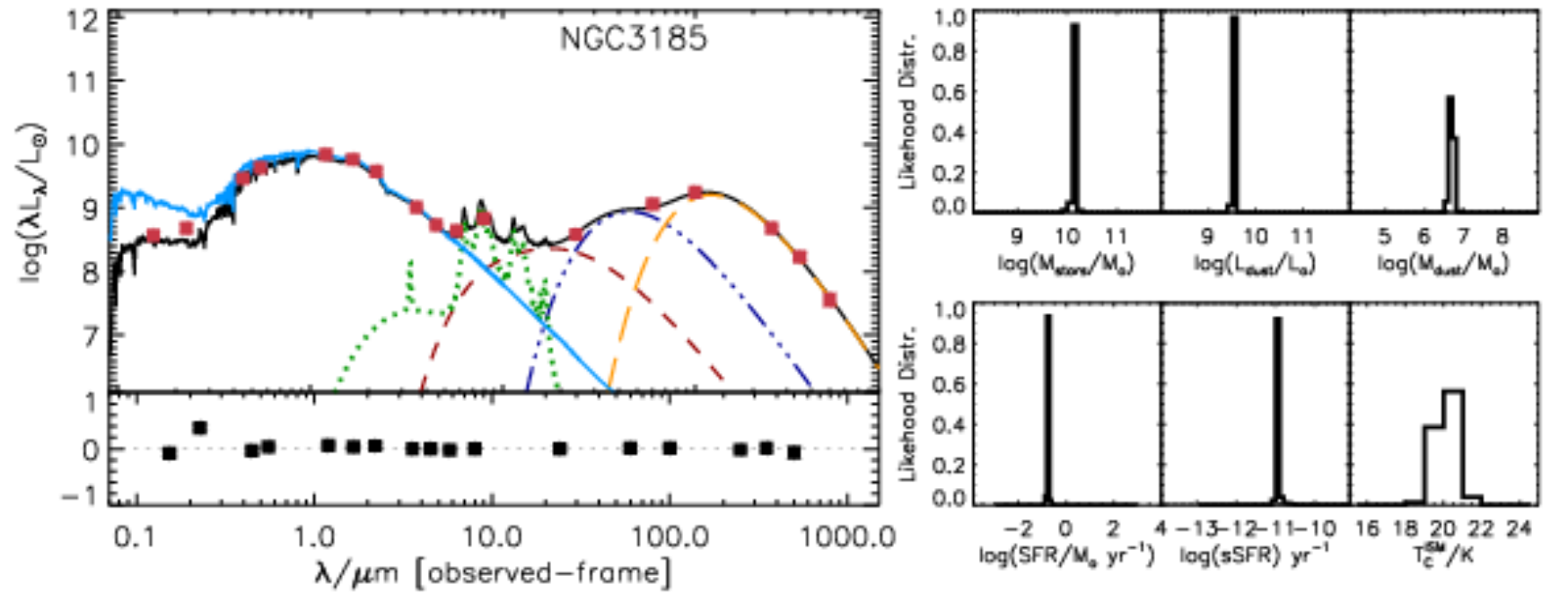}}
\caption{As Figure \ref{seds1}, but for NGC 3034 (top), NGC 3077 (middle), and NGC 3185 (bottom).}
\label{seds2}
\end{figure*}

\begin{figure*}
\centerline{\includegraphics[width=0.925\linewidth]{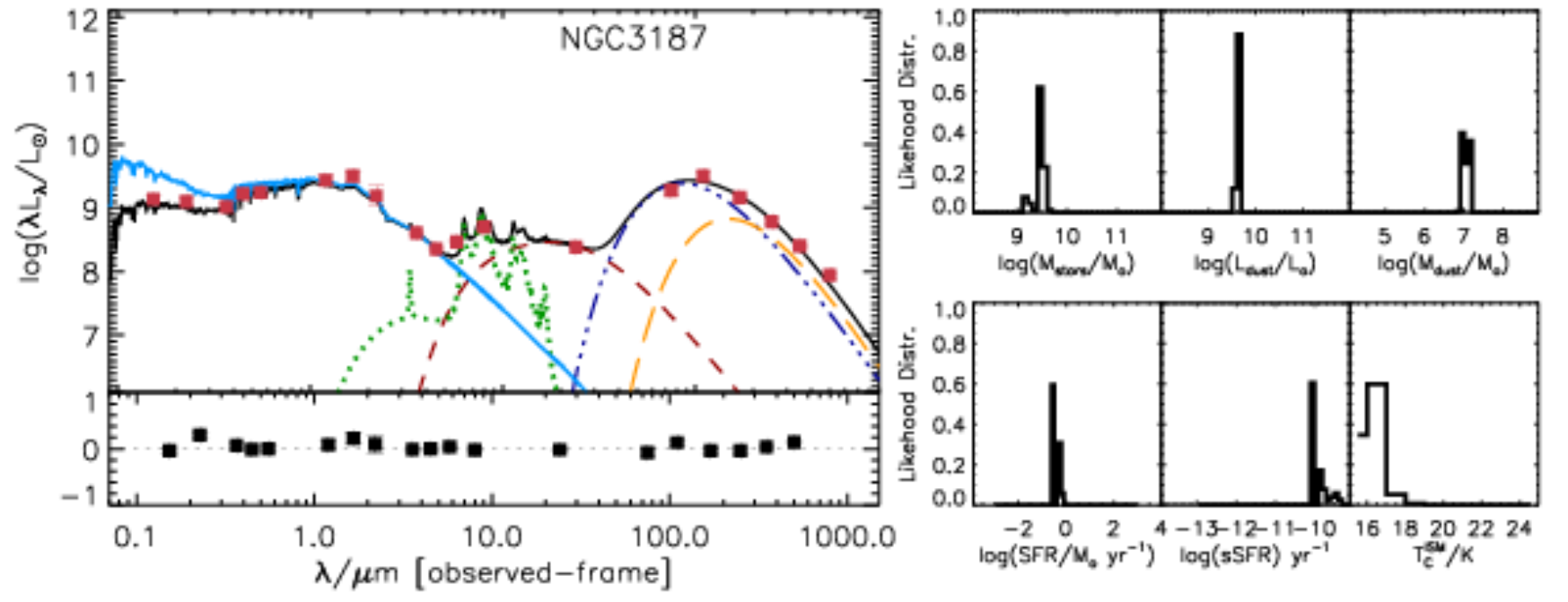}}
\centerline{\includegraphics[width=0.925\linewidth]{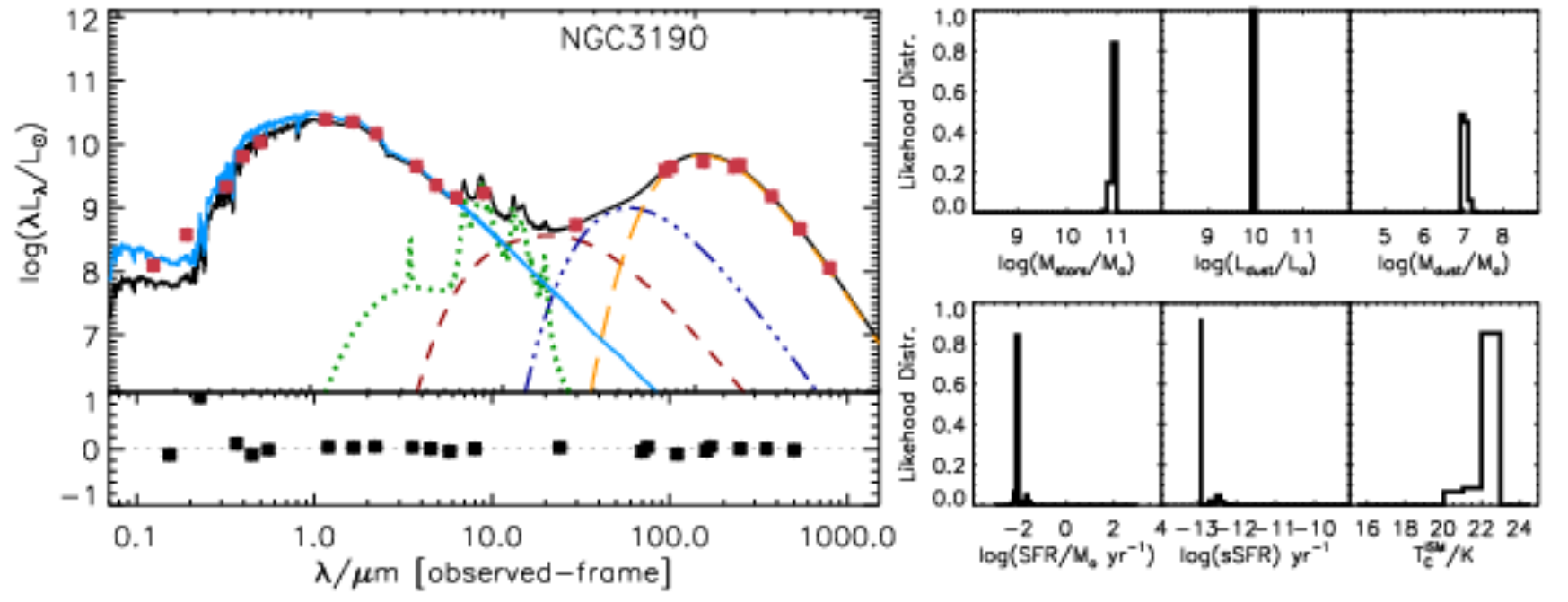}}
\centerline{\includegraphics[width=0.925\linewidth]{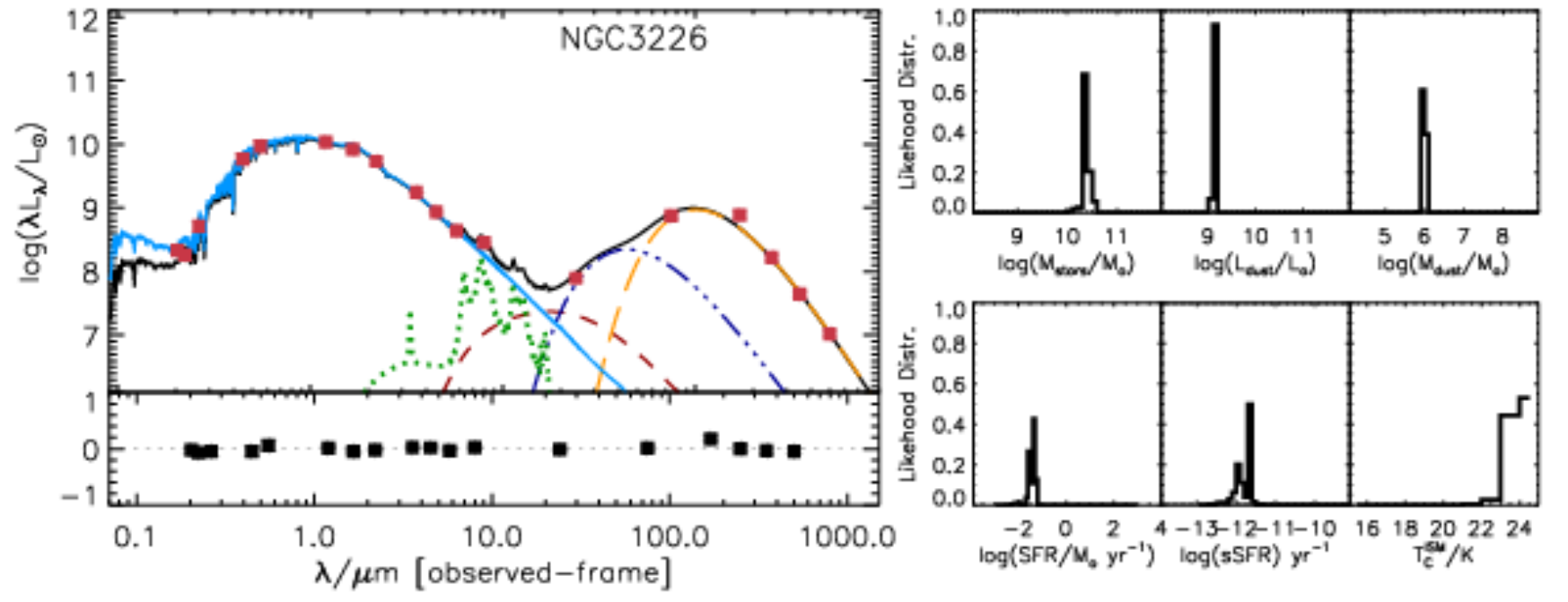}}
\caption{As Figure \ref{seds1}, but for NGC 3187 (top), NGC 3190 (middle), and NGC 3226 (bottom). }
\label{seds3}
\end{figure*}

\begin{figure*}
\centerline{\includegraphics[width=0.925\linewidth]{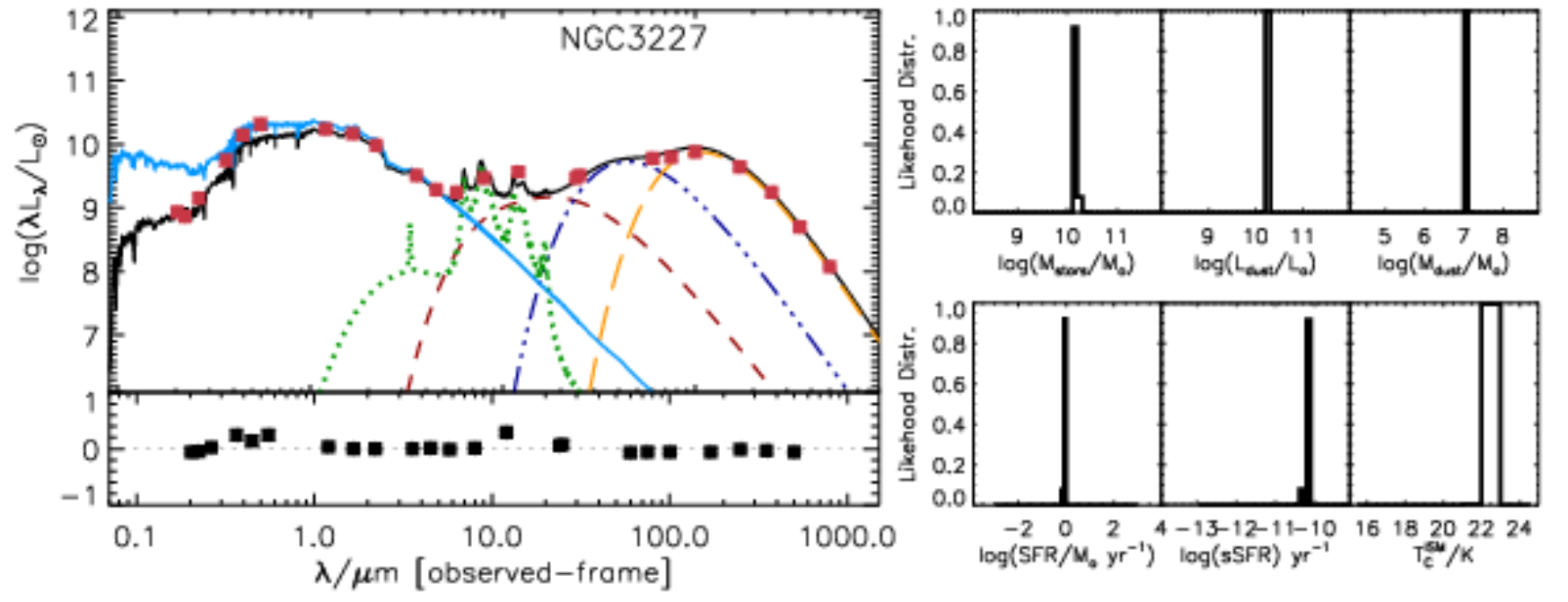}}
\centerline{\includegraphics[width=0.925\linewidth]{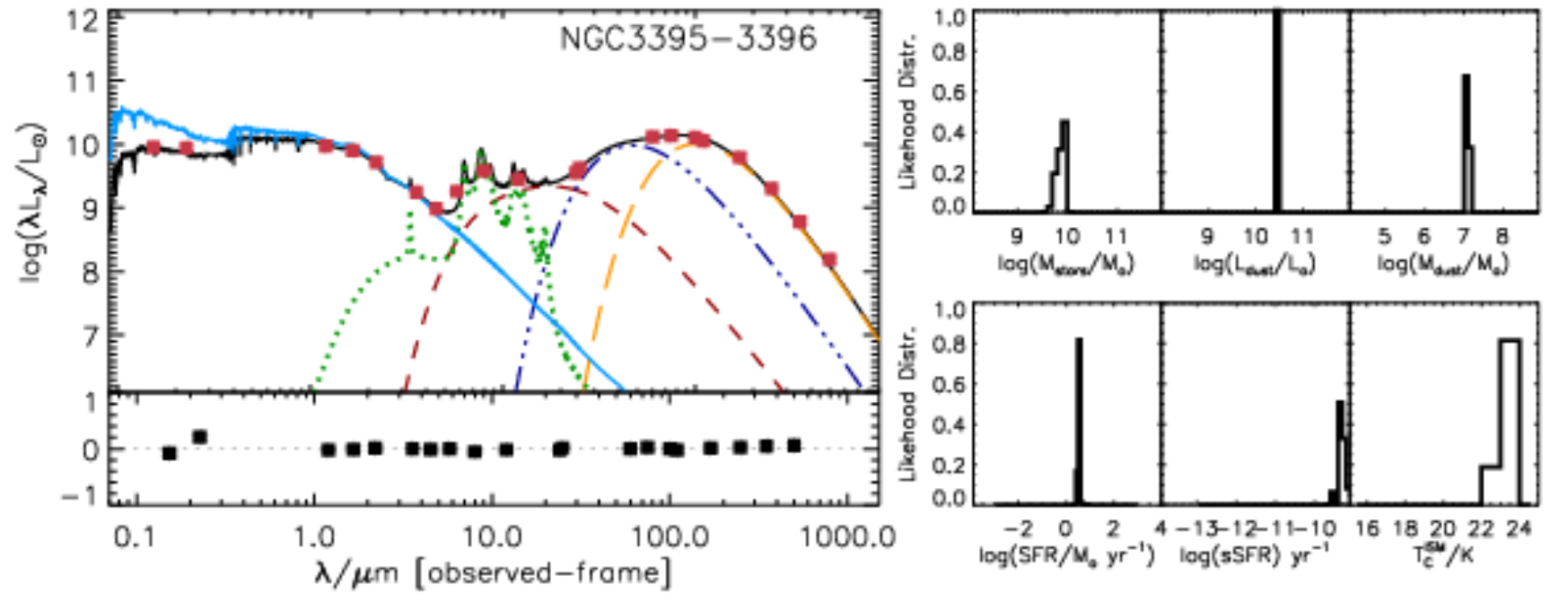}}
\centerline{\includegraphics[width=0.925\linewidth]{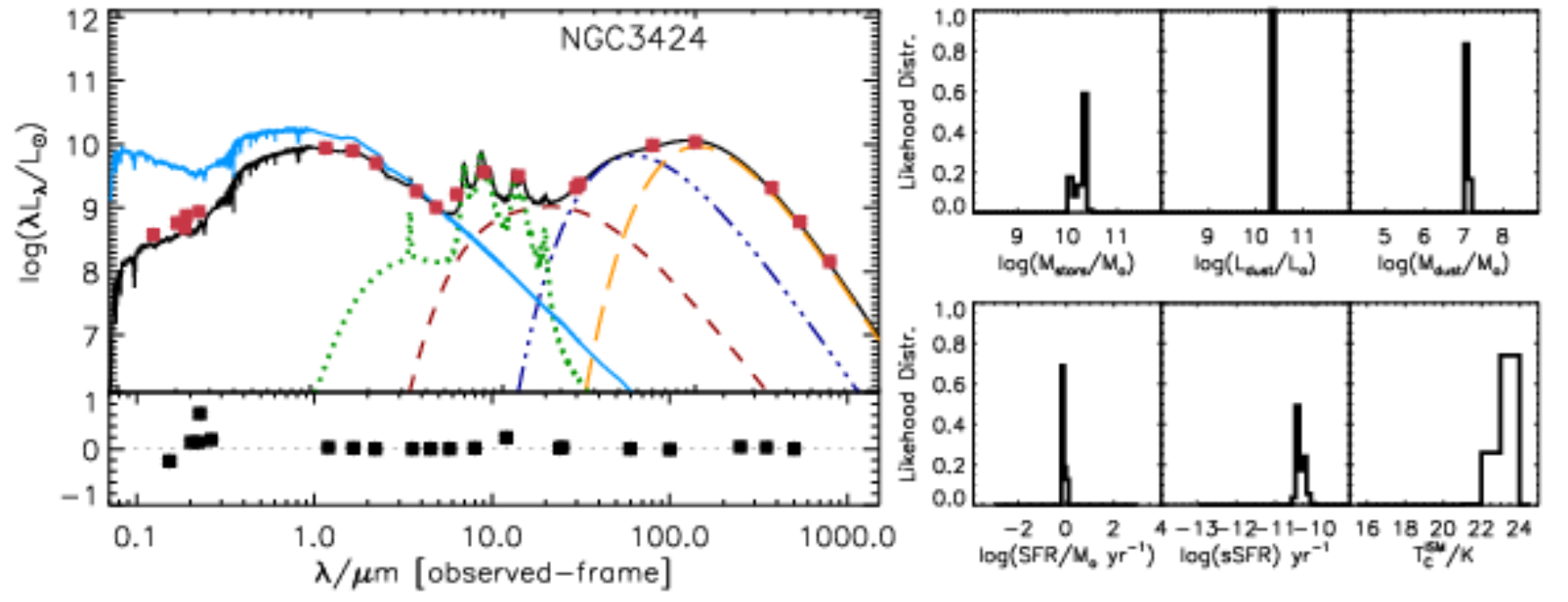}}
\caption{As Figure \ref{seds1}, but for NGC 3227 (top), NGC 3395/3396 (middle), and NGC 3424 (bottom).}
\label{seds4}
\end{figure*}

\begin{figure*}
\centerline{\includegraphics[width=0.925\linewidth]{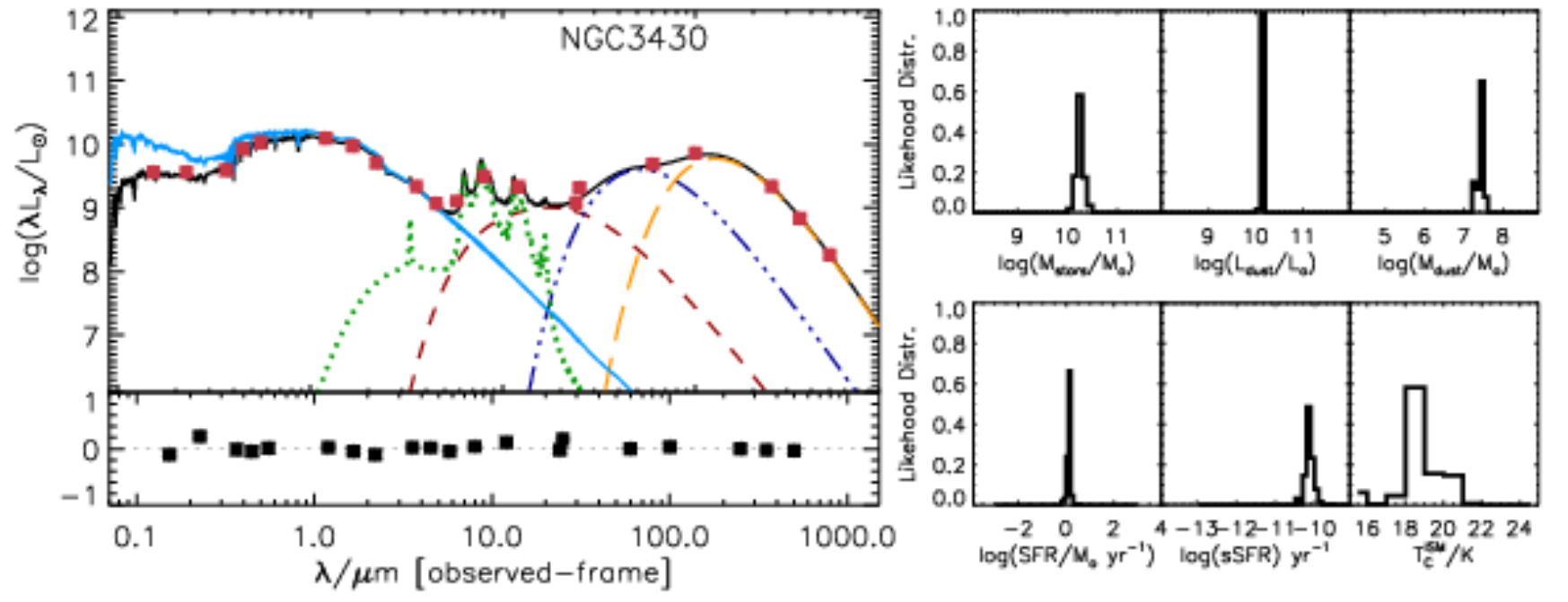}}
\centerline{\includegraphics[width=0.925\linewidth]{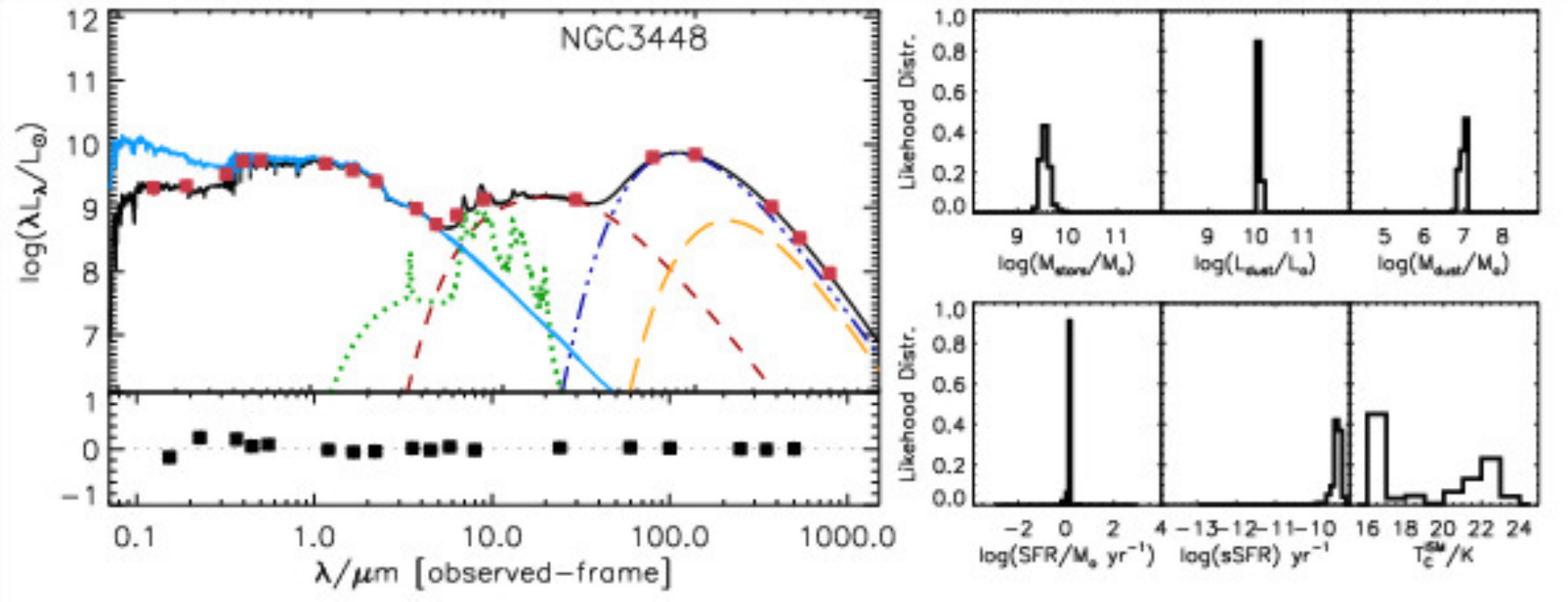}}
\centerline{\includegraphics[width=0.925\linewidth]{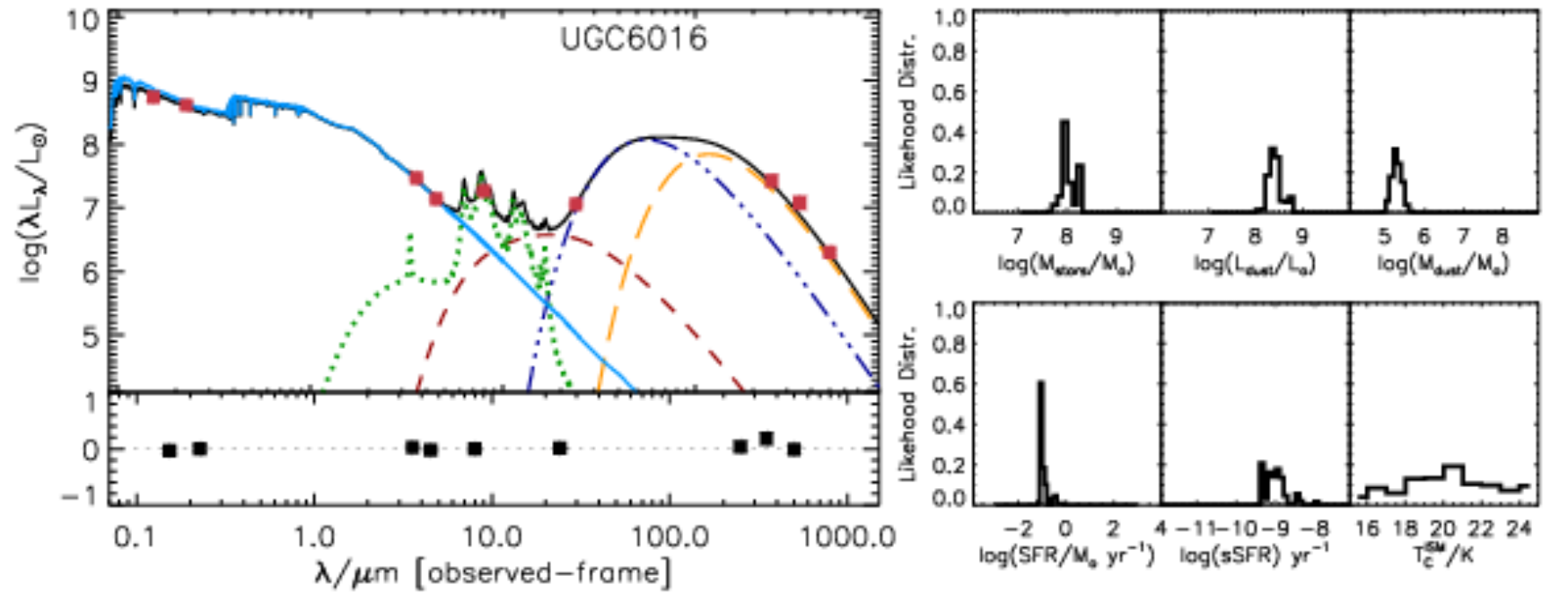}}
\caption{As Figure \ref{seds1}, but for NGC 3430 (top), NGC 3448 (middle), and UGC 6016 (bottom). 
Note that the axes of the UGC 6016 plots have smaller values than the rest of the plots.}
\label{seds5}
\end{figure*}

\begin{figure*}
\centerline{\includegraphics[width=0.925\linewidth]{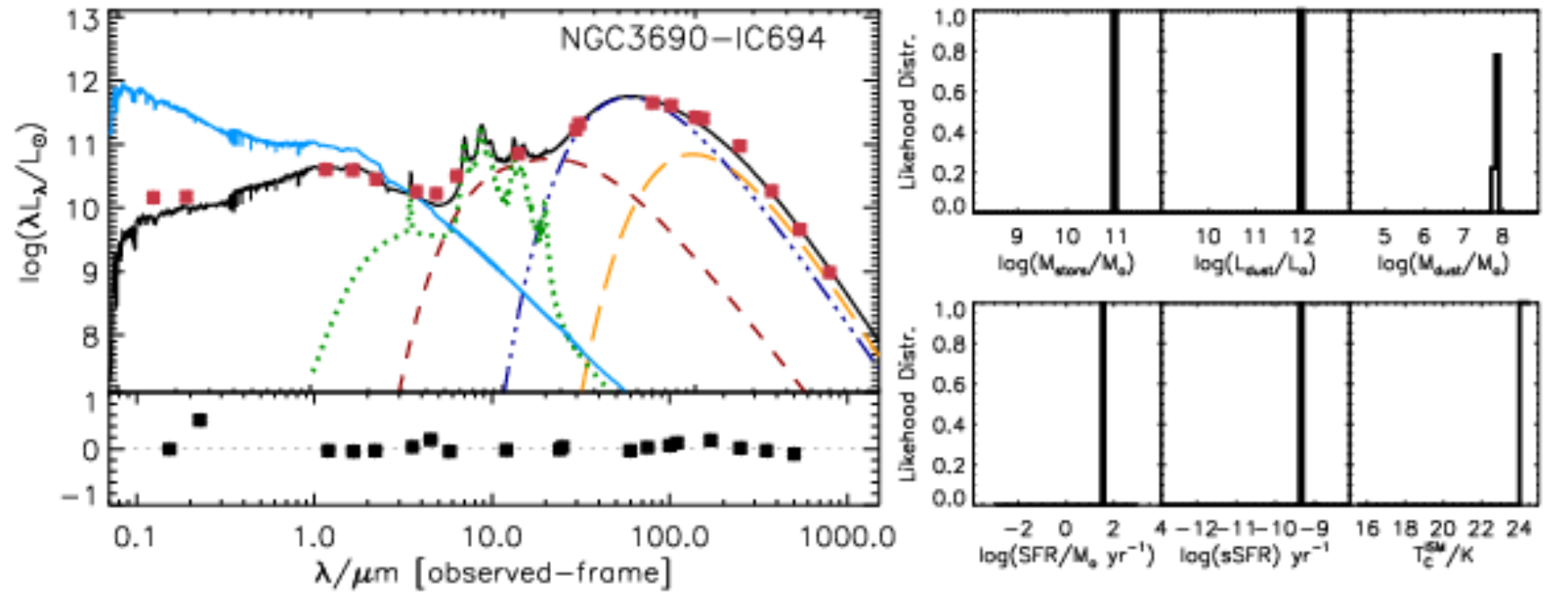}}
\centerline{\includegraphics[width=0.925\linewidth]{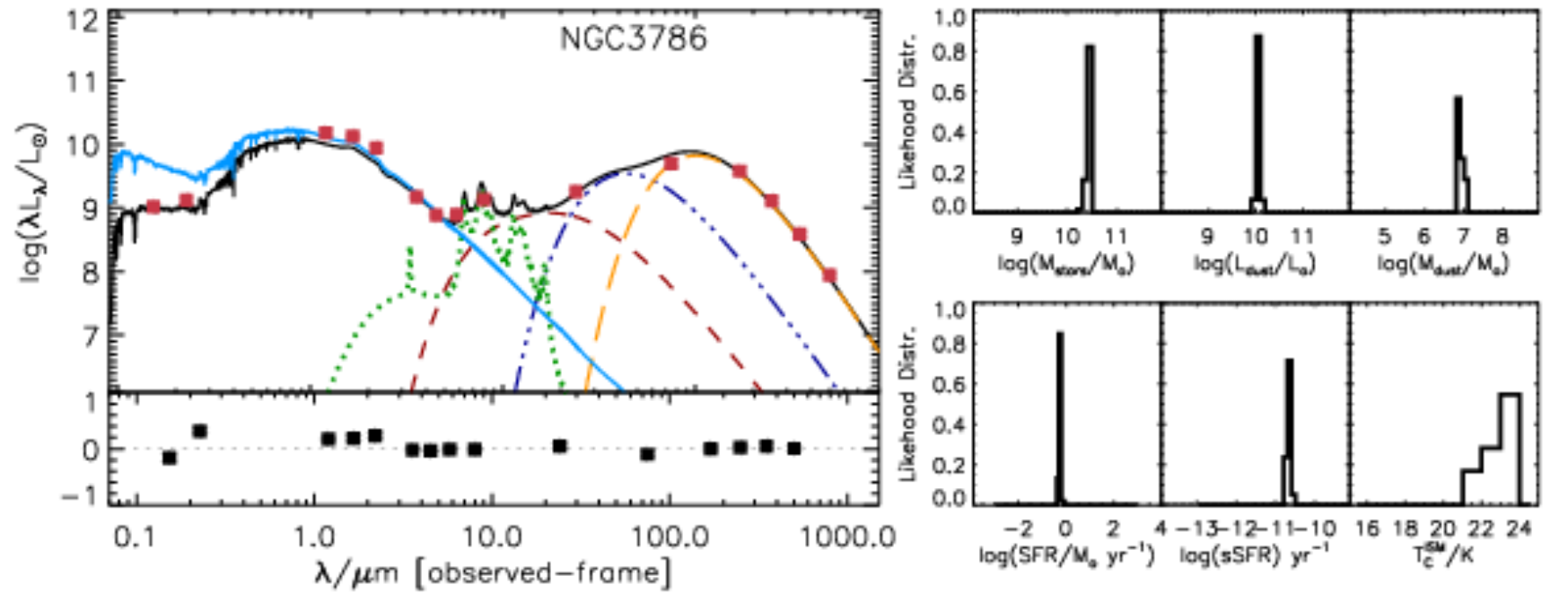}}
\centerline{\includegraphics[width=0.925\linewidth]{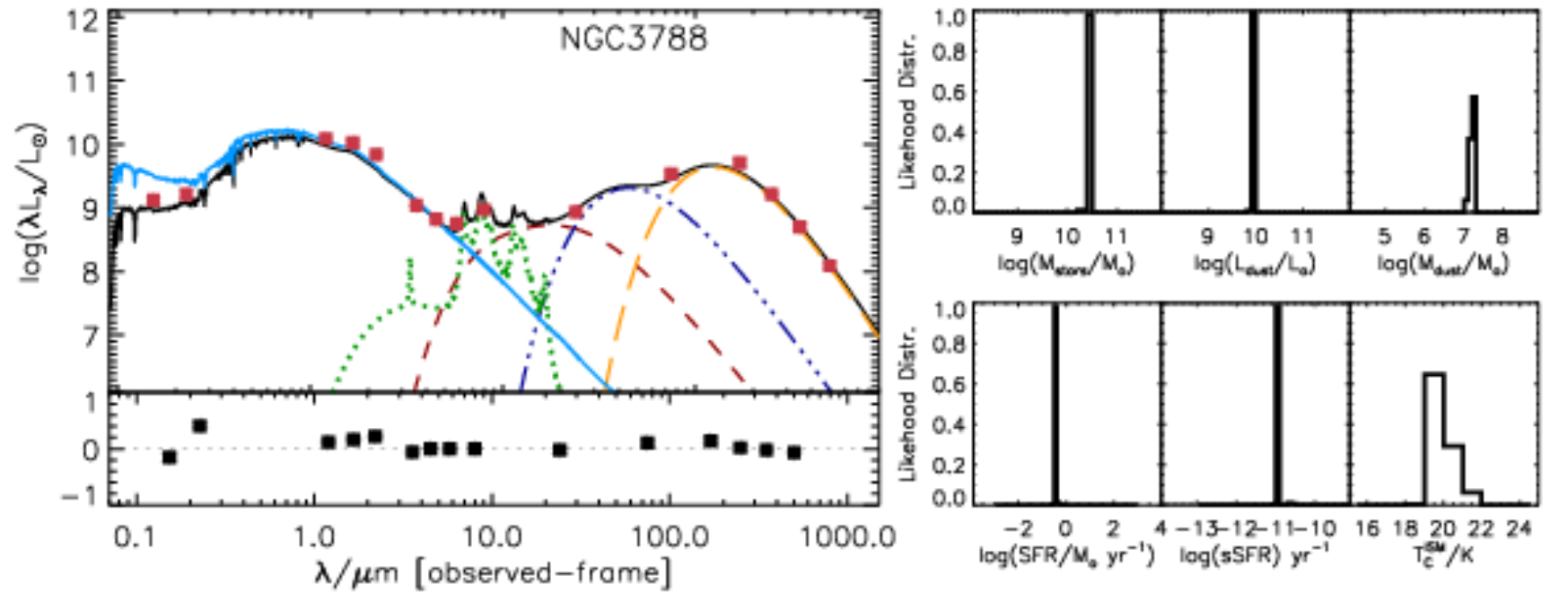}}
\caption{As Figure \ref{seds1}, but for NGC 3690/IC 694 (top), NGC 3786 (middle), and NGC 3788 (bottom). Note that
the axes of the NGC 3690 plots have larger values than the rest of the plots. }
\label{seds6}
\end{figure*}

\begin{figure*}
\centerline{\includegraphics[width=0.925\linewidth]{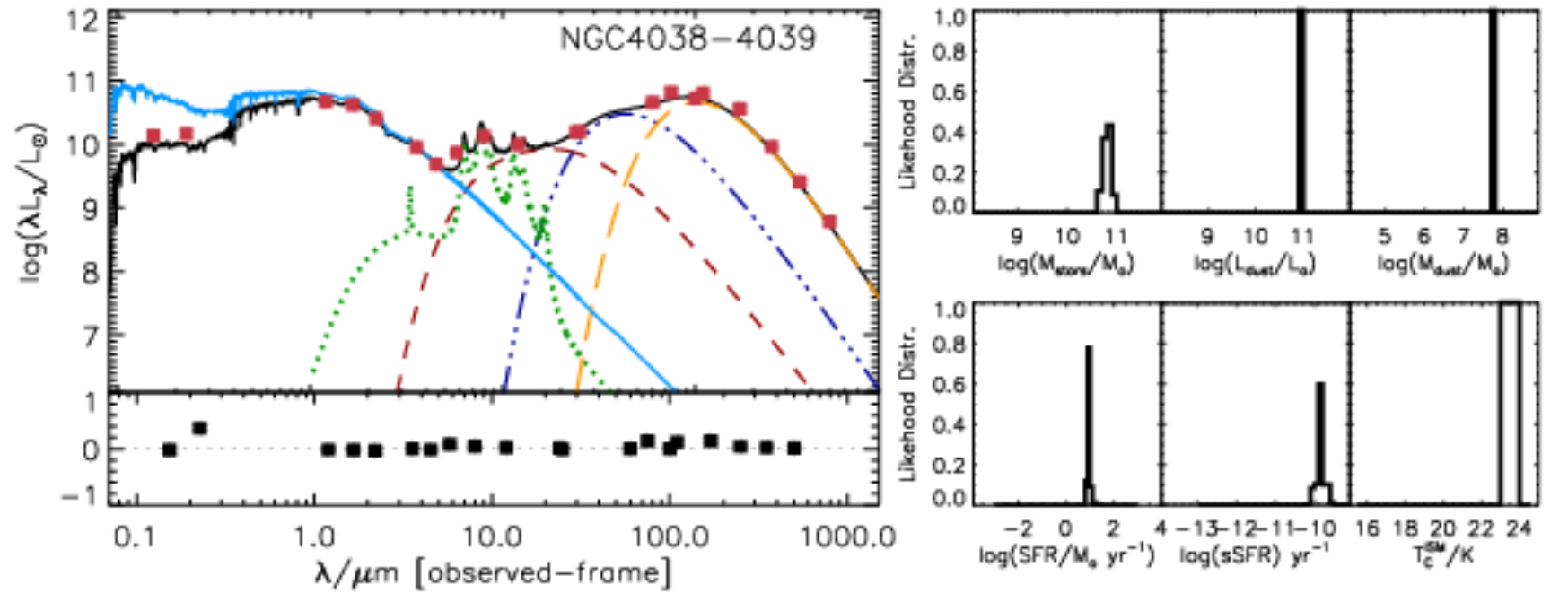}}
\centerline{\includegraphics[width=0.925\linewidth]{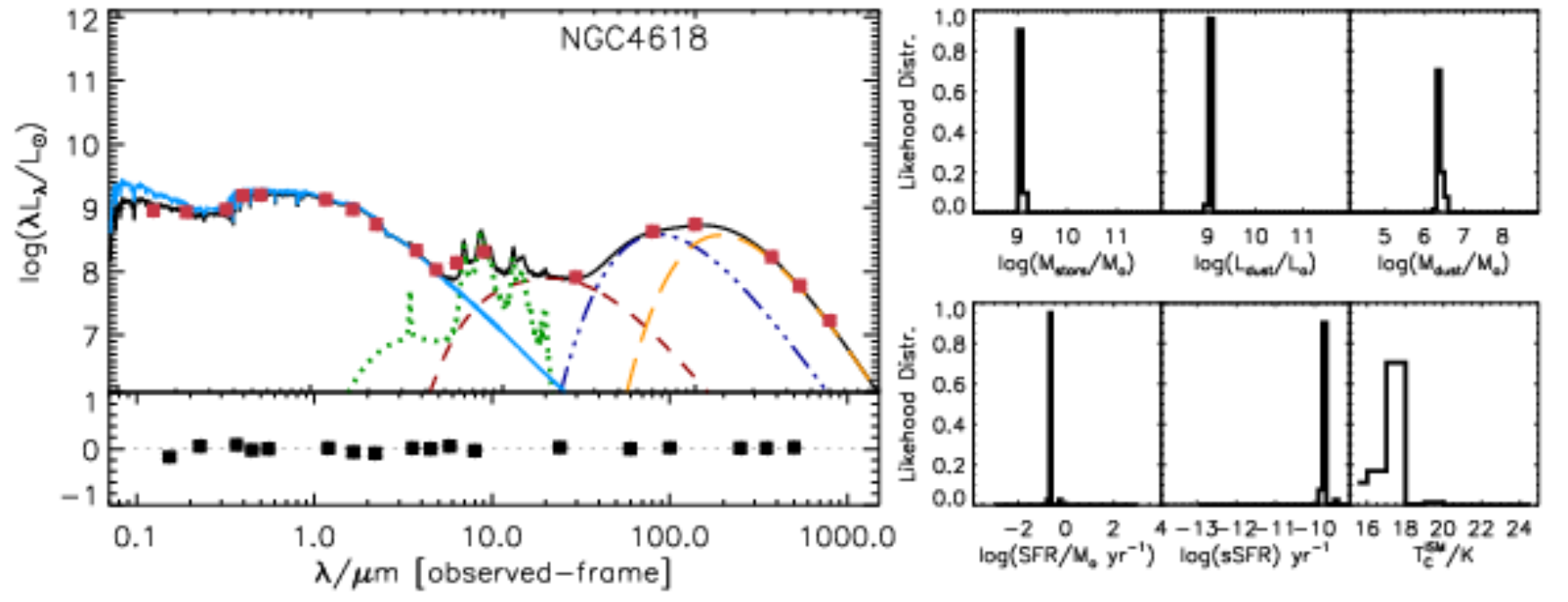}}
\centerline{\includegraphics[width=0.925\linewidth]{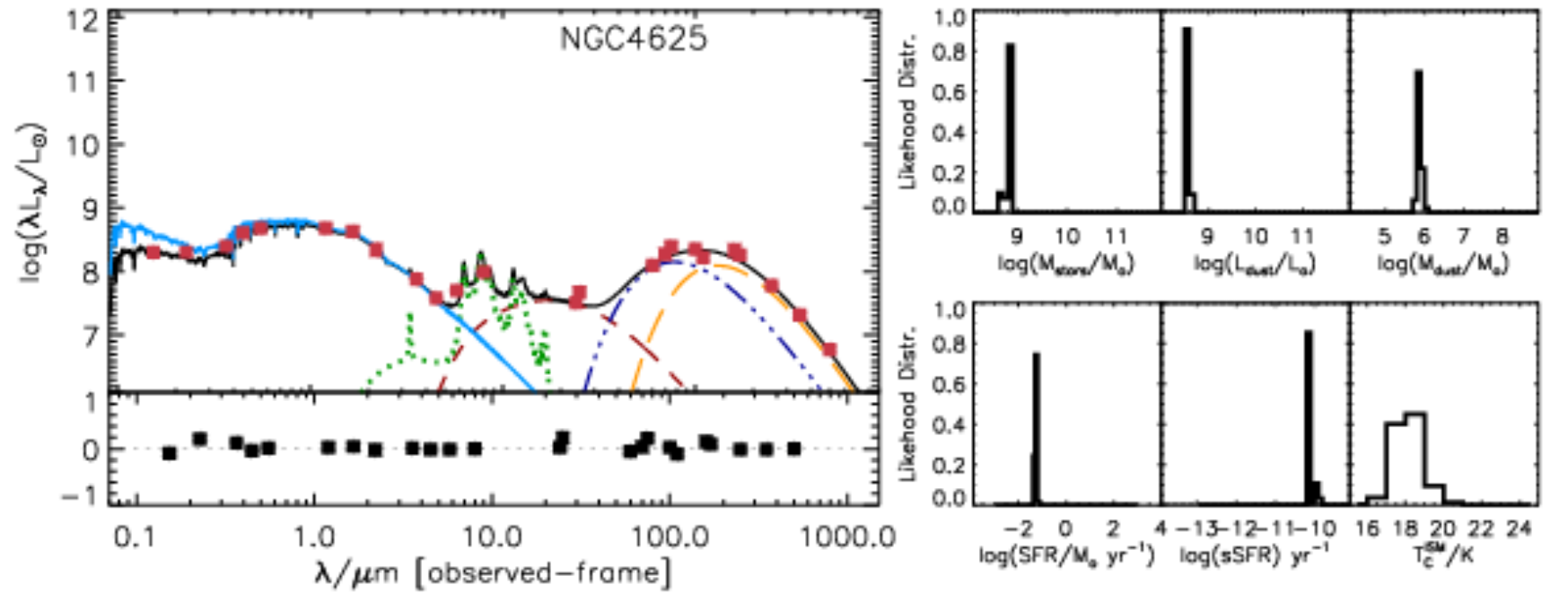}}
\caption{As Figure \ref{seds1}, but for NGC 4038/4039 (top), NGC 4618 (middle), and NGC 4625 (bottom).}
\label{seds7}
\end{figure*}

\begin{figure*}
\centerline{\includegraphics[width=0.925\linewidth]{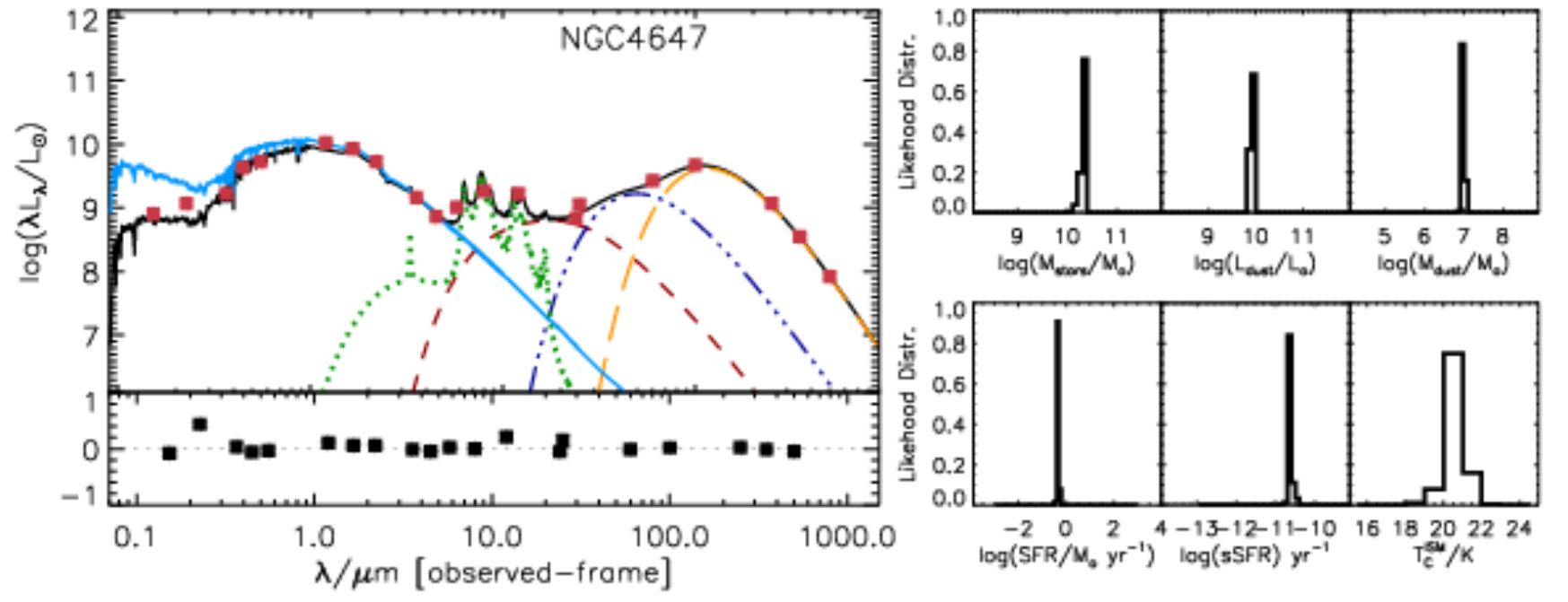}}
\centerline{\includegraphics[width=0.925\linewidth]{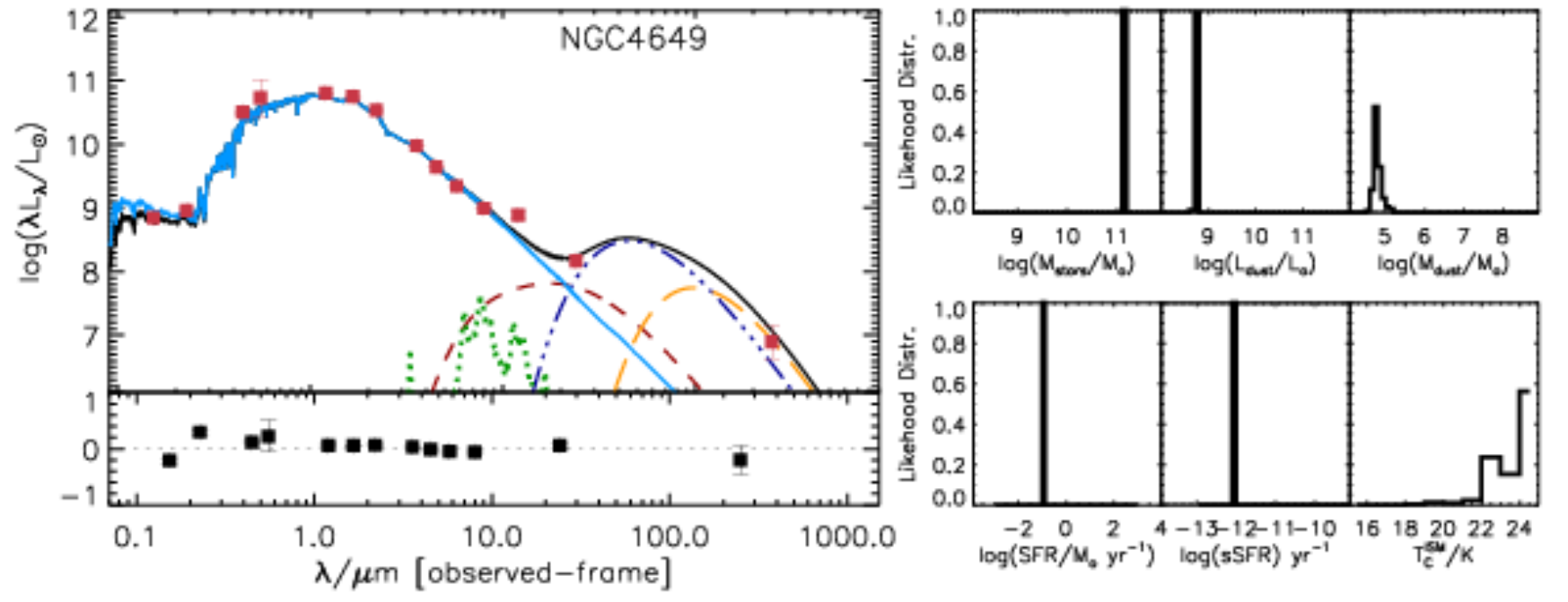}}
\centerline{\includegraphics[width=0.925\linewidth]{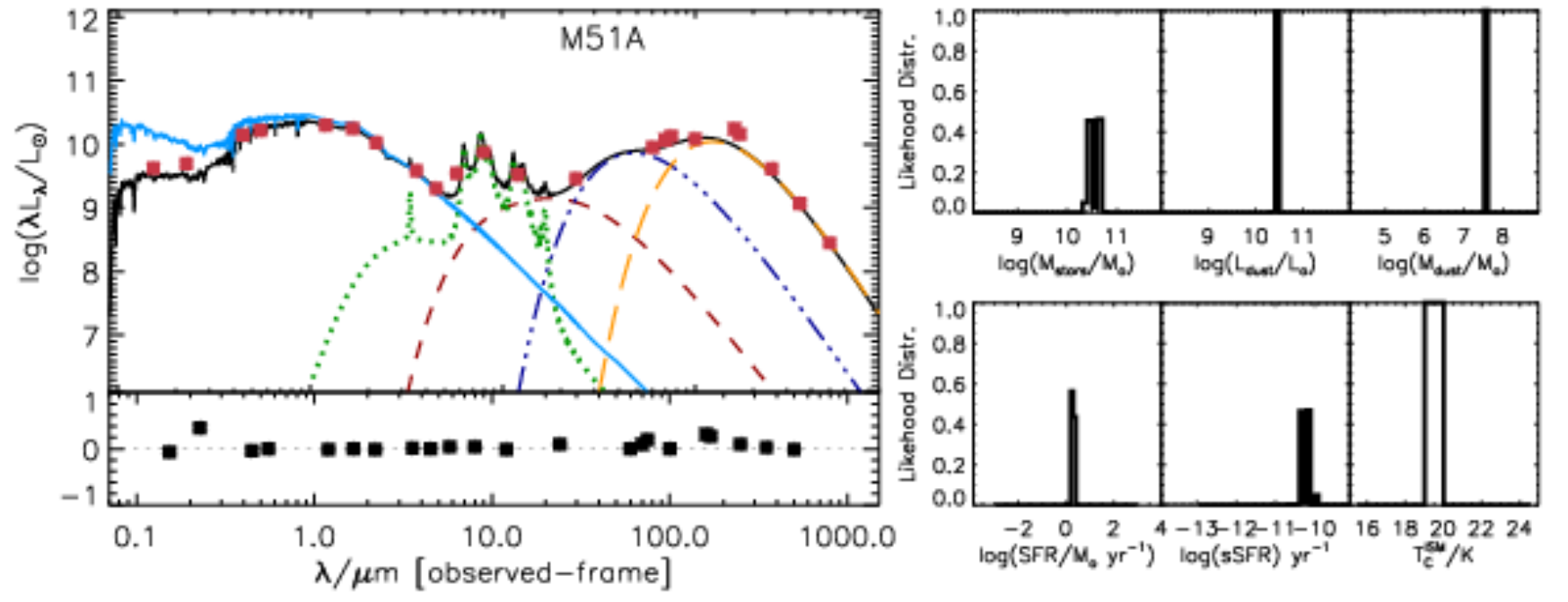}}
\caption{As Figure \ref{seds1}, but for NGC 4647 (top), NGC 4649 (middle), and M51A (bottom).}
\label{seds8}
\end{figure*}

\begin{figure*}
\centerline{\includegraphics[width=0.925\linewidth]{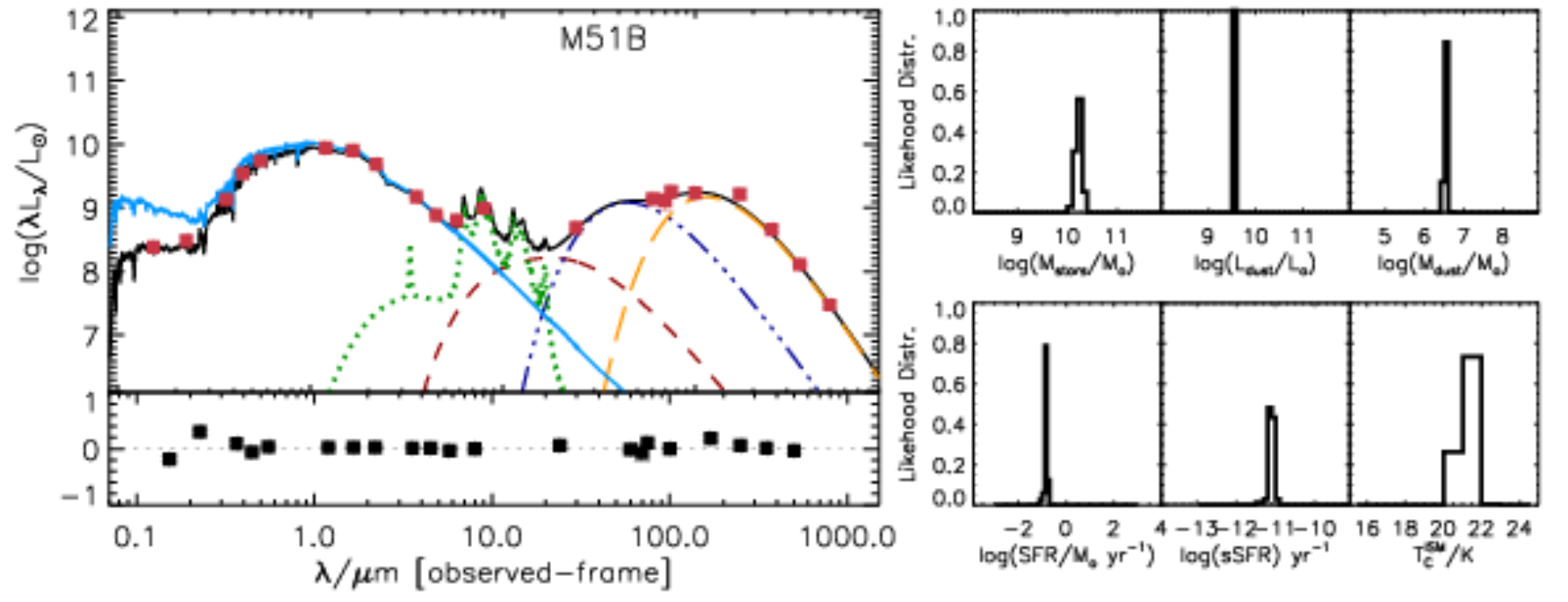}}
\centerline{\includegraphics[width=0.925\linewidth]{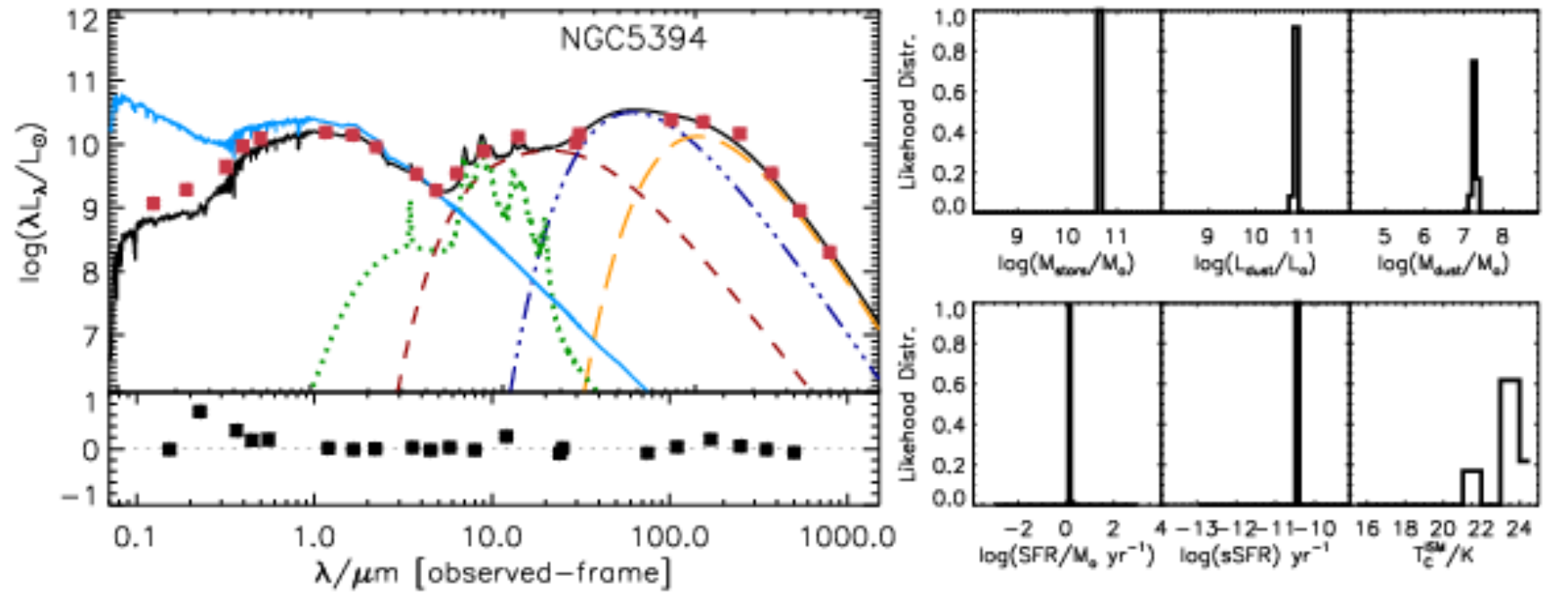}}
\centerline{\includegraphics[width=0.925\linewidth]{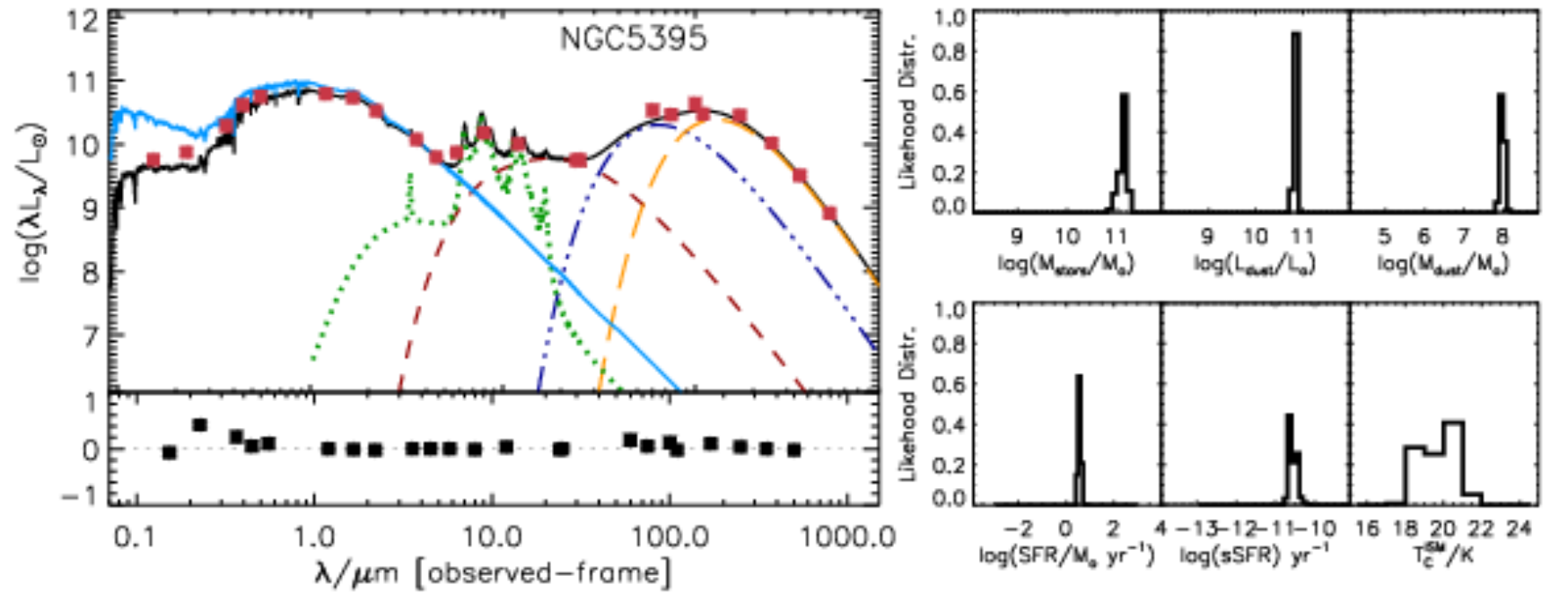}}
\caption{As Figure \ref{seds1}, but for M51B (top), NGC 5394 (middle) and NGC 5395 (bottom).}
\label{seds9}
\end{figure*}

\begin{figure*}
\centerline{\includegraphics[width=0.925\linewidth]{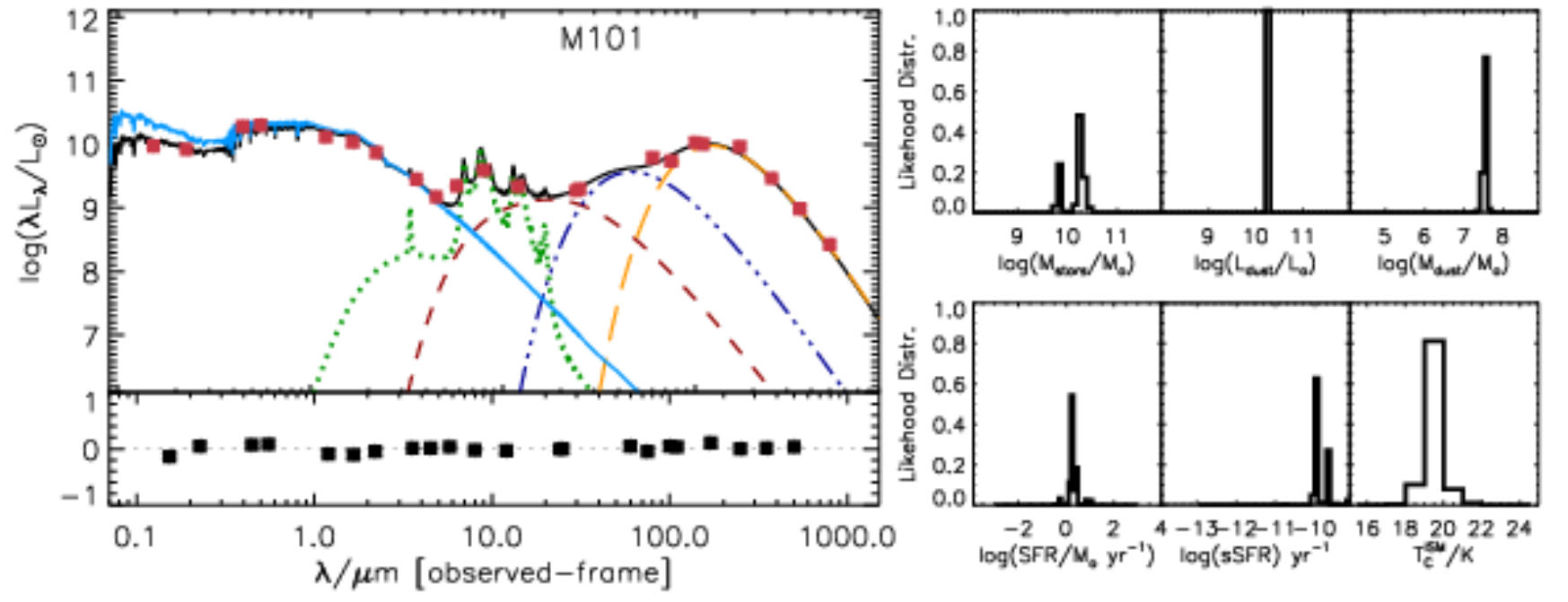}}
\centerline{\includegraphics[width=0.925\linewidth]{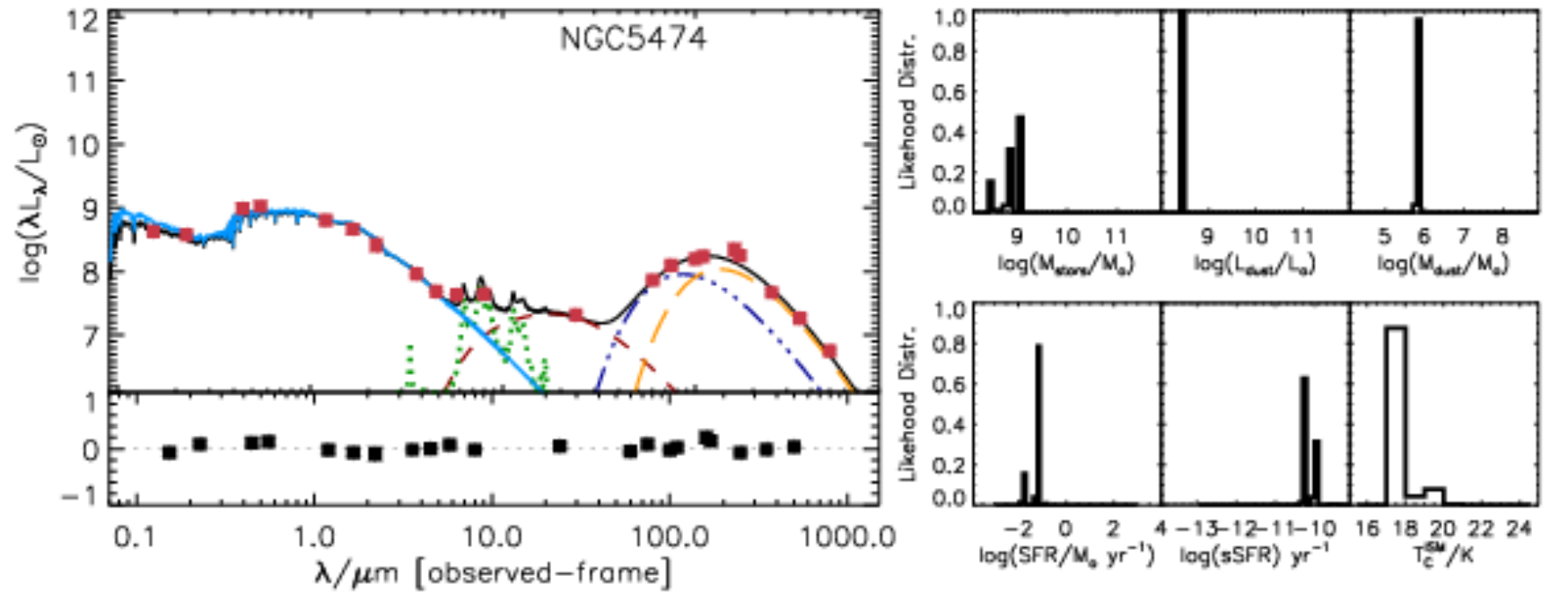}}
\caption{As Figure \ref{seds1}, but for M101 (top) and NGC 5474 (bottom).}
\label{seds10}
\end{figure*}

\section{SED FITTING WITH MAGPHYS}

\subsection{Fitting Process}
To estimate SFR, specific star formation rates (sSFRs), stellar and dust masses, 
and dust temperatures, we used the SED fitting code MAGPHYS \citep{dac08}. MAGPHYS fits 
SEDs with a combination of UV$-$NIR stellar spectral libraries from \citet{bru03} and a simple, 
physically-motivated model for IR emission from dust developed in \citet{dac08}. It models the 
ISM as a mix of diffuse dust interspersed with denser, warmer stellar birth clouds. MAGPHYS also 
includes a set of UV$-$NIR libraries that modify the \citet{bru03} population synthesis with the 
\citet{cb07} population synthesis, which provides different treatment of post-Asymptotic Giant 
Branch (AGB). We fit our UV to FIR SEDs with MAGPHYS with and without the post-AGB 
modifications and found consistent results; from here on, we only use the results with the 
earlier \citet{bru03} libraries as their treatment of the post-AGB stars is more consistent with 
current understanding \citep[e.g.,][]{zibetti12}. The IR dust libraries have five components: 
a fixed polycyclic aromatic hydrocarbon (PAH) spectrum shape derived from the M17 SW 
star-forming region \citep{madden06}, a NIR continuum associated with the PAH emission 
modeled by a modified blackbody ($\beta=1$) at 850 K, a hot MIR continuum 
modeled by the sum of two modified blackbodies ($\beta=1$) at 130 K and 250 K, a warm (30-60 K) 
dust component modeled as a modified blackbody ($\beta=1.5$), and a cold (15-25 K) dust 
component modeled as a modified blackbody ($\beta=2$). The warm dust component is assumed 
to exist both in the diffuse ISM and in denser birth clouds, while the cold dust exists only in the 
diffuse ISM. MAGPHYS determines probability distribution functions (PDFs) for the fitted parameters 
by combining UV$-$NIR and IR spectral libraries such that the energy absorbed in the UV/visible 
regime is re-emitted in the IR. It gives both the best-fit obscured SED and the associated unobscured stellar SED.

We input the measured and literature fluxes in our set of 28 filters to MAGPHYS and examined the variation 
in derived galaxy properties including dust luminosity, SFR, sSFR, stellar and dust mass, and dust 
temperatures and discuss the results below. Then we performed five additional fits: one without UV 
photometry, one without SPIRE photometry, one without any photometry at wavelengths $\lambda 
\ge$ 30\um, one without either UV or SPIRE photometry, and one with only UBV, 2MASS, and IRAC 
photometry.  For these fits for each galaxy, we examined the median and 68\% confidence interval 
for the fitted parameters to determine the influence of the particular dataset on the value of 
(and constraints on) these properties, as we elaborate below. For the non-interacting galaxies, we only
performed the fit with all of the available photometry.

\begin{figure}
\centerline{\includegraphics[width=\linewidth]{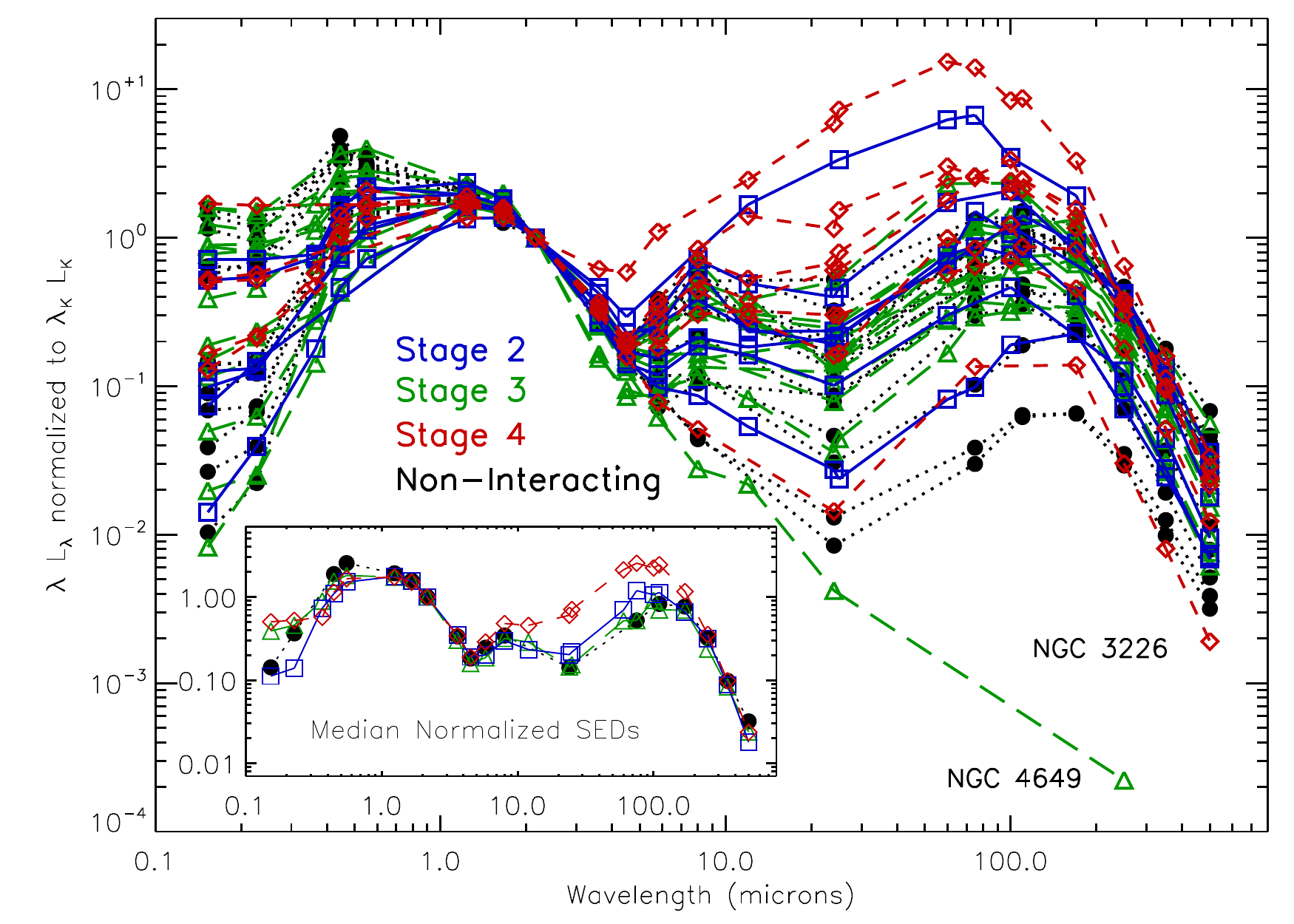}}
\caption{SEDs normalized to the 2MASS \emph{Ks} band luminosity for each galaxy. Stage 2 (weakly interacting), Stage 3
(moderately interacting), Stage 4 (strongly interacting), and non-interacting galaxies are shown respectively in blue 
squares joined with solid lines, green triangles joined with long dashes, red diamonds joined with short dashes, and
black circles joined with dotted lines, respectively. Inset, we show the median SED for each class of galaxies.
These SEDs show a tendency for Stage 4 galaxies to have more hot-warm dust emission in the 10-60\um~range 
relative to both its cold dust emission in the SPIRE bandpasses and its stellar NIR emission, whereas the ratio of
NIR to FIR emission is relatively consistent. Additionally, the more strongly interacting galaxies typically have a younger
stellar population than the Stage 2 galaxies as suggested by the relative amounts of UV to NIR emission. The two
labeled galaxies are elliptical galaxies. }
\label{sed_shape}
\end{figure}

\subsection{SED Fits} 
Figures~\ref{seds1} $-$ \ref{seds10} show the SEDs along with the best-fit models for our interacting galaxies. The
contributions of the different components of the IR model described above are also shown.
The median and 68\% confidence interval of these parameters are given in Table \ref{param}. Just as the galaxies 
exhibit a variety of UV versus IR morphologies (Figures~\ref{n2976img}$-$\ref{n5474img}), the SEDs have a 
corresponding range of relative UV, NIR, and FIR emission. For example, some galaxies (e.g., NGC 
3190 or M51B) have very little UV flux in comparison with their visible and IR flux, while others (e.g., NGC 
3187) have almost as much UV flux as IR flux. The SEDs also show a range of obscuration from the 
heavily obscured galaxies (e.g., NGC 3690) to relatively unobscured galaxies (e.g., NGC 4618), or ones 
with extended UV disks such as NGC 3430 or UGC 6016. Appendix C briefly describes each galaxy, 
discussing any particular issues regarding the photometry and the SED fitting. Note in particular that fits 
to edge-on galaxies tend to over-estimate the amount of UV obscuration and hence the model UV fluxes
tend to be too low compared to the observations.

\begin{deluxetable*}{lccccccc}[h]
\tabletypesize{\scriptsize}
\tablecaption{MAGPHYS Parameters\label{param}}
\tablewidth{0pt}
\tablehead{
\colhead{Name} &\colhead{Dust Luminosity} &\colhead{Stellar Mass} & \colhead{Dust Mass} & \colhead{SFR\tablenotemark{a}} & \colhead{Specific SFR} & 
\colhead{Cold Dust Temp.} & \colhead{Warm Dust Temp.}\\
\colhead{} &\colhead{(10$^{9}$~L$_{\odot}$)} &\colhead{(10$^{10}$~M$_{\odot}$)} & \colhead{(10$^{7}$~M$_{\odot}$)} 
& \colhead{(M$_{\odot}$ yr$^{-1}$)} & \colhead{(10$^{-11}$~yr$^{-1}$)} & \colhead{(K)} & \colhead{(K)}}
\startdata
NGC2976 			&0.86$^{+0.01}_{-0.02}$			&0.10~~$^{+0.05}_{-0.01}$	
					&0.17~~$^{+0.02}_{-0.01}$ 		&0.084~~$^{+0.009}_{-0.015}$	
					&8.51~~$^{+0.20}_{-3.14}$ 		&18.8$^{+0.1  }_{-0.2  }$ 			&48.6$^{+3.7}_{-0.1}$ \\
NGC3031 			&4.05$^{+0.05}_{-0.05}$			&2.95~~$^{+2.42}_{-0.07}$	
					&1.18~~$^{+0.24}_{-0.15}$ 		&0.373~~$^{+0.022}_{-0.044}$	
					&1.35~~$^{+0.03}_{-0.59}$ 		&18.4$^{+0.6  }_{-0.9  }$ 			&53.6$^{+3.5}_{-21.2}$ \\	 
NGC3034 			&54.95$^{+1.28}_{-1.25}$~~		&1.45~~$^{+0.03}_{-0.03}$	
					&0.86~~$^{+0.09}_{-0.18}$ 		&0.728~~$^{+0.008}_{-0.008}$	
					&5.37~~$^{+0.13}_{-0.12}$		&23.5$^{+0.8  }_{-0.7  }$ 			&51.4$^{+5.6}_{-8.0}$ \\	 
NGC3077  			&0.75$^{+0.02}_{-0.02}$			&0.068$^{+0.035}_{-0.002}$	
					&0.041$^{+0.001}_{-0.001}$ 		&0.097~~$^{+0.031}_{-0.011}$	
					&14.96~~$^{+6.17}_{-8.20}$~		&22.9$^{+0.1  }_{-0.1  }$ 			&52.5$^{+0.1}_{-0.1}$ \\					
NGC3185 			&3.29$^{+0.04}_{-0.04}$			&1.45~~$^{+0.03}_{-0.03}$	
					&0.49~~$^{+0.02}_{-0.07}$ 		&0.183~~$^{+0.002}_{-0.002}$	
					&1.20~~$^{+0.03}_{-0.03}$ 		&20.2$^{+0.4  }_{-0.4  }$ 			&57.8$^{+1.4}_{-3.7}$ \\				
NGC3187 			&4.09$^{+0.05}_{-0.05}$			&0.31~~$^{+0.08}_{-0.04}$	
					&1.02~~$^{+0.06}_{-0.05}$ 		&0.274~~$^{+0.325}_{-0.003}$	
					&8.51~$^{+10.33}_{-~0.11}$		&16.2$^{+0.4  }_{-0.3  }$ 			&32.6$^{+9.7}_{-1.7}$ \\	 
NGC3190 			&8.55$^{+0.82}_{-0.10}$			&8.32~~$^{+0.39}_{-0.37}$	
					&1.44~~$^{+0.03}_{-0.03}$ 		&0.0084$^{+0.0004}_{-0.0004}$	
					&~0.011$^{+0.001}_{-0.001}$		&22.3$^{+0.1  }_{-0.1  }$ 			&57.9$^{+0.1}_{-0.2}$ \\	 	
NGC3226 			&1.40$^{+0.10}_{-0.11}$			&2.29~~$^{+0.28}_{-0.15}$	
					&0.10~~$^{+0.01}_{-0.01}$ 		&0.045~~$^{+0.003}_{-0.017}$	
					&2.14~~$^{+0.05}_{-0.94}$		&24.1$^{+0.7  }_{-0.5  }$ 			&54.4$^{+4.0}_{-3.7}$ \\	 	
NGC3227  			&16.98$^{+0.40}_{-0.39}$~~		&1.32~~$^{+0.03}_{-0.03}$	
					&1.15~~$^{+0.03}_{-0.03}$ 		&0.836~~$^{+0.019}_{-0.019}$	
					&6.76~~$^{+0.16}_{-0.15}$ 		&22.5$^{+0.1  }_{-0.1  }$ 			&59.1$^{+0.1}_{-0.1}$ \\	 				
NGC3395/3396		&26.30$^{+0.61}_{-0.60}$~~		&0.70~~$^{+0.11}_{-0.08}$	
					&1.13~~$^{+0.39}_{-0.01}$ 		&3.289~~$^{+0.235}_{-0.148}$	
					&42.17~$^{+10.92}_{-~0.96}$~ 	&23.8$^{+0.1  }_{-1.3  }$ 			&55.8$^{+3.0}_{-0.1}$ \\	 		
NGC3424 			&20.42$^{+0.48}_{-0.46}$~~		&2.04~~$^{+0.05}_{-0.81}$	
					&1.14~~$^{+0.17}_{-0.03}$ 		&0.753~~$^{+0.163}_{-0.050}$	
					&3.80~~$^{+2.22}_{-0.09}$		&23.3$^{+0.1  }_{-0.7  }$ 			&55.4$^{+0.1}_{-4.6}$ \\	 		
NGC3430 			&13.49$^{+0.64}_{-0.61}$~~		&1.82~~$^{+0.42}_{-0.23}$	
					&2.55~~$^{+0.12}_{-0.35}$ 		&1.355~~$^{+0.165}_{-0.188}$	
					&7.59~~$^{+1.96}_{-2.22}$ 		&19.0$^{+0.6  }_{-0.8  }$ 			&55.5$^{+3.61}_{-5.1}$ \\	 			
NGC3448 			&11.22$^{+1.37}_{-0.26}$~~		&0.35~~$^{+0.07}_{-0.05}$	
					&0.91~~$^{+0.12}_{-0.16}$ 		&1.371~~$^{+0.065}_{-0.047}$	
					&37.58~~$^{+9.73}_{-7.73}$~ 		&18.7$^{+3.9  }_{-2.3  }$ 			&32.9$^{+22.2}_{-0.4}$ \\	 	
UGC6016	 			&0.25$^{+0.09}_{-0.08}$			&0.010$^{+0.009}_{-0.002}$	
					&0.020$^{+0.007}_{-0.005}$ 		&0.091~~$^{+0.048}_{-0.008}$	
					&94.41~$^{+55.22}_{-47.09}$~ 	&20.4$^{+2.7  }_{-2.7  }$ 			&41.1$^{+6.4}_{-7.2}$ \\	 	
NGC3690/IC694		&812.8~$^{+18.93}_{-18.50}$~~~	&8.13~~$^{+0.19}_{-0.19}$	
					&7.28~~$^{+0.17}_{-1.56}$ 		&35.65~~~$^{+0.83}_{-0.81}$~~
					&42.17~~$^{+0.98}_{-0.96}$~		&24.8$^{+0.2  }_{-0.1  }$ 			&58.0$^{+0.1}_{-4.3}$ \\	 		
NGC3786 			&12.59$^{+0.29}_{-1.87}$~~		&3.02~~$^{+0.07}_{-0.62}$	
					&0.72~~$^{+0.29}_{-0.02}$ 		&0.612~~$^{+0.014}_{-0.109}$	
					&2.14~~$^{+0.05}_{-0.23}$ 		&23.8$^{+0.1  }_{-2.3  }$ 			&59.6$^{+0.4}_{-1.0}$ \\			
NGC3788 			&8.07$^{+0.19}_{-0.09}$			&2.95~~$^{+0.07}_{-0.07}$	
					&1.59~~$^{+0.04}_{-0.27}$ 		&0.352~~$^{+0.008}_{-0.004}$	
					&1.20~~$^{+0.03}_{-0.03}$ 		&19.6$^{+0.7  }_{-0.1  }$ 			&58.2$^{+0.9}_{-2.0}$ \\					
NGC4038/4039		&97.72$^{+2.28}_{-2.22}$~~		&6.46~~$^{+0.62}_{-1.33}$	
					&5.15~~$^{+0.12}_{-0.12}$ 		&8.954~~$^{+0.422}_{-0.788}$	
					&13.34~~$^{+3.45}_{-2.74}$~ 		&23.7$^{+0.1  }_{-0.1  }$ 			&59.9$^{+0.1}_{-0.1}$ \\	 	
NGC4618 			&1.04$^{+0.02}_{-0.01}$			&0.11~~$^{+0.01}_{-0.01}$	
					&0.25~~$^{+0.04}_{-0.01}$ 		&0.225~~$^{+0.003}_{-0.003}$	
					&18.84~~$^{+0.44}_{-0.43}$~ 		&17.7$^{+0.1}_{-1.6}$ 			&40.8$^{+0.1}_{-7.3}$ \\	 		
NGC4625	  			&0.39$^{+0.01}_{-0.01}$			&0.063$^{+0.001}_{-0.001}$	
					&0.074$^{+0.007}_{-0.004}$ 		&0.050~~$^{+0.001}_{-0.001}$	
					&7.59~~$^{+0.18}_{-0.17}$ 		&18.4$^{+0.3}_{-1.3}$ 			&33.7$^{+3.9}_{-2.8}$ \\	 	
NGC4647 			&8.07$^{+0.19}_{-0.27}$			&2.04~~$^{+0.05}_{-0.09}$	
					&0.99~~$^{+0.02}_{-0.04}$ 		&0.470~~$^{+0.011}_{-0.011}$	
					&2.40~~$^{+0.06}_{-0.05}$ 		&20.9$^{+0.2}_{-0.1}$ 			&52.7$^{+4.3}_{-0.1}$ \\				
NGC4649 			&0.50$^{+0.01}_{-0.01}$			&13.80~$^{+0.32}_{-0.31}$	
					&0.006$^{+0.002}_{-0.001}$ 		&0.114~~$^{+0.003}_{-0.001}$	
					&0.085$^{+0.002}_{-0.002}$		&24.5$^{+0.1}_{-1.9}$ 			&59.5$^{+0.4}_{-1.6}$  \\	 
M51A				&25.70$^{+0.60}_{-0.59}$~~		&3.02~~$^{+1.77}_{-0.33}$	
					&3.78~~$^{+0.09}_{-0.09}$ 		&1.936~~$^{+0.138}_{-0.022}$	
					&6.76~~$^{+0.82}_{-2.50}$ 		&19.6$^{+0.1}_{-0.1}$ 			&54.9$^{+0.1}_{-0.1}$ \\	 	
M51B	 			&3.65$^{+0.08}_{-0.20}$			&1.95~~$^{+0.05}_{-0.36}$	
					&0.34~~$^{+0.03}_{-0.02}$ 		&0.142~~$^{+0.003}_{-0.014}$	
					&0.76~~$^{+0.09}_{-0.02}$		&21.0$^{+0.1}_{-0.3}$ 			&59.3$^{+0.1}_{-5.8}$ \\	 
NGC5394 			&63.10$^{+1.47}_{-1.44}$~~		&4.79~~$^{+0.11}_{-0.11}$	
					&1.77~~$^{+0.26}_{-0.08}$ 		&1.538~~$^{+0.018}_{-0.035}$	
					&3.39~~$^{+0.08}_{-0.08}$ 		&23.2$^{+1.0}_{-1.5}$ 			&55.6$^{+1.3}_{-2.2}$ \\	 		
NGC5395	 			&66.07$^{+4.73}_{-1.50}$~~		&15.14~$^{+0.71}_{-4.42}$	
					&9.48~~$^{+1.66}_{-1.22}$ 		&3.690~~$^{+0.403}_{-0.513}$	
					&2.69~~$^{+1.11}_{-0.55}$ 		&19.7$^{+0.7}_{-0.8}$ 			&45.3$^{+5.6}_{-9.7}$ \\		
M101		 		&18.20$^{+0.42}_{-0.82}$~~		&1.82~~$^{+0.47}_{-1.12}$	
					&3.25~~$^{+0.31}_{-0.11}$ 		&1.959~~$^{+0.905}_{-0.367}$	
					&11.89~$^{+9.25}_{-1.29}$~ 		&19.4$^{+0.3}_{-0.4}$ 			&58.2$^{+0.8}_{-2.2}$ \\	 	
NGC5474  			&0.26$^{+0.01}_{-0.01}$			&0.074$^{+0.041}_{-0.041}$	
					&0.072$^{+0.002}_{-0.002}$ 		&0.070~~$^{+0.007}_{-0.054}$	
					&6.03~~$^{+4.57}_{-0.14}$ 		&17.9$^{+0.1}_{-0.1}$ 			&30.4$^{+0.8}_{-0.1}$ 	\\
\hline
NGC925				&4.39$^{+0.10}_{-0.10}$			&0.92~~$^{+0.02}_{-0.52}$	
					&2.38~~$^{+0.03}_{-0.08}$		&0.459~~$^{+0.236}_{-0.010}$	
					&4.79~~$^{+7.10}_{-0.11}$		&15.4$^{+0.7}_{-0.1}$		&35.7$^{+6.2}_{-4.9}$	\\		
NGC1291				&2.17$^{+0.05}_{-0.02}$			&6.17~~$^{+0.75}_{-0.54}$	
					&0.89~~$^{+0.14}_{-0.03}$		&0.118~~$^{+0.014}_{-0.022}$	
					&0.13~~$^{+0.08}_{-0.06}$		&17.2$^{+0.1}_{-0.4}$		&57.4$^{+1.9}_{-1.9}$	\\	
NGC2841				&4.54$^{+0.11}_{-0.10}$			&2.95~$^{+1.41}_{-0.32}$	
					&1.98~~$^{+0.32}_{-0.15}$		&0.220~~$^{+0.044}_{-0.069}$	
					&0.54~~$^{+0.09}_{-0.22}$		&17.2$^{+0.2}_{-0.4}$		&55.0$^{+1.4}_{-1.3}$	\\	
NGC3049				&4.29$^{+0.10}_{-0.19}$			&0.11~~$^{+0.01}_{-0.01}$	
					&0.66~~$^{+0.07}_{-0.14}$		&0.459~~$^{+0.569}_{-0.036}$	
					&37.58~~$^{+63.76}_{-4.59}$		&17.3$^{+1.1}_{-0.3}$		&54.9$^{+1.7}_{-2.5}$	\\	
NGC3184				&4.86$^{+0.11}_{-0.11}$			&0.42~~$^{+0.09}_{-0.07}$	
					&1.25~~$^{+0.31}_{-0.12}$		&0.207~~$^{+0.216}_{-0.087}$	
					&2.69~~$^{+7.62}_{-1.57}$		&18.2$^{+0.3}_{-0.7}$		&52.9$^{+3.3}_{-4.3}$	\\	
NGC3521				&21.88$^{+0.51}_{-0.50}$~~		&4.57~$^{+1.05}_{-1.10}$	
					&3.52~~$^{+0.43}_{-0.56}$		&0.771~~$^{+0.104}_{-0.130}$	
					&1.35~~$^{+0.49}_{-0.56}$		&19.9$^{+0.5}_{-0.4}$		&50.6$^{+3.4}_{-4.9}$	\\	
NGC3621				&6.64$^{+0.15}_{-0.30}$			&0.84~~$^{+0.26}_{-0.18}$	
					&1.29~~$^{+0.16}_{-0.18}$		&0.598~~$^{+0.050}_{-0.059}$	
					&5.37~~$^{+1.75}_{-1.39}$		&18.6$^{+0.8}_{-0.6}$		&46.2$^{+5.4}_{-9.7}$	\\	
NGC3938				&7.80$^{+0.18}_{-0.18}$			&0.26~~$^{+0.05}_{-0.01}$	
					&1.50~~$^{+0.13}_{-0.10}$		&0.225~~$^{+0.470}_{-0.005}$	
					&8.51~~$^{+12.62}_{-0.19}$		&19.1$^{+0.2}_{-0.4}$		&46.5$^{+8.2}_{-5.6}$	\\	
NGC4236				&0.27$^{+0.01}_{-0.01}$			&0.020$^{+0.001}_{-0.001}$	
					&0.009$^{+0.001}_{-0.001}$		&0.022~~$^{+0.001}_{-0.001}$	
					&10.59~~$^{+0.25}_{-0.24}$~~	&24.6$^{+0.1}_{-0.6}$	&48.4$^{+0.8}_{-0.1}$	\\	
NGC4559				&7.36$^{+0.09}_{-0.25}$			&0.52~~$^{+0.01}_{-0.01}$	
					&2.00~~$^{+0.43}_{-0.18}$		&1.294~~$^{+0.015}_{-0.058}$	
					&23.71~~$^{+0.55}_{-0.54}$~~	&17.8$^{+0.4}_{-1.1}$		&45.8$^{+7.6}_{-8.3}$	\\	
NGC4594				&7.28$^{+0.17}_{-0.17}$			&32.36~$^{+0.75}_{-0.74}$	
					&2.77~~$^{+0.06}_{-0.38}$		&0.032~~$^{+0.001}_{-0.001}$	
					&0.011$^{+0.001}_{-0.001}$		&17.8$^{+0.4}_{-0.1}$		&59.1$^{+0.1}_{-2.3}$	\\	
NGC4736				&7.03$^{+0.42}_{-0.24}$			&1.02~~$^{+0.46}_{-0.05}$	
					&0.53~~$^{+0.06}_{-0.06}$		&0.598~~$^{+0.028}_{-0.033}$	
					&4.27~~$^{+0.14}_{-1.76}$		&21.8$^{+0.8}_{-1.0}$		&35.4$^{+12.7}_{-3.0}$	\\	
NGC4826				&4.29$^{+0.10}_{-0.10}$			&3.72~~$^{+0.27}_{-0.63}$	
					&0.31~~$^{+0.05}_{-0.05}$		&0.070~~$^{+0.016}_{-0.039}$	
					&0.09~~$^{+0.03}_{-0.13}$		&23.4$^{+0.9}_{-0.6}$		&58.4$^{+0.8}_{-3.4}$	\\	
NGC5055				&19.50$^{+0.45}_{-0.44}$~~		&5.01~$^{+0.61}_{-1.38}$	
					&4.00~~$^{+0.39}_{-0.14}$		&0.906~~$^{+0.076}_{-0.219}$	
					&1.35~~$^{+0.49}_{-0.56}$		&19.2$^{+0.1}_{-0.3}$		&54.6$^{+1.8}_{-6.0}$	\\	
NGC6946				&23.99$^{+0.56}_{-0.55}$~~		&2.19~~$^{+0.27}_{-0.49}$	
					&2.77~~$^{+0.34}_{-0.41}$		&1.629~~$^{+0.241}_{-0.350}$	
					&6.76~~$^{+1.96}_{-0.82}$		&20.7$^{+0.6}_{-0.8}$		&54.9$^{+2.4}_{-1.7}$		
\enddata
\tablecomments{Median of the PDFs determined by MAGPHYS. The uncertainty given is the 68\% confidence range.}
\tablenotetext{a}{Averaged over 100 Myr}
\end{deluxetable*}

 \section{DISCUSSION}
 \subsection{Variation in SED Shape with Interaction Stage}
We now discuss the shape of the SED as a function of the interaction stage. We first normalized each 
SED to its 2MASS \emph{Ks} luminosity. Emission in the 2MASS \emph{Ks} filter is dominated by the old stellar populations 
and hence is a good proxy for stellar mass. The comparison between the SEDs is shown in 
Figure \ref{sed_shape}. In the inset, we show the median normalized SED for each class of galaxies. 
The SED shapes between the three stages vary by approximately as much
as the variations within a stage. However, there are some significant variations, 
especially in the Stage 4 SEDs compared to the Stage 2 and 3 SEDs.  Stage 4 galaxies typically have more 
emission from the hot/warm dust than earlier interaction stages, as evidenced by the stronger 
10$-$60\um~emission relative to their stellar mass. Further, Stage 4 galaxies tend to have more 
warm dust relative to their cold dust FIR emission. In contrast to this variation in the relative MIR 
emission, all three stages have similar ratios of NIR stellar emission to FIR emission from cold 
dust. We will discuss the statistical significance of these results in the next section.

\begin{figure}
\centerline{\includegraphics[width=\linewidth]{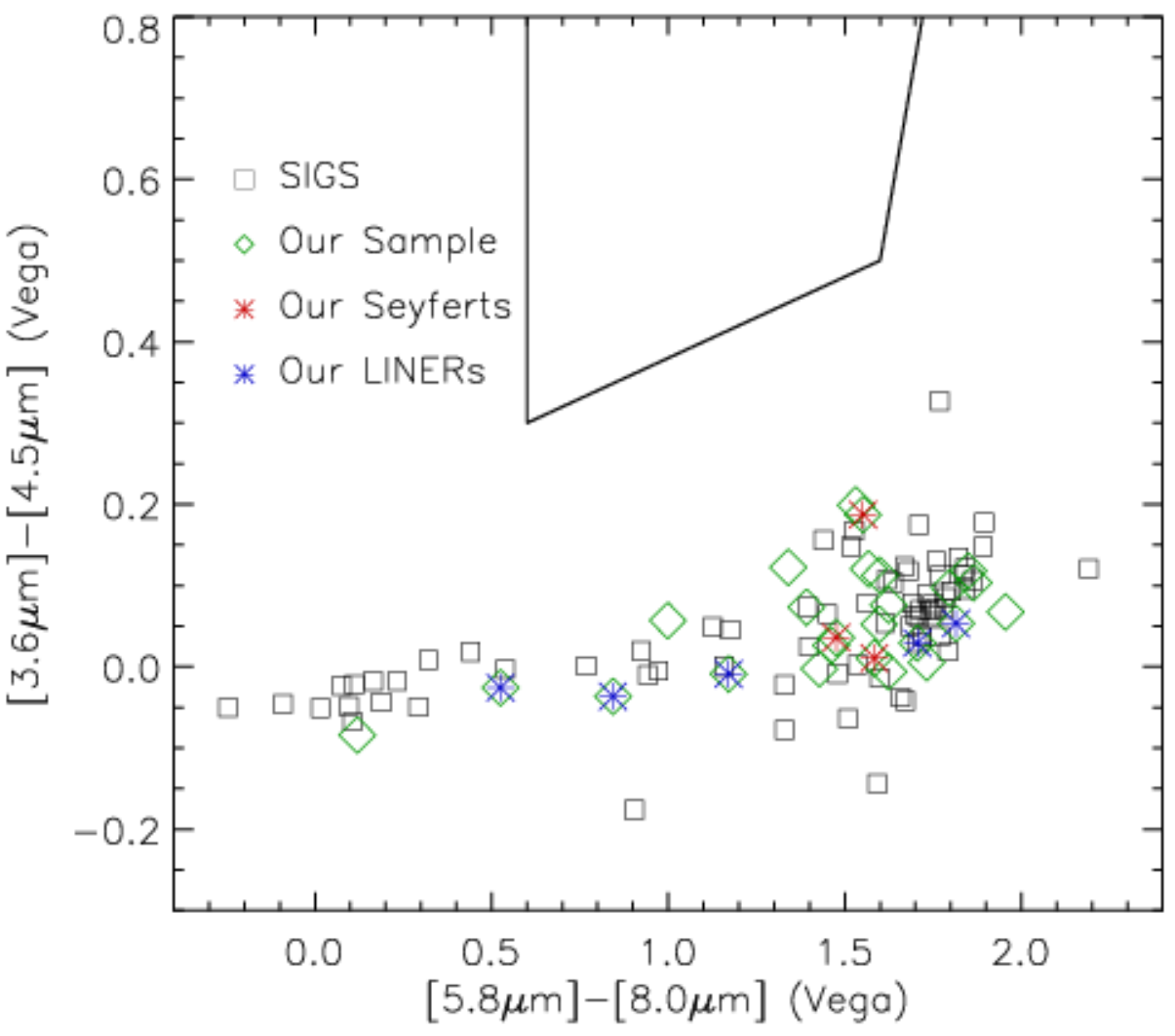}}
\caption{IRAC color-color plot of the SIGS galaxies (square) with our sample galaxies (diamonds) over-plotted, showing that
none fall in the wedge identified by Stern et al. (2005) as galaxies hosting AGN. Our sample galaxies identified optically
as Seyferts or LINERs in Keel et al. (1985) or Ho et al. (1997, 2000) and  are marked with red and blue stars, respectively.}
\label{stern}
\end{figure}

These differences in the SEDs are consistent with the results of simulations, which predict that 
an integrated SED of an interacting system becomes hotter at merger coalescence during the 
peak of starburst and AGN activity (e.g., Hayward \etal~2011, 2012a; Younger \etal~2009; 
Narayanan \etal~2010a, 2010b). However, the increase in temperature in this sample of galaxies is
unlikely to be driven purely by AGN activity. None of the galaxies in our sample have typical AGN 
colors, based on their location outside the region in IRAC color space defined by \citet{stern} (Figure \ref{stern}). 
Further, while five of the sample galaxies are classified as Seyfert galaxies and 
three are classified as low-ionization nuclear emission-line region (LINER) galaxies, they are  
found in all three stages. We used the software DECOMPIR\footnote{http:sites.google.com/site/decompir}
\citep{mul11} to estimate the AGN contribution to the 8$-$1000\um~and 8$-$35\um~emission for
these nine galaxies based on the 8-500\um~photometry. We give the individual contributions 
in the descriptions in the Appendix C; the range of the contribution to the total IR is up to 10\% with
some larger, more uncertain values up to 25\%. Further, we do not find significant differences in 
the AGN contribution to either the total infrared luminosity or mid-infrared luminosity, where the 
Stage 4 SEDs are typically brighter, between the stages. Hence, the effect of AGN on the SEDs of 
the sample galaxies is modest and does not affect our conclusions.
 
The more strongly interacting systems demonstrate a tendency to have, on average, 
younger stellar populations, resulting in stronger UV emission relative to their NIR emission. 
Since the UV emission has only been corrected for Milky Way extinction, additional intrinsic 
extinction could increase this effect. Ignoring NGC 4649, a large elliptical that has very little 
MIR-FIR emission, the UV bands reflect this tendency and show a large amount of variation 
between galaxies, which is likely due to the different amounts of dust attenuation and the 
sensitivity of the UV to recent star formation history. Comparing to the stellar mass proxy of 
2MASS \emph{Ks}, Stage 4 galaxies typically have a 1:2 luminosity ratio between emission 
in the \emph{GALEX} bands and 2MASS \emph{Ks} band, whereas Stage 2 galaxies typically 
have a 1:10 luminosity ratio.

 \subsection{Variations in Galaxy Parameters with Interaction Stage}
\begin{figure*}
\centerline{\includegraphics[width=0.75\linewidth]{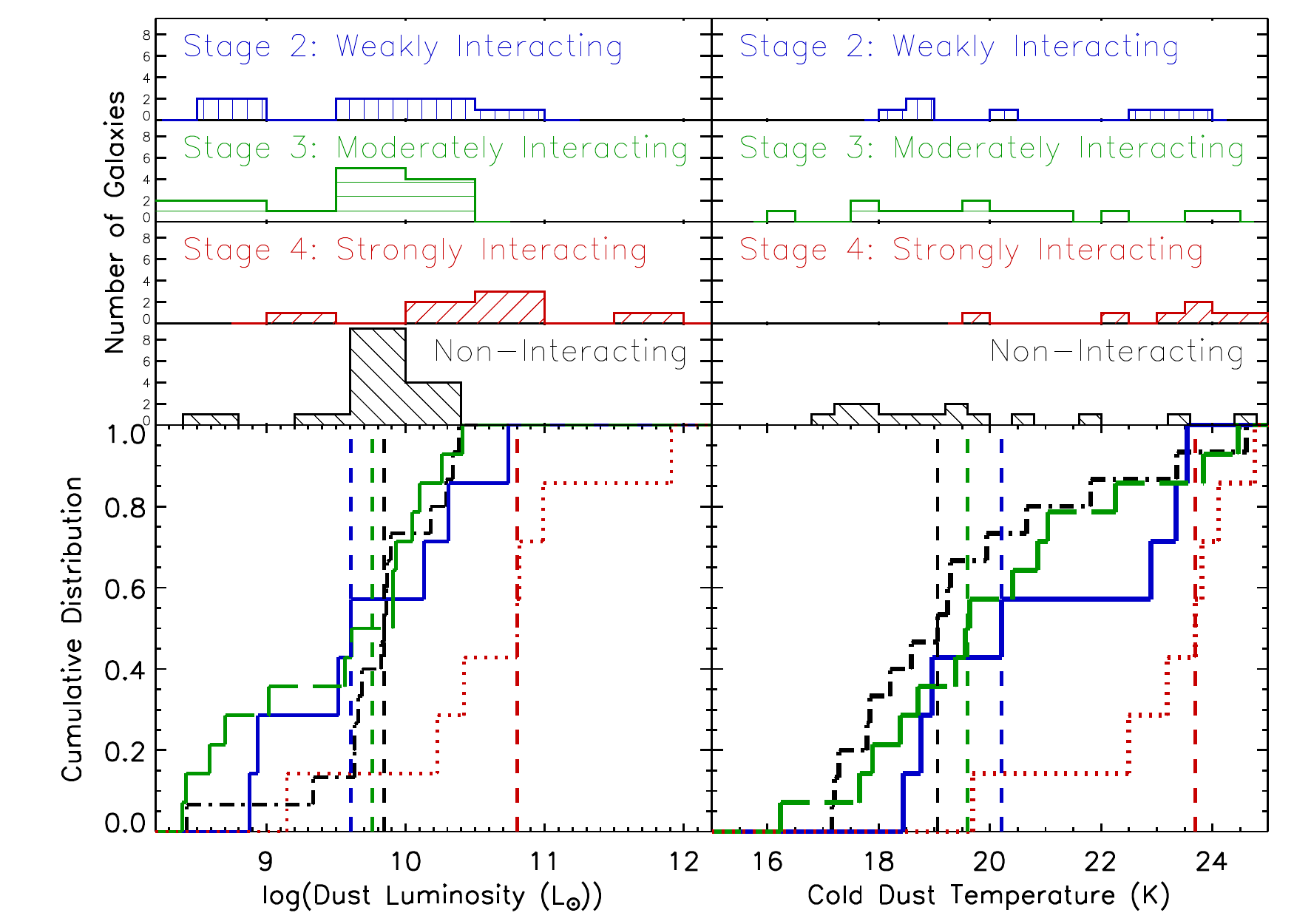}}
\caption{Histograms (top) and cumulative distributions (bottom) of the dust luminosity (left) and
cold dust temperature (right) derived by MAGPHYS for
the three interaction stages defined by the classification system of \citet{dopita02}, where 
Stages 2-4 are weakly (blue, solid), moderately (green, long dashed), and strongly 
(red, short dashed) interacting, respectively. Non-interacting galaxies are shown in black dot-dashed lines. 
There is a difference in both luminosity and temperature between the Stages 3 and 4 populations and 
between Stage 4  and non-interacting populations as defined by a K-S Test (see Table \ref{ks-out}). 
The vertical dotted lines give the median value for each stage. The median dust luminosity is lowest for 
the `weakly interacting' Stage 2 galaxies and increases by more than an order of magnitude for the `strongly interacting' Stage 
4 galaxies. The 15-25 K dust temperature is noticeably higher in the Stage 4 galaxies. }
\label{ldust_dist}
\end{figure*}

\begin{figure*}
\centerline{\includegraphics[width=0.75\linewidth]{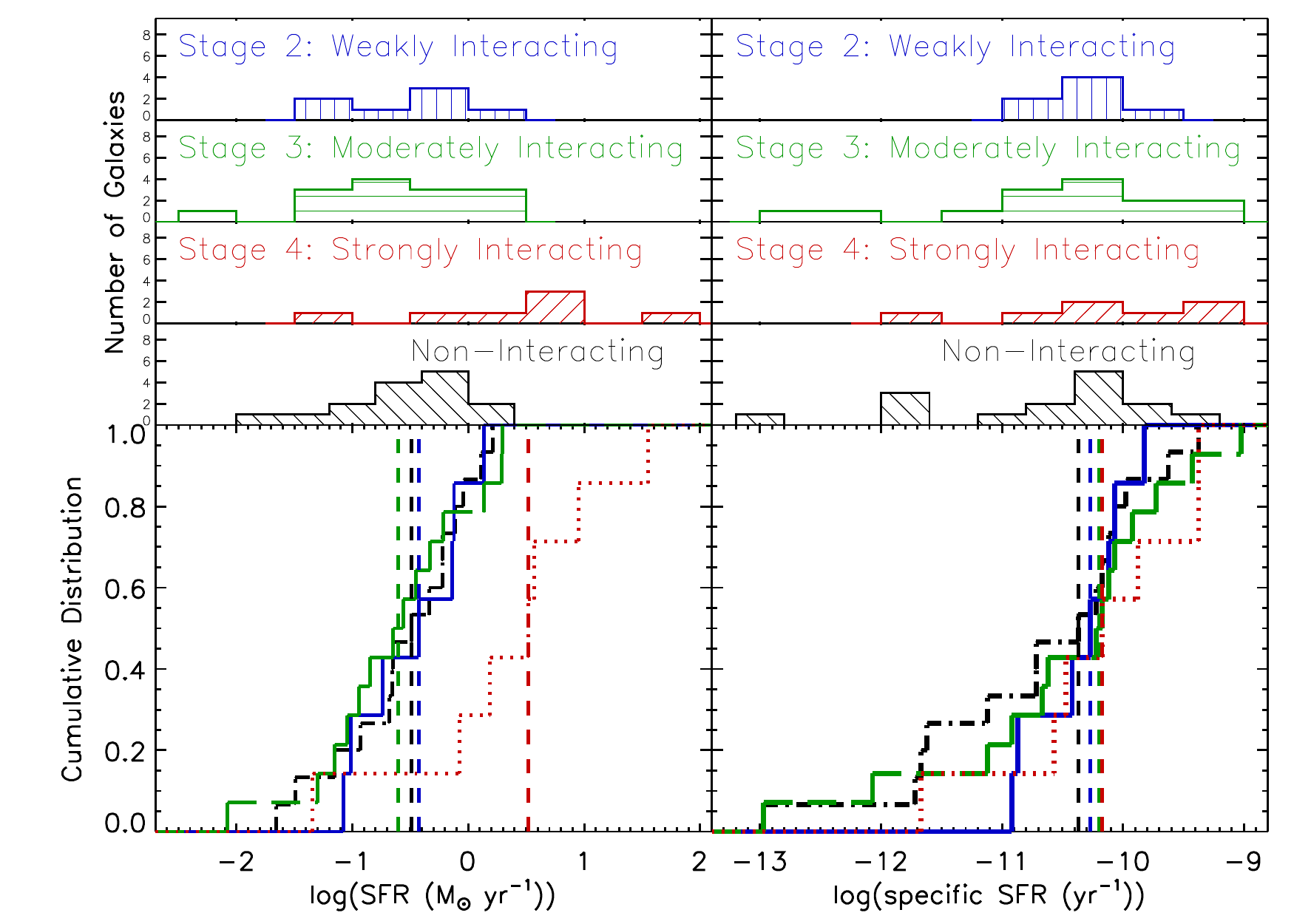}}
\caption{Histograms (top) and cumulative distributions (bottom) of SFR (left) and sSFR (right) 
derived by MAGPHYS for each interaction stage and the control sample shown with
the same color and linestyles as Figure \ref{ldust_dist}. SFR shows an 
increase between non-interacting through moderately interacting galaxies and the Stage 4, `strongly
interacting' systems, an evolution similar to that of dust luminosity. In contrast, the sSFR distributions
are very similar, as is supported by the results of the K-S tests.}
\label{sfr_dist}
\end{figure*}

Figures \ref{ldust_dist}-\ref{sfr_dist} show the histograms and  cumulative distributions for the three interaction 
stages defined in \S2.2 for four parameters whose PDFs are shown in Figures  \ref{seds1}-\ref{seds10}
compared to the non-interacting galaxies. Table \ref{ks-out} gives the results of the 
Kolmogorov-Smirnov \citep[K-S; e.g., ][]{press} tests performed
to determine the likelihood that the samples for the different stages originate from
a common parent population for each parameter. With the number of galaxies in our sample, we do not
have the same statistical power that the full SIGS sample will have (Brassington et al. 2013, in prep.).

The mass of dust is one of four parameters derived from the SED that differs with marginal statistical 
significance between the Stage 3 and Stage 4 galaxies ($p=0.02$). Both the mass of warm dust and
of cold dust likewise differ.  The warm dust mass also differs between the Stage 4 galaxies and the
non-interacting galaxies. In contrast, the distribution of dust mass in the ISM of non-interacting 
galaxies falls between the distributions of Stage 2/3 and Stage 4 galaxies. The dust mass in 
the ISM and stellar mass also each show only small increases in their median values as the 
interaction sequence progresses, and their cumulative distributions are not significantly 
different.  Indeed, both the stellar mass ($p=0.96$) and total dust mass ($p=0.96$) for Stage 2 
and Stage 3 galaxies are consistent with coming from the same populations. 
These trends are consistent with the SED shapes described in \S 5.1,  where the Stage 4 
SEDs typically show more emission from the hot/warm dust relative to the cold dust 
emission and stellar emission.

The dust luminosity (Figure \ref{ldust_dist}, left) shows marginally statistically significant differences 
between Stage 3 and Stage 4 galaxies ($p=0.01$) and between the non-interacting galaxies and the 
Stage 4 galaxies ($p=0.02$). The median dust luminosity increases with 
interaction stage by over an order of magnitude between Stage 2 and Stage 4. \citet{elbaz11} defined an 
``IR main sequence'' of galaxies in which the ratio of total IR luminosity to 8\um~luminosity has a Gaussian 
distribution. We examined this ratio for our sample and found good agreement with the expected 
distribution.  The only interacting galaxy that lies off this relation, by roughly an order of magnitude, is the large 
elliptical NGC 4649, as would be expected. \citet{elbaz11} also defined two modes of star formation: a 
``normal'' mode exemplified by the galaxies on the IR main sequence and a ``starburst'' mode with excess 
sSFR in comparison. Our sample's agreement with the IR main sequence indicates that our set of 
interacting galaxies do not contain systems with significantly increased sSFR.
 
 We also considered the evolution of the cold and warm dust temperatures. The cold dust 
temperature (Figure \ref{ldust_dist}, right) is the third parameter showing evidence for
 differences between the Stage 3 and Stage 4 ($p=0.01$) and between the non-interacting
 galaxies and the Stage 4 galaxies ($p=0.01$). The cold dust temperature's median value varies in a similar 
manner to the IR luminosity, increasing between Stages 3 and 4 but relatively constant 
between Stages 2 and 3; the median value of the warm dust temperature is by contrast fairly 
constant. Only Stage 4 does not span the range of the 15$-$25 K cold dust temperature, while
in Stage 2 and Stage 4 the warm dust temperatures are confined to the 45$-$60 K range. 
 The similarity in evolutionary trend in the IR luminosity and 
cold dust temperature is likely due to the predominance of cold dust mass and luminosity 
in the total dust estimates. We might expect a similar trend to be exhibited in the temperature 
of the warm dust primarily present in the stellar birth clouds, however the warm dust 
temperature is less well constrained in MAGPHYS than the cold dust temperature. The 
cold dust contribution and warm dust temperature are both correlated with the warm dust 
mass (which drives the MIR continuum intensity and the shape of the SED). Therefore, 
the warm dust temperature PDFs tends to be broader, with a 68\% range that is typically 
four to five times the size of the cold dust 68\% confidence interval. However, since the 
cold dust mass is typically over $\sim80$\% of the total dust mass, the total dust mass is 
still fairly well constrained.

The SFR, shown in Figure \ref{sfr_dist} (left), shows an increase in median value with interaction stage, like the
the dust mass and dust luminosity, and has a marginally significant probability the same population 
did not yield the Stage 4 galaxies as well as the Stage 2 ($p=0.03$), Stage 3 ($p=0.02$), and non-interacting
($p=0.02$) galaxies. Since 
the warm (30$-$60 K) dust, primarily heated by young stars with ages less than 10 Myr, contributes
significantly to the total dust luminosity, it makes sense that these three parameters show similar evolution
over the interaction stages. However, an increase in SFR between the different stages could be also 
attributed to our Stage 4 galaxies simply being larger with greater gas reservoirs. To test this, we also 
examined the evolution of sSFR over the interaction stages (Figure \ref{sfr_dist}, right). In contrast to SFR, 
we do not find much difference in the median values of the sSFR, and the cumulative distributions
are very similar in both width and normalization. We therefore do not see enhanced sSFR
in more evolved mergers,  consistent with the ratios of total IR luminosity to 8\um~luminosity 
being due to a ``normal'' mode of star formation for our whole sample. We also do not see differences
between the distributions of the sSFR of the three interacting galaxy samples and of the non-interacting galaxies.
Xu et al. (2010) and Yuan et al. (2012) examined star formation in a sample of local major mergers and both
found that that the sSFR distributions of galaxies in spiral-spiral interactions and non-interacting systems were 
unlikely to originate from the same population (based on a K-S test: $p=0.03-0.04$). However, they also found a 
mass dependence in the enhancement of sSFR; only those galaxies with 
$M_{*} > 10^{10.5} M_{\odot}$ were found to have significant enhancements. Our sample shows a similar trend, but 	
we only have six galaxies with $M_{*} > 10^{10.5} M_{\odot}$ of which one is an elliptical, and they are spread across
Stages 3 and 4.

We also examined the star formation efficiency, which we define as the ratio of SFR to warm (30$-$60 K) birth cloud 
dust mass, a proxy for molecular gas mass. While the ratio of dust mass to gas mass is not necessarily 
the same between galaxies, this ratio provides a means of estimating the star formation efficiency 
under the assumption of constant gas-to-dust ratio. This star formation efficiency allows us to compare
the SFR taking into account the variable gas reservoirs. We find that regardless of stage, the star 
formation efficiency ranges over more than three orders of magnitude and the cumulative distributions 
show no evidence of originating from different populations. This result agrees with the findings of Casasola
et al. (2004) who found similar star formation efficiencies, defined as the ratio of FIR luminosity (a proxy for
SFR) and molecular hydrogen mass, for interacting and non-interacting galaxies. 

\begin{deluxetable*}{lcccccccc}
\tabletypesize{\scriptsize}
\tablecaption{K-S Probabilities\label{ks-out}}
\tablewidth{0pt}
\tablehead{
\colhead{Parameter} &\multicolumn{3}{c}{With Non-Interacting} &\multicolumn{3}{c}{Between Stages} \\
\colhead{} &\colhead{Stage 2} &\colhead{Stage 3 } & \colhead{Stage 4} 
&\colhead{Stages 2 and 3} &\colhead{Stages 3 and 4} & \colhead{Stages 2 and 4} }
\startdata
Dust Luminosity						&	0.238	&	0.224	&	\textbf{0.017}	&	0.767	&	\textbf{0.008}	&	0.129		\\
Total Dust Mass						&	0.095	&	0.072	&	0.462		&	0.964	&	\textbf{0.022}	&	0.423		\\
Stellar Mass							&	0.366	&	0.655	&	0.569		&	0.964	&	0.271		&	0.129		\\
SFR									&	0.980	&	0.983	&	\textbf{0.017}	&	0.964	&	\textbf{0.022}	&	\textbf{0.028}	\\
sSFR								& 	0.569	& 	0.953	& 	0.606	 	& 	0.964	&	0.768		&	0.883		\\
Cold Dust Temperature					&	0.337	&	0.417	&	\textbf{0.006}	&	0.767	&	\textbf{0.008}	&	0.423		\\
Warm Dust Temperature					& 	0.569	& 	0.237	& 	0.106		& 	0.271	&	0.271		& 	0.129		\\
Cold Dust Mass						&	0.095	&	0.072	&	0.462		&	0.964	&	\textbf{0.022}	&	0.423		\\
Total Warm Dust Mass					&	0.682	&	0.678	&	\textbf{0.019}	&	0.767	&	\textbf{0.022}	&	0.129		\\
Birth Cloud Warm Dust Mass				&	0.912	&	0.678	&	\textbf{0.019}	&	0.492	&	\textbf{0.022}	&	0.129		\\
SFR/Dust Luminosity					&	0.238	&	0.224	&	0.999		&	0.492	&	0.271		&	0.883		\\
SFR/Birth Cloud Warm Dust Mass			&      	0.119      	&      	0.478      	&      	0.462	      	&      	0.492	&   	0.492		& 	0.883		\\
Dust Mass/Stellar Mass					&	0.644	&	0.916	&	0.857		&	0.767	&	0.271		&	0.423		\\
Birth Cloud /Cold ISM Dust Mass			&	0.719	&	0.224	&	0.037		&	0.767	&	0.271		&	0.129		\\
\enddata
\tablecomments{The probabilities given in this table are the probability that the same population yielded 
the compared samples, so small values indicate a common population is unlikely. The marginally significant 
differences are given in bold. The total dust mass is composed of warm dust in the stellar birth clouds and the 
diffuse ISM and cold dust in the diffuse ISM.}
\end{deluxetable*}

All the apparent variations with interaction stage come with a few caveats. First, with only thirty-one 
galaxies, our sample has limited statistical power to identify significant variations, especially with half 
the galaxies in Stage 3. Analysis of trends in SFR and sSFR with the full SIGS sample (Brassington 
\etal~2013, in preparation), which covers the stages much more uniformly, will have greater 
statistical power (albeit these comparisons lack the \emph{Herschel} SPIRE data and SED analysis
that provides more accurate SFR measurements). Second, our classification scheme permits us to 
examine parameter variations with respect to the strength of the interaction. While this sequence crudely 
mimics an interaction, the dynamics of two colliding galaxies often includes multiple encounters prior to final coalescence, 
modifying the level of star formation at intermediate stages \citep[e.g.,][]{tor12}, as well as the 
intensity of the final burst (e.g., Hopkins \etal~2008, 2009). As a result, interacting systems often do 
not progress linearly through the interaction stages defined by our classification system.

\begin{figure*}
\centerline{\includegraphics[width=\linewidth]{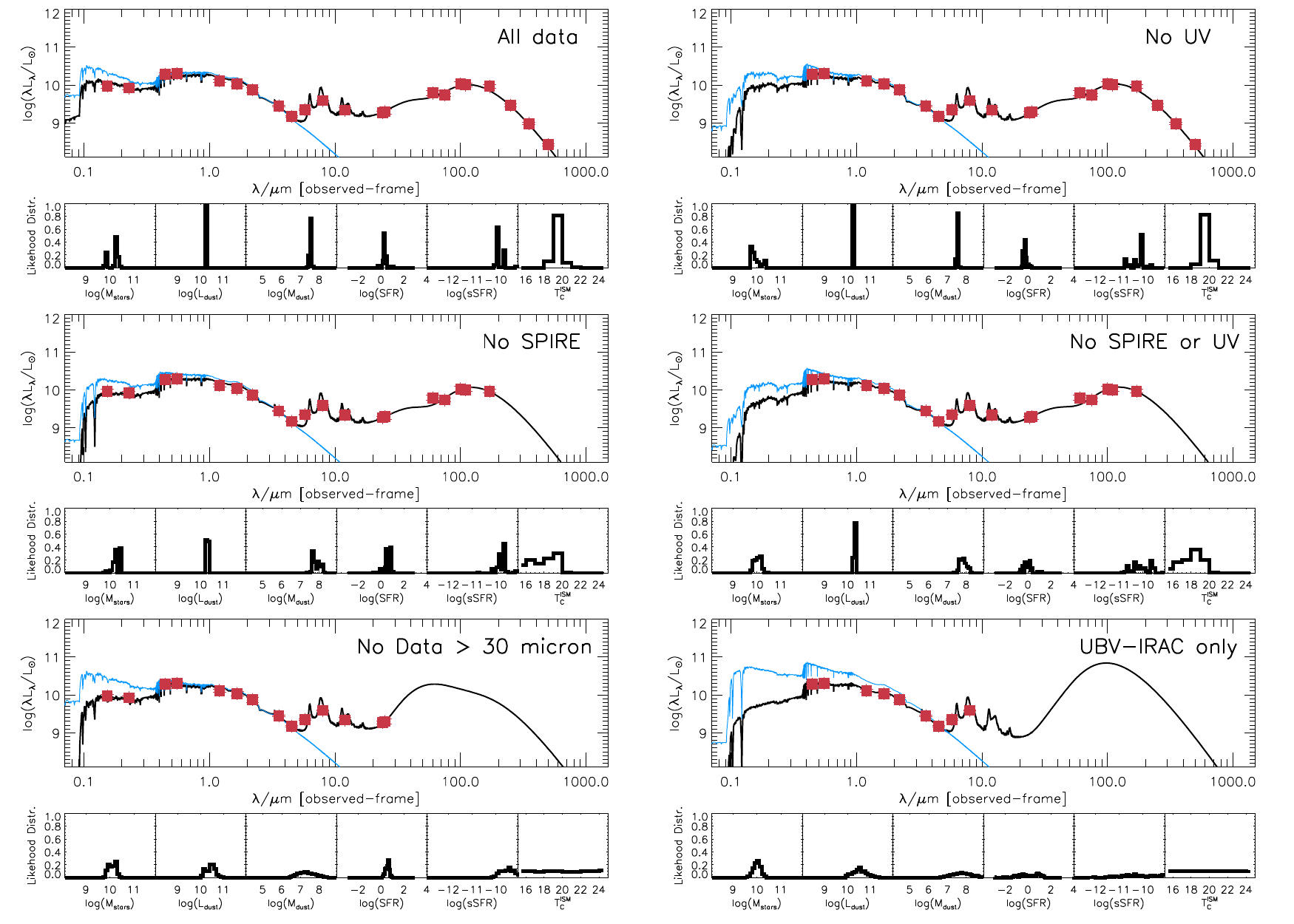}}
\caption{MAGPHYS fits with the 6 different data sets for M101, as a representative 
example of the set of fits done for each galaxy, showing the increasingly constrained 
parameters (from lower right to top left) as more data are used. UV photometry (present 
in the left panels) is crucial in constraining sSFR, while SPIRE data (present in the top 
two panels) is essential for the determination of cold dust temperature and dust mass.
The lower right panel demonstrates the limited constraints that ground based photometry 
alone can provide.   }
\label{msed}
\end{figure*}

\subsection{Relative Importance of Specific Data Sets in Constraining Galaxy Parameters}
For each galaxy, we ran six MAGPHYS fits to measure the relative importance of UV, SPIRE, and MIR-FIR
data in constraining the value of the derived SFR; stellar mass; sSFR; and dust temperatures, luminosity, 
and masses. We did this by comparing fits with all available data with fits using a subset of the complete
dataset in order to determine if and how the absence of a particular dataset results in a systematic 
over- or under-estimation of these parameters. Figure \ref{msed} shows a representative example: 
the best-fit SEDs for all six fits for M101 as well as the accompanying PDFs for the parameters of interest.

\begin{figure}
\centerline{\includegraphics[width=\linewidth]{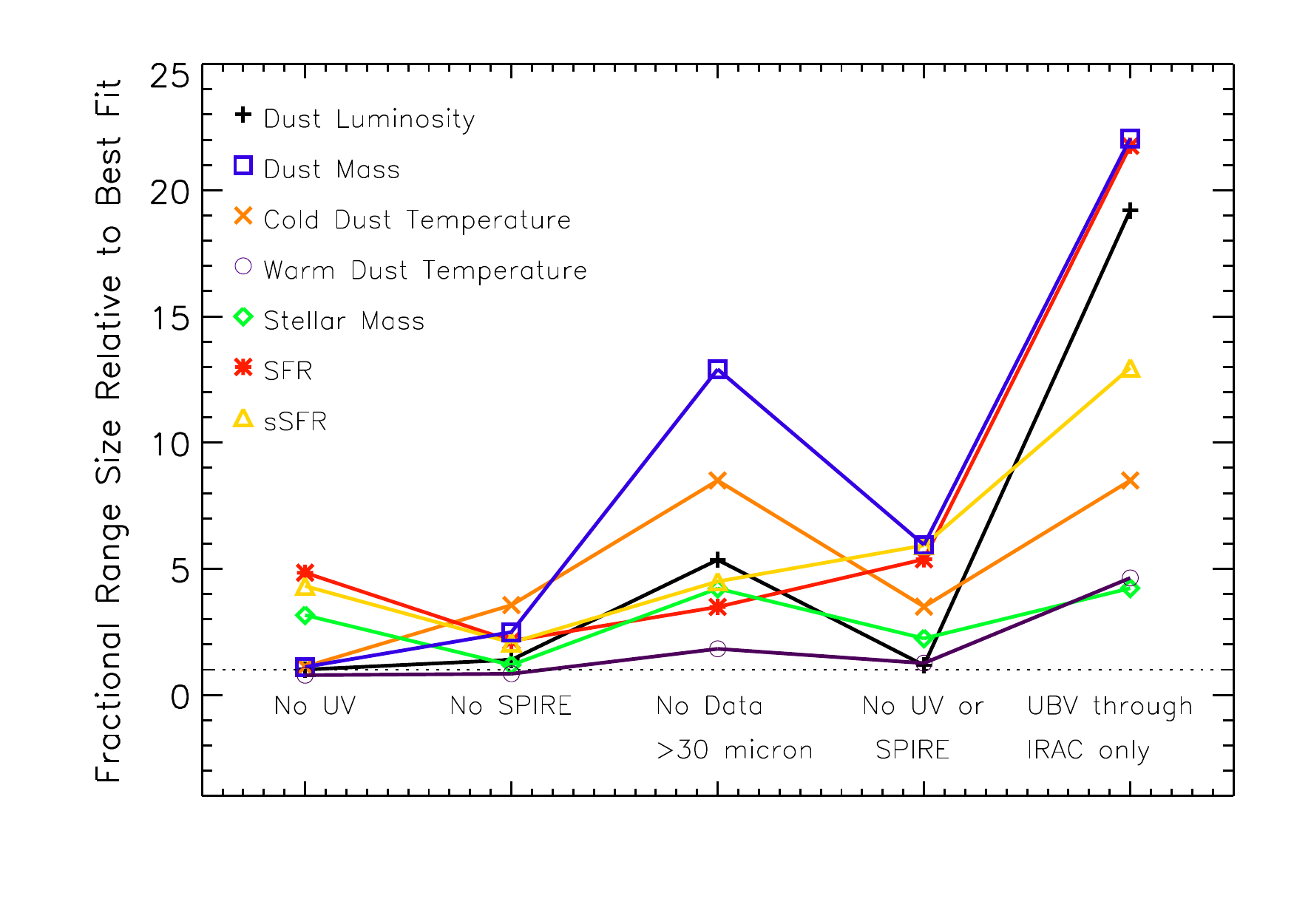}}
\caption{Parameter confidence intervals increase with the omission of data. Points indicate the median size 
of the 68\% confidence interval in the probability distribution function (PDF) for each parameter for the 
whole set of galaxies relative to the range when all of the data are used in the fit. Using all available data, 
the median 68\% ranges are: 6.3$\times10^{8}$~L$_{\odot}$ (dust luminosity), 1.8$\times10^{6}$~M$_{\odot}$ 
(dust mass), 0.80~K (cold dust temperature), 4.4~K (warm dust temperature), 2.1$\times10^{9}$~M$_{\odot}$ 
(stellar mass), 0.047~M$_{\odot}$~yr$^{-1}$ (SFR), and 1.3$\times10^{-11}$~yr$^{-1}$ (sSFR). UV data are 
crucial for the determination of sSFR, whereas dust luminosity, mass, and 15-25 K temperature are best 
constrained by SPIRE data with further constraints applied by photometry from 30-200\um.}
\label{prange}
\end{figure}

\begin{figure}
\centerline{\includegraphics[width=\linewidth]{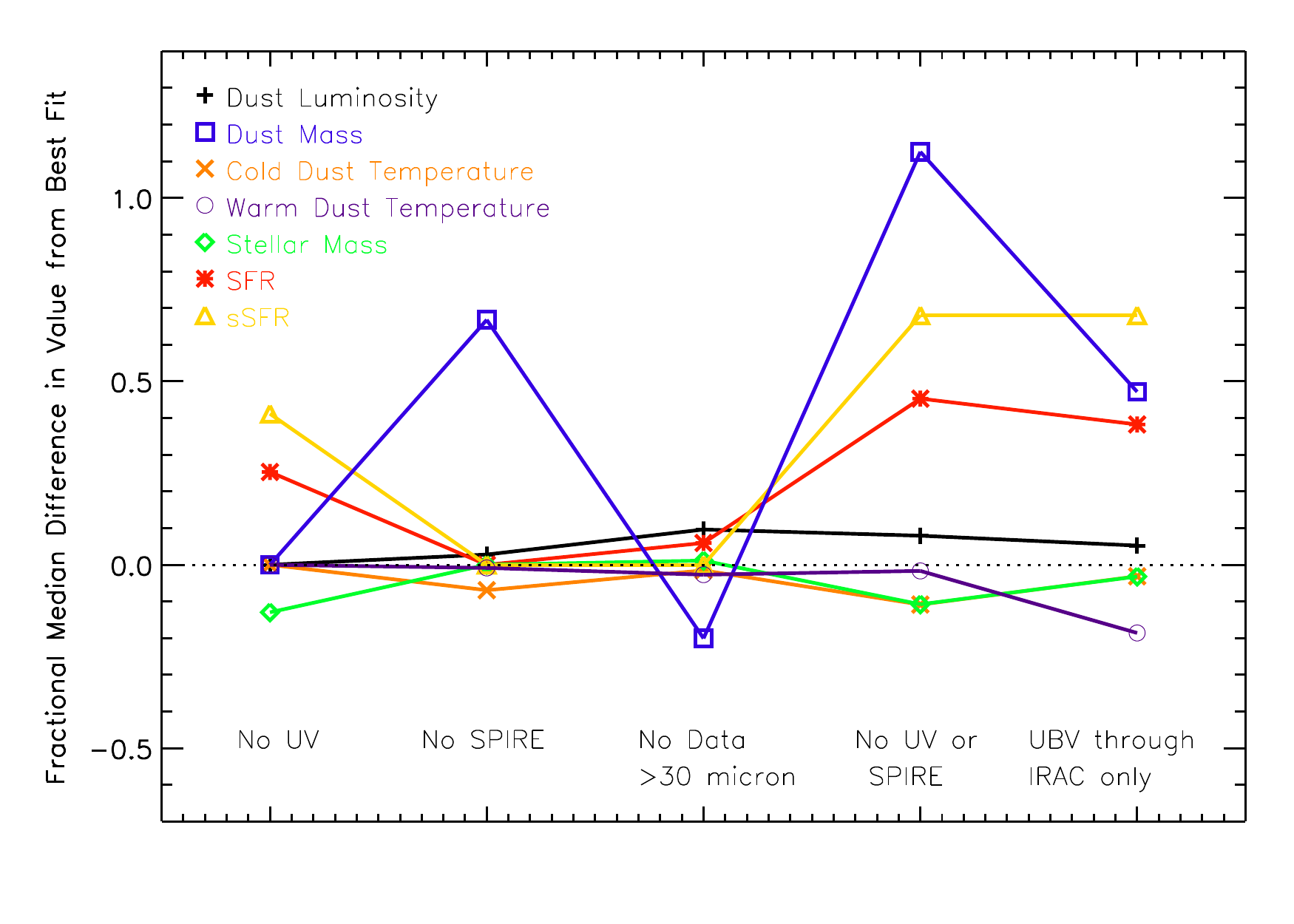}}
\caption{Median fractional difference for each parameter relative to its value 
that parameter when all of the data are used in the fit. The absence of UV data 
results in sSFR $\sim$40\% higher than in fits using all the data. Dust mass, made up
primarily of cold dust, is very sensitive to changes in the cold dust temperature. }
\label{pdiff}
\end{figure}

As expected, the SFR is significantly constrained by UV data. This is demonstrated in Figure \ref{prange}, 
where the median 68\% confidence interval size is a factor of $\sim$4 larger in the absence of UV 
photometry. While stellar mass is fairly well constrained by the UBV-IRAC data alone, the absence of 
UV data tends to result in younger stellar  population templates being selected by MAGPHYS. This 
effect can be seen in the differences in the UV slope and the strength of the Lyman and Balmer breaks  
in the various panels of Figure \ref{msed}. When a younger stellar population template is selected, a 
smaller fraction of the stellar emission is assumed to originate from late-type stars, resulting in a 
tendency to estimate the stellar mass $\sim10-20$\% lower than when all the data are used.  This can be 
seen in Figure \ref{pdiff} where we plot the median fractional difference in the value of the galaxy 
properties for fits with incomplete data sets. UV photometry constrains both the SFR and stellar 
mass, and it is also the most important wavelength regime to constrain the sSFR. The absence of 
UV data also tends to result in an over-estimation of the SFR resulting in an estimate of the sSFR 
$\sim$40\% higher than in fits using all available data.

Herschel data are particularly crucial in constraining the cold dust temperature. PACS data 
typically outline the peak of the IR emission, but in cases of the coldest dust temperatures, 
PACS 170\um~is typically too indeterminate and it is only in conjunction with the SPIRE 
250\um~data that the cold dust temperature is reasonably constrained. In contrast, as 
expected, warm dust temperature is not well constrained by the SPIRE data. Dust luminosity is 
typically well-estimated with a combination of IRAS and PACS data, only becoming about a 
factor of $\sim$2 more uncertain in absence of SPIRE data.

SPIRE observations are crucial for constraining the dust mass, whose 68\% confidence 
interval would be at least a factor of 3 larger without SPIRE data. When MIR data at 
wavelengths $\lambda \ge$ 30\um~are likewise absent, the dust mass becomes almost 
completely unconstrained as the SED contains little information about the dust emission. 
The dust mass estimate is the most sensitive to the absence of specific datasets. Interestingly, 
the dust mass estimated by MAGPHYS is $\sim$60\% higher when SPIRE data are excluded as
compared with when all data are used; however, when photometry at wavelengths $\lambda
\ge 30$\um~is excluded, the dust mass is estimated $\sim$20\% lower than when all data are used. 
The over-estimates in dust mass are correlated with the under-estimates in the cold dust 
temperature. Since, cold dust tends to make up the bulk of the dust mass and because cold 
dust mass varies as $T^{-6}$ (assuming it is modeled as a $\beta=2$ modified blackbody), a 5\% change in the 
dust temperature results in a 30-40\% difference in the dust mass. \citet{aniano12} recently showed 
similar results for NGC 628 and NGC 6946, where fits undertaken only with data at wavelengths 
$\lambda \le$ 170\um~tend to overpredict the emission in the SPIRE bands and the associated 
cold dust mass. When all data at wavelengths $\lambda \ge$ 30\um~are omitted, MAGPHYS relies
primarily on more common UV-NIR dominated sources.

\begin{figure}
\centerline{\includegraphics[width=\linewidth]{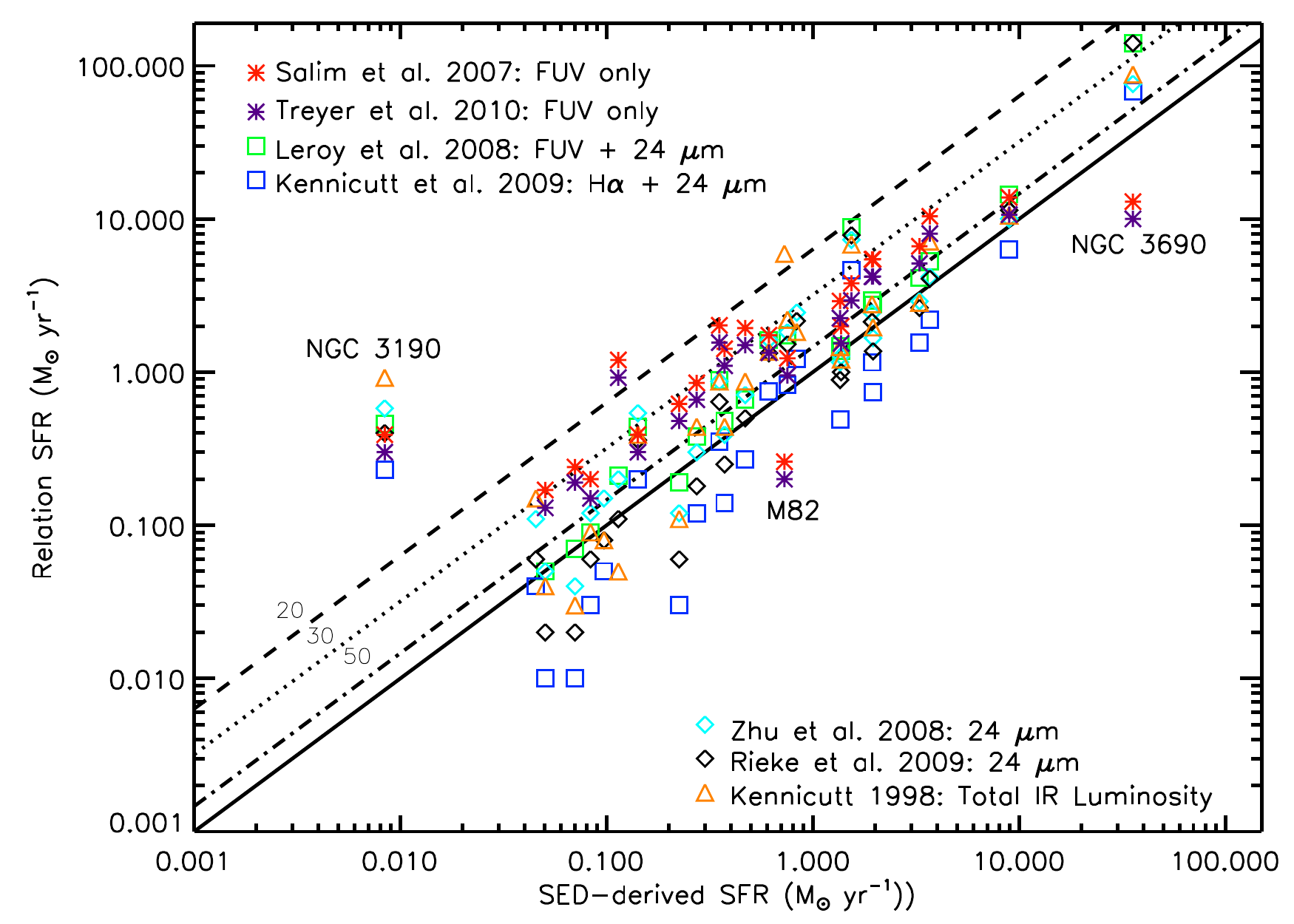}}
\caption{Comparison of the SED-derived SFR to those from various SFR proportionality relations 
from the literature. The solid line shows where the points should lie if the relations agree with the 
SED-derived values, which is the SFR averaged over 100 Myr. The  dotted, dashed, and dot-dashed
lines shows SFR averaged over 20, 30, and 50 Myr, respectively. SFR is modeled as an exponential burst
with a peak value such that the average over 100 Myr is given by the MAGPHYS SFR and decay times of
20, 30, and 50 Myr. While the relations depending solely or partially on IR emission agree well with the 
MAGPHYS SFR, the FUV relations are typically high and agree better with a SFR modeled as an 
exponential burst with a width of 30-50 Myr. 
}
\label{sfrcomp}
\end{figure}

\subsection{Comparison of the SFR Derived from the SED to Monochromatic and Broad-band Relations}

Because the SFR is such a key physical parameter, many statistical heuristic attempts have 
been developed to estimate it from simple observables such as the UV flux (as measured by 
GALEX)  or H$\alpha$,  the 24\um~flux, the total IR flux, and combinations of UV and MIR fluxes 
\citep[e.g.,][]{cal12}. We compared the SFR derived by MAGPHYS to the expected SFR based on a wide 
range of relations (Figure \ref{sfrcomp}): from the FUV relations of \citet{sal07} and \citet{tre10}, from 
the MIPS 24\um~emission relations of \citet{rie09} and Zhu \etal~(2008, as given in Calzetti \etal~2010), 
from the relations combining UV and 24\um~emission of \citet{ler08} and \citet{ken09}, and from 
the total IR luminosity relation of \citet{ken98}. We correct the FUV emission from dust using the 
prescription given in \citet{sal07} and use the dust luminosity derived by MAGPHYS for the total 
IR luminosity.  We assume a \citet{cha03} initial mass function (IMF) and therefore have applied 
correction factors of 1.06 for relations that assume a \citet{krou01} IMF and 1.60 for relations that 
assume a \citet{sal77} IMF, following Calzetti (2012) and Schiminovich \etal~2007.  Figure \ref{sfrcomp} shows the resulting values 
plotted against the SFR determined by MAGPHYS.

One trend is quickly apparent: the SFRs estimated from UV photometry alone tend to be high compared
with the SED-derived SFR, whereas the relations based solely or partially on IR photometry agree 
fairly well, at least for SFR greater than $\sim$ 0.1 M$_{\odot}$ yr$^{-1}$. Median differences are of 
0.7-0.9 M$_{\odot}$ yr$^{-1}$ for the FUV relations. Interestingly, our most active galaxy, 
NGC 3690/IC694, shows the inverse trend, as does a prototype starburst M82. This may indicate that the 
correction for dust is insufficient for these systems. The outlier of NGC 3190 (Fig. \ref{sfrcomp}) is due 
to the low SFR associated with the best MAGPHYS model for this galaxy that significantly 
underestimated its UV emission, likely due to the edge-on geometry of the system.

The apparent over-estimation of the SFR by the FUV relations is rooted in the time over which 
the SFR is estimated in MAGPHYS. FUV emission is dominated by star formation within the 
past 50 Myr, although mid-to-late B stars can also contribute a significant fraction \citep{cal12}. 
In contrast, the IR relations typically assume that a fraction of the stellar light is absorbed to heat 
dust, and as a result, while the youngest and hottest stars dominate the heating of hot dust, the 
accumulation of low-mass stars contribute significantly to the heating of the more diffuse dust. 
Hence, the IR relations represent star formation over a longer timescale. The SFRs derived by 
MAGPHYS are averaged over the last 100 Myr, which is more consistent with the timescales 
associated with the IR-dependent SFR relations. Figure \ref{sfrcomp} also shows the SFRs 
averaged over shorter time periods with SFR modeled as an exponential decay with a peak 
value such that the average over 100 Myr is the MAGPHYS SFR value and decay times of 20, 
30, and 50 Myr, respectively. The SFRs estimated from the FUV emission agree better with an 
exponential decay star formation history with a width between 30 and 50 Myr, which is the
expected timescale of a starburst episode.  

\section{SUMMARY}
We modeled the FUV-FIR SEDs of fourteen groups of thirty-one interacting galaxies, 
typically with 15$-$25 flux points,  to determine the most probable evolution of dust 
luminosity, star formation rate, specific star formation rate, dust mass, stellar mass, and 
dust temperature.  The systems were classified as either weakly, moderately, or 
strongly interacting (Stages 2-4 in the Dopita \etal~(2002) scheme). The broad similarities in SED shape 
between different stages emphasize one key conclusion from this study: as the interaction 
progresses, and even as bursts of star formation may occur, the changes are most clearly 
seen not in the distribution of energy broadly but in minor and subtle changes to the SED 
shapes. Bulk SED properties change little, and only gradually, in typical interactions. 
Strongly interacting galaxies typically have SEDs characterized by stronger MIR 
emission relative to both their NIR and FIR emission and more UV emission relative to 
their NIR emission. 

There are marginally statistically differences (as determined by a 
K-S Test) in the derived galaxy properties: dust luminosity and mass, SFR, 
and cold dust temperature increase from Stage 3 to Stage 4,  SFR  
increase from Stage 2 to Stage 4, and dust luminosity, SFR, and cold dust temperature
increases from the non-interacting galaxies to the Stage 4 galaxies. 
In contrast, the sSFR does not show variations with 
interaction stage. The relative constancy of the sSFR between the different stages 
suggests that this lack of evolution is not due to uncertainty in stage classification or in 
the association of interaction stage and progress along the interaction. Rather, our set 
of interacting galaxies shows no clear evidence for a burst of star formation prompted by 
the interactions or that such effects occur on timescales such that we see enhancements in 
both stellar mass and SFR, leaving the sSFR relatively stable. 
This suggests a need to be circumspect about this canonical activity 
during the early stages of galaxy mergers.

Different wavelengths have different effects in constraining galaxy parameters in the MAGPHYS 
SED analysis. UV data are important in accurately determining stellar population age and that they 
are important contributors to determining accurately the (specific) star formation rate. The stellar 
mass is primarily determined by UBV-IRAC data. SPIRE data are crucial in determining the dust 
mass; in its absence the cold dust temperature tends to be underestimated, because the location 
of the peak is much less constrained without data longward of 170\um. Cold dust mass, which
tends to make up the bulk of the dust mass, goes as $T^{-6}$ (assuming $\beta$=2), so the change 
in the dust mass is large for even a small underestimate in the dust temperature. The possible
contributions to the SED from AGN are modest for this sample and do not affect our conclusions.

The SFRs derived by MAGPHYS agree reasonably well with simple relations based solely or partially on 
IR photometry. Relations based on corrected FUV emission tended to overestimate SFR compared 
to the SED-derived SFR, which is averaged over 100 Myr. The SFR estimated 
from FUV can best be understood if it  represents an exponential decay star formation history 
with a width of 30-50 Myr.

The complete SIGS sample will bring a significant increase in statistical power in determining
galaxy property trends. In addition, testing the accuracy of MAGPHYS against simulations of 
interacting galaxies will help improve the diagnostic power of SEDs. In a future paper, we will 
examine what kind of simulated interactions and their parameters best reproduce observed 
systems and their SEDs.  Further, we will test how well MAGPHYS recovers galaxy 
parameters as a function of their interaction details.

\acknowledgements 
We are grateful to Gregory Snyder, Maria Koutoulaki, and Flora Stanley for their classifications of our galaxies, 
Steven Willner for his classifications and helpful comments, Volker Tolls for his assistance in processing Herschel 
data, Don Neill and Susan Neff at the \emph{GALEX} helpdesk for the reprocessing of the observation of NGC 
3690/IC 694, Erik Hoversten for his assistance with regards to the UVOT corrections, Christopher Klein for his 
preparation of the IRAC observations, and Joseph Hora for his assistance with galaxies on the edge of SPIRE 
observations.  This work was based on archival data obtained 
from the \emph{Spitzer} Science Archive, the Mikulski Archive for Space Telescopes (MAST), the \emph{Swift} 
data archive, and the \emph{Herschel} Science Archive. This research has made use of the NASA/IPAC 
Extragalactic Database (NED), which is operated by the Jet Propulsion Laboratory, California Institute of 
Technology, under contract with the National Aeronautics and Space Administration.This work was 
supported in party NASA grant NNX10AD68G and NASA JPL RSA 1369566. AZ acknowledges support 
from AR-12011X Chandra grant and EU IRG grant 224878. 

\appendix

\section{CLASSIFICATION SCHEME}
\begin{figure*}[h]
\centerline{\includegraphics[width=0.9\linewidth]{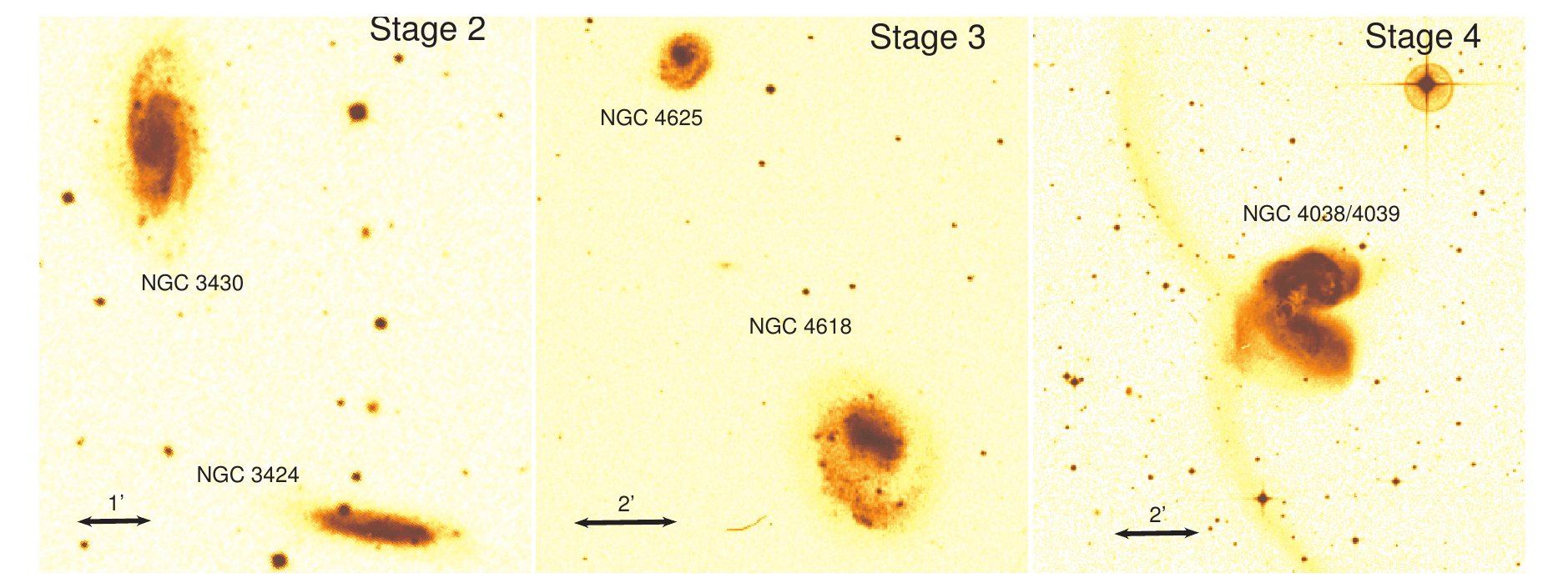}}
\caption{Representative examples of Stage 2-4 (left-right) galaxies, showing increasing morphological distortion.}
\label{class_scheme}
\end{figure*}

The classification methodology used to determine the interaction stage of each system is based on the Dopita et al. (2002) 
classification scheme. Stage 1 galaxies are non-interacting and Stage 5 galaxies are post-merger or coalescence systems.
Stages 2-4 are weakly, moderately, and strongly interacting systems, based on their degree of morphological distortion. In
Figure \ref{class_scheme}, we show a representative system from our sample. The Stage 2 galaxies show little distortion 
and the galaxies are typically well separated. The Stage 3 galaxies have a range of proximity and show some degree of 
distortion and the Stage 4 galaxies show significant morphological distortion and are typically close even in projection.

\section{PROCESSING CONCERNS}
\subsection{PACS and High Pass Filters}
In the course of processing the PACS photometry, we initially used the PhotProject pipeline scripts to make mosaics. When 
we compared the measured fluxes to values obtained by the MIPS and \emph{IRAS} instruments at the same wavelengths, 
we found some were more than a factor of 2 too low, with M101 having the greatest disagreement. 
This  disagreement is the result of the range of angular extents of the galaxies. The PhotProject 
pipeline includes a high pass filtering algorithm that removes a significant portion of the extended emission in the field and 
therefore affects nearby galaxies with more extended emission much more egregiously. As a result,  we found it essential 
for nearby galaxies to use the Madmap pipeline. 

\subsection{Swift UVOT and Coincidence Losses}
As we noted in \S 3.3.2, the UVOT telescope is vulnerable to coincidence losses. When multiple photons arrive in the same 
pixel within the same frame, only a single photon is counted. Losses become greater than 1\% when the count rate 
exceeds 0.007 counts per second per pixel. While corrections for these losses and the uncertainties involved is well 
determined for point sources \citep{poole08, kuin08}, there is a lack of similar understanding for extended sources. 
Although some of the sample galaxies, notably the starburst galaxies M82 and Arp 299, have count rates high enough 
over most of their surface that would result in significant coincidence losses, even for the rest of the sample have  
regions of high fluxes, typically limited to the nuclear  area and star forming regions. We therefore opted to analyze only 
those galaxies lacking \emph{GALEX} photometry as well as an additional test case, as described in \S 3.3.2. Fortunately for the two 
galaxies having UVOT, but lacking \emph{GALEX}, only the nucleus and a few small regions are bright enough to require 
corrections. We excluded the regions with high count rates and then measured them independently as point sources, 
correcting them for coincidence losses using the method of \citet{poole08}. These corrections always account for less than 
1.5\% of the count rate. Based on our test case of NGC 3424, the agreement between the \emph{GALEX} and UVOT photometry is
excellent.

\section{NOTES ON INDIVIDUAL GALAXIES}
\noindent
Group 1: \newline
\indent
NGC 2976: A small galaxy on the outskirts of the M81 group, NGC 2976 has two active, infrared
bright regions at the edges of its disk, whose presence is almost completely missed in the UV and NIR. \newline
\indent
NGC 3031: Also known as M81, this galaxy is in our closest system and has very clear spiral arms 
and filamentary structure seen in all the images across our whole filter range. Its nucleus
is a LINER. The best-fit decomposition of the IR spectrum performed with DECOMPIR has a 4\% AGN
contribution to the total IR emission and a 16\% contribution in the 8$-$35\um~range.\newline
\indent
NGC 3034: M82 is one of the nearest starburst galaxies and has  a strong galactic
outflow perpendicular to its stellar disk, which is visible in both the UV and \emph{Herschel} images. Its IRAC 5.8\um, IRAC 
8.0\um, and MIPS 24\um~images are saturated. Its
SED is among the worst fit by MAGPHYS with the current set of models. This may be due in part to the high obscuration - 
its fit has a $<\tau_{V}>=2.9$ - or perhaps to the inclusion of some emission from the outflow, 
which MAGPHYS does not model. We measured the contribution of the outflow above and below the disk
of the galaxy within the aperture and found that it contributes $\sim$20-30\% of the UV emission and $\sim$10-20\%
of the emission at wavelengths greater than 150\um, which does not fully account for the discrepancy in the UV bands. We selected
the 3.6\um-derived aperture to minimize the contribution of the emission from the outflow. The outflow's UV emission is primarily scattered
light from the disk, while its IR emission is from the dust in the outflow. Since MAGPHYS cannot model such a feature, its inclusion would
tend to bias the best-fit model. \newline
\indent
NGC 3077: Another small galaxy in the M81 group, NGC 3077 lies behind a nearby bright star that prevented it
from being observed in the UV by \emph{GALEX} and UVOT. Its MIR-FIR images shows evidence of tidal stripping in 
the asymmetric structure and southwest lobe. 

\noindent
Group 2: \newline
\indent
NGC 3185: This Seyfert galaxy shows a circum-galactic ring of star-forming material in both UV and IR images. 
The DECOMPIR decomposition requires a 3\% and a 12\% contribution of the AGN to the total IR luminosity and
8$-$35\um~luminosity respectively.
Its disk shows up in the FIR image as two bright regions at the opposite ends of the galaxy.
It is the most distant member of the triplet it forms with NGC 3190 and NGC 3187 on the sky, but it and NGC 3190 have 
much closer recessional velocities than NGC 3187 (1217 km s$^{-1}$ and 1271 km s$^{-1}$ vs 1581 km s$^{-1}$). 
Group 2 is the only compact group in our sample, although the SIGS sample has several others. Tzanavaris \etal~(2010) 
found a bi-modality for the sSFR distributions of compact groups depending on the slope of the IRAC photometry between
4.5\um and 8.0\um.  NGC 3185 has a negative IRAC slope and its sSFR places it in the relatively quiescent population as expected
by Tzanavaris \etal~(2010).
\newline
\indent
NGC 3187: This galaxy has a pair of tidally elongated arms, which are best detected in the UV. This galaxy did not have archival
\emph{IRAS} fluxes. NGC 3187 has a positive IRAC slope and its sSFR places on the edge of the distribution of galaxies with positive
slopes, argued by Tzanavaris \etal~(2010) to indicate active star formation. \newline
\indent
NGC 3190: The most massive of the three galaxies in this group, NGC 3190 is a nearly edge on LINER and has a dust lane, 
that appears prominently in absorption in UV and is correspondingly bright in IR emission. Its SED shows  
particularly low UV relative to its visible emission, but presumably the geometry of this dust lane  
explains the poor fit to the UV and the disagreement in the estimates of its SFR. The DECOMPIR best-fit requires no 
contribution from an AGN. NGC 3190 has a negative IRAC slope, but the low SFR value derived by MAGPHYS places it in
the quiescent category of Tzanavaris \etal~(2010).

\noindent
Group 3: \newline
\indent
NGC 3226:  As an elliptical galaxy, NGC 3226's emission is dominated by its stars, although there appears to be 
a faint tidal feature in the 8\um~emission directed roughly to the north. NGC 3226 is particularly faint in the MIR relative to
its NIR emission and only has upper limits in the MIR-FIR from \emph{IRAS}. It has a LINER nucleus, but the decomposition
of its IR spectrum indicates no significant contribution form an AGN. 
\newline
\indent
NGC 3227: Along with its smaller companion NGC 3226, NGC 3227 was not observed by \emph{GALEX}, so we use 
UVOT data instead. Its IRS spectrum shows [Ne {\sc v}] emission consistent with a Seyfert nucleus. Its DECOMPIR fits 
are the worst of the nine galaxies, but have the highest AGN contribution 
of 15$-$25\% to the total IR luminosity and 45$-$60\% of the MIR luminosity. 
 
\noindent
Group 4: \newline
\indent
NGC 3395/3396: The apertures of this pair of galaxies overlapped sufficiently to make determination of the emission
belonging to each galaxy problematic. We therefore opted to treat the system as a combined system. This pair of galaxies is
distantly associated with Group 5. This group does not have UBV photometry. The SED shows little attenuation in 
the UV and strong 60\um~emission relative to the 100\um~emission. 

\noindent
Group 5: \newline
\indent
NGC 3424: This edge-on galaxy's central region becomes increasingly bright relative to  its disk with increasing wavelength. 
Similarly to NGC 3190, the other nearly-edge on galaxy, its NUV-FUV slope is quite steep and not particularly well fit by 
MAGPHYS. We used this galaxy as a test case for analyzing UVOT photometry and found good agreement with the 
\emph{GALEX} photometry. It lacks UBV photometry.  \newline
\indent
NGC 3430: Seen nearly face-on, this galaxy has a large and extended UV disk. This system provides a nice example of a system fairly 
early in the interaction sequence and hence fairly undisturbed morphologically. The IR peak of its SED is not very well constrained as 
this galaxy was not observed by PACS and did not have MIPS 70\um~and 160\um~fluxes available.

\noindent
Group 6: \newline
\indent
 NGC 3448: This is the larger member of this pair of dwarf galaxies. A bridge of emission extends from NGC 3448 in the direction of 
 UGC 6016, seen most prominently in the NUV. \newline
 \indent
 UGC 6016: This dwarf galaxy is very faint in the IR, but has a large but diffuse UV envelope. Due to its lack of 
 significant detections in several of our filters and in ancillary \emph{IRAS}, we cannot constrain the galaxy parameters very tightly. Its 
 distance is also the most uncertain of the sample as it did not have a distance modulus or a recessional velocity in the PSCz catalog.
 
 \noindent
 Group 7: \newline
 \indent
 NGC 3690/IC 694: Also known as Arp 299, this system is the most active in our sample, 
 with the highest amount of star formation and is our only LIRG, 
 showing a corresponding large amount of UV attenuation. Its  8\um~image is saturated, and it 
 does not have ancillary UBV fluxes.  
 
 \noindent
 Group 8: \newline
 \indent
 NGC 3786/88: The galaxies in this pair have very similar UV$-$NIR fluxes, but NGC 3788, which is more edge-on, 
 has higher fluxes in the \emph{Herschel} bands. However, their best-fit SEDs have similar infrared luminosities, likely due to the 
 relatively high contribution of the warm dust in the model for NGC 3786 compared to NGC 3788. Neither galaxy has ancillary
 \emph{IRAS} or UBV fluxes. NGC 3786 is a Seyfert galaxy, with a bright nuclear region showing [Ne {\sc v}] emission in its
 IRS spectrum and a partial ring of star-forming regions. Its decomposition requires
 an AGN contribution of 7\% and 26\% to the total IR luminosity and MIR luminosity respectively.
 
 \noindent
 Group 9: \newline
 \indent
 NGC 4038/4039: Also known as the Antennae, this pair of galaxies is one of our most evolved systems and, because we 
 cannot separate them, we treat it as a single entity.  Its clumpy distribution of star forming regions are clearly apparent at 
 8\um~and in the UV bands; the two nuclei are most clearly seen in 2MASS and IRAC, a reflection of the relative PAH 
 strengths to warm dust.
 
 \noindent
 Group 10: \newline
 \indent
 NGC 4618: Paired with NGC 4625 in a roughly equal-mass dwarf galaxy interaction, NGC 4618 has an off-center nucleus 
 with a single arm curving to the south-west, features seen in all the images. \newline
 \indent
 NGC 4625:  While relatively compact in the infrared, NGC 4625 has a faint, diffuse set of flocculent set of spiral arms 
 observed best in the NUV band. 
 
 \noindent
 Group 11: \newline
 \indent
 NGC 4647: This spiral galaxy is located at the edge of the IRAC field, which complicated the measurement of its flux because its 
 aperture, determined on the NUV image, extends past the edge of the IRAC image. We had to manually correct for
 the edge pixels without flux. We also do not have ancillary UBV fluxes. \newline
 \indent
 NGC 4649: A large elliptical galaxy also known as M60, NGC 4649 is very faint in the mid-infrared and absent in the 
 far-infrared, a dramatic contract to its companion. It has the lowest sSFR of the sample. 
 
 \noindent
 Group 12: \newline
 \indent
 M51A: The Whirlpool Galaxy is the larger galaxy in this well-studied system. It has quite consistent morphology
 across the wavelengths, but with more inter-arm filamentary emission and greater extent in the UV than in the IR. The
 decomposition of its IR spectrum is best fit without an AGN contribution, despite its LINER nucleus. This is one of the systems where the flux 
 from the smaller galaxy was subtracted from the aperture of the large galaxy.
 \newline
 \indent
 M51B: The smaller companion to the Whirlpool Galaxy, M51B is dominated by early-type stars and 
 has very little UV emission. Its FIR emission is confined to its nucleus.
  Its MIPS 160\um~measurement was a factor  of three lower 
 than the PACS measurement, and we opted to omit it from the fits.  

 \noindent
 Group 13: \newline
 \indent
 NGC 5394/95: The smaller galaxy of our most distant pair, NGC 5394 shows a beautiful pair of tidal tails, especially in the UV. 
 NGC 5395 has a LINER nucleus, which requires only a small AGN contributions of 3\% and 12\% to the galaxy's
 total IR and MIR luminosities respectively in the DECOMPIR fits.  This is the other system where the small galaxy flux needed to be subtracted from the large galaxy aperture.
 
\noindent
Group 14: \newline
\indent
M101: Another well-studied galaxy, its \emph{GALEX}, \emph{Spitzer}, and \emph{Herschel} images show clumpy star forming region 
structures along the spiral arms. Its large size made determining a single unbroken aperture with SExtractor complicated. As a result,
some of the outermost UV emitting regions were not included. \newline
\indent
NGC 5474: A small companion to M101, its core is  offset to the north from its center.

\end{document}